\PassOptionsToPackage{dvipsnames}{xcolor}
\documentclass[epjc3,twocolumn,nopacs]{svjour3}
\usepackage{amsmath,amsfonts,amssymb,cite,orcidlink,physics,siunitx,verbatimbox,xspace}
\usepackage[utf8]{inputenc}
\usepackage[T1]{fontenc}
\hypersetup{colorlinks=true,linkcolor=RedViolet,citecolor=RedViolet,filecolor=RedViolet,urlcolor=RedViolet}

\smartqed  
\journalname{Eur. Phys. J. C}

\makeatletter
\def\cl@chapter{\@elt {theorem}}
\makeatother
\usepackage{cleveref}

\newcommand{\squark}{{\tilde q}}

\newcommand{\gluino}{{\tilde g}}
\newcommand{\gaugino}{{\tilde \chi}}

\newcommand{\slepton}{{\tilde l}}

\newcommand{\resummino}{{\emph{Resummino}}\xspace}
\definecolor{darkmagenta}{rgb}{0.55, 0.0, 0.55}

\begin{document}

\title{Electroweak superpartner production at 13.6 TeV with \resummino}

\author{
  Juri Fiaschi\thanksref{e1,addr1}\orcidlink{0000-0003-0860-9569}
  \and\
  Benjamin Fuks\thanksref{e2,addr2}\orcidlink{0000-0002-0041-0566}
  \and\
  Michael Klasen\thanksref{e3,addr3}\orcidlink{0000-0002-4665-3088}
  \and\
  Alexander~Neuwirth\thanksref{e4,addr3}\orcidlink{0000-0002-2484-1328}
}

\thankstext{e1}{E-mail: {\color{RedViolet}fiaschi@liverpool.ac.uk}}
\thankstext{e2}{E-mail: {\color{RedViolet}fuks@lpthe.jussieu.fr}}
\thankstext{e3}{E-mail: {\color{RedViolet}michael.klasen@uni-muenster.de}}
\thankstext{e4}{E-mail: {\color{RedViolet}alexander.neuwirth@uni-muenster.de}}

\institute{Department of Mathematical Sciences, University of Liverpool, Liverpool L69 3BX, United Kingdom\label{addr1}
          \and
          Laboratoire de Physique Th\'eorique et Hautes \'Energies (LPTHE), UMR 7589, Sorbonne Universit\'e et CNRS, 4 place Jussieu, 75252 Paris Cedex 05, France\label{addr2}
          \and
          Institut  für  Theoretische  Physik,  Westfälische  Wilhelms-Universität  Münster,  Wilhelm-Klemm-Straße 9, 48149 Münster, Germany\label{addr3}
}

\date{Received: date / Accepted: date}

\maketitle

\vspace*{-8.0cm}
  \noindent {\small\texttt{MS-TP-23-14}}\\
  \noindent {\small\texttt{LTH 1339}}
\vspace*{7.0cm}

\begin{abstract}
  Due to the greater experimental precision expected from the currently ongoing LHC Run 3, equally accurate theoretical predictions are essential. We update the documentation of the \resummino package, a program dedicated to precision cross section calculations for the production of a pair of sleptons, electroweakinos, and leptons in the presence of extra gauge bosons, and for the production of an associated electroweakino-squark or electroweakino-gluino pair. We detail different additions that have been released since the initial version of the program a decade ago, and then use the code to investigate the impact of threshold resummation corrections at the next-to-next-to-leading-logarithmic accuracy. As an illustration of the code we consider the production of pairs of electroweakinos and sleptons at the LHC for centre-of-mass energies ranging up to \SI{13.6}{\tera\electronvolt} and in simplified model scenarios. We find slightly increased total cross section values, accompanied by a significant decrease of the associated theoretical uncertainties. Furthermore, we explore the dependence of the results on the squark masses.
\end{abstract}

\section{Introduction}
\label{sec:intro}
The Run~3 data taking period of the LHC has started in 2022 at an unprecedented centre-of-mass energy of $\sqrt{S} =\SI{13.6}{TeV}$. The two main general-purpose LHC experiments ATLAS and CMS are expected to collect an integrated luminosity of about \SI{300}{\per\femto\barn}, which will complement the \SI{140}{\per\femto\barn} already collected at \SI{13}{TeV} during Run~2. This increase in luminosity and centre-of-mass energy will thus make it possible to further explore extensions of the Standard Model (SM) of particle physics, such as supersymmetry (SUSY) and its minimal incarnation dubbed the Minimal Supersymmetric Standard Model (MSSM)~\cite{Nilles:1983ge, Haber:1984rc}. The latter is one of the most appealing options for new physics, as it answers various open questions of the SM and provides explanations to several of its conceptual limitations. These include, among others, the presence of dark matter in the universe, the unification of the strong and electroweak forces at large energy scales, and the stabilisation of the mass of the Higgs boson with respect to radiative corrections.

Experimental searches at the LHC have mainly focused on signatures arising from the production and decay of squarks and gluinos, as these particles can be strongly (and thus generally copiously) produced. For a long time, the signatures of these QCD-sensitive superparticles were consequently expected to be the first visible sign of supersymmetry in LHC data. However, with the associated mass limits being now deeply in the TeV regime, searches for electroweakinos and sleptons received more attention and became equally important. Subsequently, accurate theoretical calculations of signal cross sections and key kinematic distributions for all supersymmetric processes became imperative, and in  particular for processes in which at least one non-strongly interacting superpartner is present in the final state.

The \resummino program~\cite{Fuks:2013vua} has been developed in this context. It consists of a public tool computing precision predictions including soft-gluon radiation resummation effects for the production of a pair of sleptons, electroweakinos, and for the associated production of one electroweakino and either one squark or one gluino. Moreover, \resummino can also be used to calculate cross section predictions for the neutral-current or charged-current production of a pair of leptons in the presence of extra gauge bosons. So far, its predictions have been used by both the ATLAS and CMS collaborations in order to extract bounds on sleptons and electroweakinos. In particular, the most stringent constraints on simplified models inspired by the MSSM enforce viable slepton and electroweakino masses to be larger than about 700~GeV and 800--1200~GeV respectively, for a not too heavy lightest SUSY particle (see {\it e.g.} \cite{ATLAS:2019lff, ATLAS:2021yqv, CMS-PAS-SUS-21-008}). The exact values of these mass limits depend on the details of the search channels, and bounds can always be evaded by either compressing the particle spectrum (thus increasing the mass of the lightest SUSY state) or reducing the branching ratio in the final state of interest (by allowing numerous potential decay modes for a given SUSY particle). 

The production cross section of a pair of electroweakinos at hadron colliders has been studied in numerous renowned works, in which fixed-order predictions at leading order (LO)~\cite{Barger:1983wc, Dawson:1983fw} and next-to-leading order (NLO)~\cite{Beenakker:1999xh} have been considered. Further precision was obtained by matching these fixed-order results with either parton showers~\cite{Baglio:2016rjx, Frixione:2019fxg} (NLO+PS) or the threshold resummation of the next-to-leading logarithms~\cite{Debove:2010kf, Fuks:2012qx, Fiaschi:2018hgm} (NLO+NLL). Furthermore, approximate next-to-next-to-leading-order predictions have been recently matched with threshold resummation at the next-to-next-to-leading logarithmic accuracy~\cite{Fiaschi:2020udf} (aNNLO+NNLL). Similarly, slepton pair production total cross sections are known at LO~\cite{Baer:1993ew, Dawson:1983fw}, NLO~\cite{Beenakker:1999xh}, NLO+PS~\cite{Jager:2012hd, Frixione:2019fxg}, NLO+NLL~\cite{Bozzi:2007qr, Fuks:2013lya, Fiaschi:2018xdm} and aNNLO+NNLL~\cite{Fiaschi:2019zgh}.

Whereas bounds on viable mass regimes for squarks and gluinos imply that their pair production is now phase-space suppressed, their single production with a typically lighter electroweakino is still relevant and could even provide the best insights on both supersymmetric masses and interactions. Associated signals include in particular the associated production of hard jets with missing transverse energy, a signature well studied at the LHC in the context of searches for dark matter~\cite{Feng:2005gj, Bai:2010hh}. Associated production rates are known at LO~\cite{Dawson:1983fw} and NLO~\cite{Berger:1999mc, Berger:2000iu}, and the matching of these fixed-order results with threshold resummation has been recently achieved~\cite{Fuks:2016vdc, Fiaschi:2022odp}. Finally, fixed-order predictions for charged-current and neutral-current lepton pairs in the presence of additional gauge bosons are known at next-to-next-to-leading-order (NNLO)~\cite{Gavin:2010az, Gavin:2012sy}, and matched results at NLO+NLL accuracy are available as well~\cite{Fuks:2007gk, Jezo:2014wra}.

\resummino takes advantage of these developments of the last few decades. It combines the LO calculations for slepton-pair, electroweakino-pair, associated gluino-electroweakino and squark-electroweakino production available from~\cite{Bozzi:2004qq, Bozzi:2007me, Debove:2008nr} with the associated NLO SUSY-QCD corrections obtained in~\cite{Bozzi:2007qr, Debove:2010kf, Fuks:2016vdc, Fiaschi:2022odp}, and with the aNNLO QCD corrections of~\cite{Fiaschi:2019zgh, Fiaschi:2020udf}. These fixed-order predictions are next matched with the threshold resummation of soft gluon radiation to all orders and at varied accuracies~\cite{Fuks:2012qx, Fuks:2013lya, Fuks:2016vdc, Fiaschi:2019zgh, Fiaschi:2020udf, Fiaschi:2022odp}, according to the standard formalism introduced in~\cite{Sterman:1986aj, Catani:1989ne, Catani:1990rp, Kidonakis:1997gm, Kidonakis:1998bk, Vogt:2000ci} or the collinear-improved one of \cite{Kramer:1996iq, Catani:2001ic, Kulesza:2002rh, Almeida:2009jt}. NLO+NLL implementations of rates associated with lepton pair production in the presence of extra gauge bosons are available as well \cite{Fuks:2007gk, Jezo:2014wra}. In addition, the code can be employed to achieve NLO+NLL cross section computations in which fixed-order predictions are matched with soft gluon resummation in the small transverse-momentum ($p_T$) regime~\cite{Debove:2009ia, Bozzi:2006fw} following the formalism of \cite{Collins:1981va, Collins:1981uk, Collins:1984kg}, or jointly at small $p_T$ and close to threshold~\cite{Debove:2011xj, Bozzi:2007tea} following the formalism of \cite{Li:1998is, Laenen:2000de, Laenen:2000ij}. 

In the remainder of this manuscript, we begin by briefly outlining in section \ref{sec:resum} the threshold resummation formalism that is implemented in \resummino, and that is relevant for electroweakino-pair, slepton-pair and squark-electroweakino and gluino-electroweakino total production cross section calculations at the LHC. For interested readers, this short description is further complemented by  additional details including self-contained analytical formulas in~\ref{app:Resummation}. The installation and running of the \resummino code is described in section \ref{sec:running} and in \ref{app:code}-\ref{app:cli}. In section \ref{sec:num} and \ref{app:table}, we make use of \resummino to compute and document for the first time total production rates obtained for the LHC Run~3, operating at a centre-of-mass energy of \SI{13.6}{TeV}. We consider simplified model scenarios in which all superparticles are decoupled, excepted for those produced in the final state. Our results highlight how theoretical uncertainties are reduced relative to the perturbative order of the fixed-order and resummed component of the matched predictions and how they compare with predictions at a centre-of-mass energy of \SI{13}{TeV}. Moreover, we explore next-to-minimal scenarios, and discuss the impact of internal squark masses on the predictions for configurations in which squarks are not decoupled but only slightly heavier than the lighter electroweakinos. We summarise our work in section \ref{sec:summary}.
\section{Threshold resummation at aNNLO+NNLL}
\label{sec:resum}
Within its first public release, the \resummino package was suitable for NLO+NLL resummation calculations in the threshold regime, as well as at small transverse momentum or in both regimes simultaneously. Since then, the implementation  of threshold-resummed cross sections has been updated so that corrections up to aNNLO+NNLL could be calculated. For that reason, we provide below a description of the formalism used for threshold resummation up to aNNLO+NNLL, and we refer instead to \cite{Fuks:2013vua} for details on $p_T$ and joint resummation.

We consider the production of a pair of heavy particles $i$ and $j$ in hadronic collisions through the process $AB\to ij$ (in which $A$ and $B$ stand for the initial hadrons). Whereas soft and collinear divergences originating from real and virtual corrections cancel in the perturbative expansion of the associated production cross section, logarithmic terms remain due to the different phase spaces inherent in the different ingredients of the calculation~\cite{Kinoshita:1962ur, Lee:1964is}. These logarithmic contributions encode the effects of soft and collinear emission from either initial-state or final-state coloured particles. As the partonic energy approaches the production energy threshold, they become large and can hence spoil the convergence of fixed-order calculations. 

This can be cured through QCD resummation techniques that rely on dynamical and kinematic factorisation of the cross section to account for soft and collinear radiation to all orders~\cite{Collins:1989gx}. In practice, these factorisation properties are exploited in Mellin space. The hadronic differential cross section $\dd \sigma_{AB}/\dd M^2$ is expressed in terms of the Mellin variable $N$, that is conjugate to the quantity $\tau=M^2/S$ (with $M$ being the invariant mass of the produced particles in the Born process and $S$ the hadronic centre-of-mass energy). This hadronic cross section can be written as a product of the densities $f_{p/H}$ of parton $p$ in hadron $H$ (in Mellin space, where the corresponding Mellin moments are obtained relative to the momentum fraction $x$), and of the corresponding partonic cross section $\sigma_{ab\to ij}$ (whose Mellin moments are derived relative to the variable $z=M^2/s$ with $s$ being the partonic centre-of-mass energy),
\begin{equation}
\begin{split}
	&M^2 \frac{\dd \sigma_{AB\to ij}}{\dd M^2} (N-1) = \sum_{a,b}\ f_{a/A}(N,\mu_F^2)\\ 
	&\qquad \qquad \times f_{b/B}(N,\mu_F^2)\  \sigma_{ab\to ij}(N,M^2,\mu_F^2,\mu_R^2)\,.
\label{eq:HadFacN}
\end{split}
\end{equation}
Here the logarithms depend on the Mellin variable $N$ and they become large in the large-$N$ limit, and $\mu_F$ and $\mu_R$ stand for the usual factorisation and renormalisation scales. 

After accounting for soft-gluon emission to all orders, the partonic cross section can be re-expressed in a closed exponential form scaled by a hard function $\mathcal H$~\cite{Sterman:1986aj, Catani:1989ne, Catani:1990rp, Kidonakis:1997gm, Kidonakis:1998bk, Vogt:2000ci},
\begin{equation}
\begin{split}
	&\sigma^{\text{res.}}_{ab\to ij}(N, M^2, \mu_F^2, \mu_R^2) = 
   \mathcal H_{ab \to ij}(M^2,\mu_F^2, \mu_R^2)\\
	&\qquad\qquad \times \exp\Big[ G_{ab\to ij}(N,M^2,\mu_F^2, \mu_R^2) \Big]\,.
	\label{eq:G}
\end{split}
\end{equation}
The $N$-independent hard function $\cal H$ can be written in terms of the LO Mellin-transformed cross section $\sigma^{(0)}_{ab\to ij}$ and the hard matching coefficient $C_{ab\to ij}$,
\begin{equation}
\begin{split}
	\mathcal H_{ab \to ij}(M^2, \mu_F^2, \mu_R^2) \!=\! \sigma^{(0)}_{ab\to ij} C_{ab\to ij} (M^2,\mu_F^2, \mu_R^2)\,,
	\label{eq:H}
\end{split}
\end{equation}
where the coefficient $C_{ab\to ij}$ can be computed perturbatively,
\begin{equation}
  {C}_{ab\to ij} (M^2\!, \mu_F^2, \mu_R^2) \!=\!  \sum_{n=0} \left(\frac{\alpha_s}{2\pi}\right)^{\!\!n}\! {C}_{ab}^{(n)} (M^2\!, \mu_F^2, \mu_R^2)\,.
\label{eq:Cab}\end{equation}
The hard matching coefficients ${C}_{ab}^{(n)}$ are then derived from fixed-order predictions in Mellin space at a given order in the strong coupling $\alpha_s$. They correspond to the ratio of the finite $N$-independent pieces of the N$^n$LO correction terms over the LO cross section,
\begin{equation}
    {C}_{ab}^{(n)} (M^2, \mu_F^2, \mu_R^2) = \left(\frac{2\pi}{\alpha_s}\right)^n \left[\frac{\sigma^{(n)}_{ab\to ij}}{\sigma^{(0)}_{ab\to ij}}\right]_{\textrm{N-ind.}}
	\label{eq:C}
\end{equation}

The soft and collinear gluon radiation contributions appearing in \eqref{eq:G} are included in the so-called Sudakov exponent $G_{ab\to ij}$. They depend on the quark/gluon nature of the initial state and can be written at NNLL accuracy as a sum of leading logarithmic (LL) terms $G^{(1)}_{ab}$, NLL terms $G^{(2)}_{ab\to ij}$ and NNLL terms $G^{(3)}_{ab\to ij}$. This yields
\begin{equation}
\begin{split}
  & G_{ab\to ij}(N,M^2, \mu_F^2, \mu_R^2) \approx \underbrace{L\ G^{(1)}_{ab}(N)}_{\text{LL}} \\
  & \qquad + \underbrace{G^{(2)}_{ab\to ij}(N,M^2,\mu_F^2, \mu_R^2)}_{\text{NLL}} \\
  & \qquad + \underbrace{\alpha_s G^{(3)}_{ab\to ij}(N,M^2,\mu_F^2, \mu_R^2)}_{\text{NNLL}}\,, \label{eq:HtimesG}
\end{split}
\end{equation}
with $L=\ln(N e^{\gamma_E})$. 

For all processes implemented in \resummino, we provide in \ref{app:Resummation} explicit expressions for the resummation coefficients $G_{ab\to ij}^{(n)}$ and the hard matching coefficients ${C}_{ab\to ij}^{(n)}$. In principle, all above expressions should be generalised and include an index referring to a given irreducible colour representation, and the total rate given in \eqref{eq:G} should embed a sum over all possible structures emerging from the colour decomposition of the LO partonic cross section in Mellin space~\cite{Beenakker:2013mva}. However, all the processes considered, namely the production of a pair of colourless particles (electroweakino-pair and slepton-pair production) and that of the associated production of one colourless and one coloured particle (electroweakino-squark and electroweakino-gluino production), only feature a single colour representation. All analytical expressions therefore simplify, and the colour representation indices can be omitted.

In order to obtain meaningful predictions over the entire phase space, the resummed cross section has to be matched to its fixed-order counterpart. This is achieved by subtracting from the sum of the resummed total rate $\sigma_{ab}^{\text{res.}}$ and the fixed-order one $\sigma_{ab}^{\text{f.o.}}$ their overlap, that is given by the expansion $\sigma_{ab}^\text{exp.}$ of the resummed result at the same order in $\alpha_s$ as that of the fixed-order calculation,
\begin{align}
	\sigma_{ab} = \sigma_{ab}^\text{f.o.} + \sigma_{ab}^{\text{res.}}  - \sigma_{ab}^\text{exp.}\,.
\label{eq:match}\end{align}
In this expression, fixed-order predictions are computed in physical space, whereas the resummed component and its expansion to given order in $\alpha_s$ are calculated in Mellin space. This therefore requires to convolve the associated partonic cross sections with Mellin-transformed parton distribution functions (PDFs). In \resummino, this is achieved by employing the parametrisation of the MSTW collaboration~\cite{Martin:2009iq} to fit the used PDFs in $x$-space, and thereby obtain an analytic formula for its expression in Mellin space.\footnote{This procedure has been shown to lead to good fits of various global PDF sets, such as the NLO sets of the CT14~\cite{Dulat:2015mca},  CT18~\cite{Hou:2019efy}, MMHT2014~\cite{Martin:2009iq}, MSHT20~\cite{Bailey:2020ooq} and NNPDF4.0 \cite{NNPDF:2021njg} PDF fits.} Finally, an inverse Mellin transformation has to be performed in order to go back to physical space. This is achieved by choosing a distorted integration contour in the complex plane inspired by the principal value and minimal prescription procedure~\cite{Contopanagos:1993yq,Catani:1996yz}. Along this contour, the Mellin variable $N$ is parameterised by
\begin{equation}
    N = C y \ \exp\big[ \pm i \varphi \big] \quad\text{with}\ \varphi \in [\pi/2, \pi] \ \text{and}\  y \geq 0\,,
\end{equation}
so that all singularities that may appear in the inverse transform process are correctly taken care of. The parameter $C$ is indeed chosen so that the poles in the PDF Mellin moments originating from the Regge singularity are on the left of the contour, and that stemming from the Landau pole of the running of $\alpha_s$ is on its right.

\section{Running \resummino}
\label{sec:running}
\subsection{Installation}

The \resummino package is a high-energy physics program whose source code is written in C++, that is publicly available for download at \url{https://resummino.hepforge.org/}, and that is licensed under the European Union Public Licence v1.1. It can be compiled with the GNU compiler collection (GCC) and a working installation of \texttt{CMake} (version~3.0 or more recent) \cite{cmake}. It requires as external dependencies the GNU Scientific Library \texttt{GSL} (version 2.0 or more recent) \cite{Gough2009-ji} and the \texttt{Boost} package (version 1.70.0 or more recent) \cite{boost}. The former allows \resummino\ to make use of the \verb+VEGAS+ Monte Carlo routines for numerical integration~\cite{Lepage:1980dq}, and of the Levenberg-Marquardt algorithm \cite{10.2307/43633451, 10.2307/2098941} to fit non-linear functions like parton distribution functions~\cite{slhaea}. On the other hand, the \texttt{Boost} package is a dependency necessary to read supersymmetric spectra encoded in the SLHA format~\cite{Skands:2003cj, Allanach:2008qq} through the \texttt{SLHAea} library~\cite{slhaea}. In addition, \resummino must be linked to \textsc{LHAPDF} (version 6.0 or more recent)~\cite{lhapdf} for PDF handling, and \textsc{LoopTools} (version 2.15 or more recent)~\cite{Hahn:1998yk} for the evaluation of one-loop integrals.

The code can easily be installed by typing the following commands in a shell, once the \resummino zipped sources have been downloaded:
\begin{verbatim}
    unzip Resummino-X.Y.Z.zip
    cd Resummino*
    mkdir build
    cd build
    cmake .. -B . [options]
    make
    make install
\end{verbatim}
The last step is optional, and it is only relevant to install the code system-wide. The \verb+cmake+ command can be cast with options dictating the behaviour of \resummino relative to its two external dependencies \textsc{LHAPDF} and \textsc{LoopTools}. The program is equipped with \textsc{LHAPDF} version~6.2.3 and \textsc{LoopTools} version~2.15, and the usage of those built-in libraries can be enforced through the \verb+cmake+ options \texttt{-DBUILD\_LHAPDF=ON} (default: \texttt{OFF}) and \texttt{-DBUILD\_LOOPTOOLS=ON} (default: \texttt{ON}) respectively. Different versions of these packages can be used by providing information about the path to existing installations through the options \texttt{-DLHAPDF=PATH} and \texttt{-DLOOPTOOLS=PATH} if libraries and headers are installed in the same folder, or through \texttt{LHAPDF\_INCLUDE\_DIR}/\texttt{LHAPDF\_LIB\_DIR} and \texttt{LOOPTOOLS\_INCLUDE\_DIR}/\texttt{LOOPTOOLS\_LIB\_DIR} if not.

The \resummino binary, called \verb+resummino+, is located in the folder \texttt{build/bin}. After completion of compilation, users can test their local installation by running (from the installation folder)
\begin{verbatim}
    ./bin/resummino --help
    ./bin/resummino input/resummino.in
\end{verbatim}
The source files of the code are collected in the folder \texttt{src} whose architecture and  structure is described in \ref{app:code}. The folder \texttt{input} includes exemplary input files, whereas the folder \texttt{external} contains necessary external packages such as \textsc{LoopTools} and \textsc{LHAPDF}. Finally, the folder \texttt{scripts} provides a set of \texttt{docker} and \texttt{apptainer} scripts related to the usage of portable installations of \resummino.

\subsection{Running \resummino and input parameters}\label{sec:input}

\resummino can be run as exemplified at the end of the previous section, by typing in a shell, from the installation folder, the command
\begin{verbatim}
    ./bin/resummino <some-input-file>
\end{verbatim}
The keyword \verb+<some-input-file>+ provides the path to a configuration file indicating what to calculate and how. Details on all the options available to write such an input configuration file are briefly provided in the rest of this subsection, and they are additionally extensively documented in~\ref{app:input}. Moreover, an example input file \verb+input/resummino.in+ is shipped with the code, and the output obtained from running the code with this input file is described in~\ref{app:output}.

\subsubsection{Calculation, process and collider settings}\label{subsec:ctype}

A configuration for \resummino includes a list of equalities fixing the values of certain variables acting on how the code functions. This file begins with a definition of the process to be considered, including details on the collider environment. Users must specify the type of colliding particles (protons or antiprotons), the centre-of-mass energy, and the nature of the final-state particles. This is achieved through the self-explanatory variables \texttt{collider\_type} (to be set to \texttt{proton-proton} or \texttt{proton-antiproton}), \texttt{center\_of\_mass\_energy} (to be given in GeV), \texttt{particle1} and \texttt{particle2} (provided through their Particle Data Group (PDG) identifiers~\cite{ParticleDataGroup:2022pth}). For example, the following settings,
\begin{verbatim}
    collider_type = proton-proton
    center_of_mass_energy = 13600
    particle1 = 1000024
    particle2 = -1000024
\end{verbatim}
would correspond to using \resummino for calculations relevant to LHC proton-proton collisions at a centre-of-mass energy of \SI{13.6}{TeV}, and leading to the production of a pair of charginos ($pp\to \tilde\chi_1^+\tilde\chi_1^-$ at $\sqrt{S}=\SI{13.6}{TeV}$).

\begin{table}
\renewcommand{\arraystretch}{1.3} \setlength{\tabcolsep}{10pt}\centering
\begin{tabular}{@{\extracolsep{\fill}}cccc@{}}
\verb+result+ &  \verb+total+, \verb+m+ & \verb+pt+, \verb+ptj+ & \verb+total+, \verb+m+  \\ order &  NLO+NLL & NLO+NLL& aNNLO+NNLL  \\ \hline
$pp\to\gaugino\gaugino$ &  \cite{Debove:2010kf,Fuks:2012qx,Fiaschi:2018hgm}  & \cite{Debove:2009ia,Debove:2011xj}  & \cite{Fiaschi:2020udf}\\ \hline
$pp\to\slepton\slepton$ &  \cite{Bozzi:2007qr,Fiaschi:2018xdm}    & \cite{Bozzi:2006fw,Bozzi:2007tea}  & \cite{Fiaschi:2019zgh}\\ \hline
$pp\to ll$               &  \cite{Jezo:2014wra}    & \cite{Jezo:2014wra} & \checkmark\\ \hline
$pp\to\gaugino\squark$  &  \cite{Fiaschi:2022odp}   &  \\ \hline
$pp\to\gaugino\gluino$  &  \cite{Fuks:2016vdc}      &  \\
\end{tabular}
\caption{List of processes included in \resummino, given together with the orders of the different calculations supported by the code (defined as the value of the \texttt{result} variable discussed in the text) and the associated references.}
\label{tab:processes}
\end{table}

Information on the observable to calculate is passed through the variable \verb+result+, that can be set to \verb+total+ (total cross section according to the threshold resummation formalism), \verb+pt+ or \verb+ptj+ (differential cross section at fixed transverse momentum $p_T$ according the $p_T$ or joint resummation formalism respectively), or \verb+m+ (differential cross section at fixed invariant mass $M$ according to the threshold resummation formalism). Differential calculations make use of the variables \verb+pt+ and \verb+M+ to get the numerical values to employ for $p_T$ and $M$, respectively, that are provided in GeV. For example,
\begin{verbatim}
    result = total
\end{verbatim}
defines a total cross section calculation, whereas
\begin{verbatim}
    result = ptj
    pt     = 50
\end{verbatim}
estimates $\dd\sigma/\dd p_T$ for a final-state transverse momentum of \SI{50}{GeV}, using the joint resummation formalism (including therefore integration upon the invariant mass $M$). In addition, for the Drell-Yan production of a lepton pair, an invariant-mass cut is required to regularise phase-space integration when total rates are evaluated. Information on such a cut is passed through the parameters \texttt{Minv\_min} and \texttt{Minv\_max}. The list of calculations supported in \resummino is summarised in table~\ref{tab:processes}, together with the key references documenting their implementation.

The numerical precision of the performed calculation can be controlled through the input variables \texttt{precision} and \texttt{max\_iters} acting on the \texttt{VEGAS} algorithm. The former is related to the relative precision of the numerical integration process, whereas the latter refers to the maximum number of integration iterations (excluding the warm-up phase) allowed before stopping the calculation. The settings
\begin{verbatim}
    precision = 0.01
    max_iters = 5
\end{verbatim}
enforces a numerical precision of 1\% and imposes that at most five \texttt{VEGAS} iterations are used (even if the desired precision is not reached).

Next, details about the PDF set to be used in LO and higher-order calculations are provided through a few variables. These variables define the name of the \textsc{LHAPDF} set of parton densities to employ (through the self-explanatory variables \texttt{pdf\_lo} and \texttt{pdf\_nlo}\footnote{See the webpage \url{https://lhapdf.hepforge.org/pdfsets} for the list of possible names.}), and the exact identifier of the set member considered (specified as an integer through the variables \texttt{pdfset\_lo} and \texttt{pdfset\_nlo}). The cross section values spanned after considering all set members included in a given \textsc{LHAPDF} set (obtained by means of multiple runs of \resummino) then allow for PDF error assessments following standard Hessian or Monte Carlo prescriptions~\cite{Butterworth:2015oua}. For instance, including in the input file
\begin{verbatim}
    pdf_lo     = CT14lo
    pdfset_lo  = 0
    pdf_nlo    = CT14nlo
    pdfset_nlo = 0
\end{verbatim}
enforces the usage of CT14 parton densities~\cite{Dulat:2015mca} for all calculations. The central LO set \texttt{CT14lo} will be used for LO predictions, whereas both higher-order fixed-order predictions and resummed ones will rely on the central NLO set \texttt{CT14nlo}.

As briefly discussed in section~\ref{sec:resum}, Mellin-transformed PDF are obtained from a fit of the PDFs considered to the MSTW parameterisation of~\cite{Martin:2009iq}. This relies on a logarithmic sampling of Bjorken-$x$ values from a minimum value $x_{\rm min}$ to 1, and on a fitting method using weighted least squares as achieved by the Levenberg-Marquardt algorithm (handled through the \texttt{GSL} library). The minimum value $x_{\rm min}$ can be provided through the input variable \texttt{xmin}, and the weights used in the fitting procedure are given through the variables \texttt{weight\_valence}, \texttt{weight\_sea} and \texttt{weight\_gluon} for valence quark, sea quark and gluon PDFs respectively. The default option corresponds to
\begin{verbatim}
    weight_valence = -1.6
    weight_sea     = -1.6
    weight_gluon   = -1.6
    xmin           = auto
\end{verbatim}

Finally, the value of the renormalisation and factorisation scales is taken to be $\mu_F = \mu_R = (m_1+m_2)/2$ for the computation of total cross sections and $p_T$ distributions, with $m_1$ and $m_2$ being the masses of the two final-state particles. For invariant-mass distributions, the choice $\mu_F = \mu_R = M$ is adopted instead, unless the \texttt{FIXED\_SCALE} flag is turned on in the file \texttt{src/resummino.cc}. As required for the evaluation of scale uncertainties, users have the possibility to multiply these values by specific factors via the input variables \texttt{mu\_f} and \texttt{mu\_r}. For instance, a calculation corresponding to $\mu_F = (m_1+m_2)/4$ and $\mu_R = (m_1+m_2)$ would be achieved through
\begin{verbatim}
    mu_f = 0.5
    mu_r = 2.0
\end{verbatim}

\subsubsection{Model free parameters} \label{sec:params}
Information on the SUSY spectrum and interactions is passed to \resummino through a standard SLHA~\cite{Skands:2003cj, Allanach:2008qq} file whose path is specified in the input file through the variable \verb+slha+. In the case of the charged-current or neutral-current Drell-Yan process in the presence of additional gauge bosons, information on the SM electroweak parameters and on the $W'$ and $Z'$ properties is provided through a file encoded following an SLHA-like structure, and whose path is passed through the variable \verb+zpwp+ as described in~\cite{Jezo:2014wra}. An example of such a file and its structure is provided in \verb+input/ssm.in+.

By default, the electroweak input parameters are determined from the information included in the provided SLHA file.\footnote{In the case of non-consistent information in the blocks \texttt{SMINPUTS} and \texttt{MASS}, priority is given to the former.} For SUSY calculations \resummino makes use of tree-level formulas, 
\begin{equation}\begin{split}
	&G_F = \SI{1.166379e-5}{GeV^{-2}}\,, \\
	&\sin^2 \theta_w = 1-\frac{M_W^2}{M_Z^2}\,, \\
	&g_2 = 2 M_W \sqrt{\sqrt 2 G_F}\,.
\end{split}\end{equation}
following the prescription introduced in \texttt{Prospino}~\cite{Beenakker:1996ed}. In these expressions, $M_W$ and $M_Z$ stand for the $W$-boson and $Z$-boson masses, $G_F$ is the Fermi constant, $g_2$ is the weak coupling constant, and $\theta_w$ is the electroweak mixing angle. All Yukawa couplings of light quarks (for a number of active quark flavours $n_f=5$) are set to zero, and the strong coupling constant $\alpha_s$ is computed using the value returned by \textsc{LHAPDF}. Consequently, \resummino does not use the values of $\alpha_s$ and $\alpha$ included in the \textsc{SMINPUTS} block of the SLHA input file. For computations in the presence of extra gauge bosons, the code uses instead
\begin{equation}\begin{split}
    &\sin^2_{\theta_W} = 1 - \frac{M_W^2}{M_Z^2}\,, \\
    &g_1 = \sqrt{4\pi\alpha_{\rm em}} / \cos\theta_W\,, \\
    &g_2 = \sqrt{4\pi\alpha_{\rm em}} / \sin\theta_W\,,
\end{split}\end{equation}
where $g_1$ stands for the hypercharge coupling constant. Users have the possibility to change the electroweak parameter scheme through the different macros implemented in the file \texttt{src/params.cc}. For instance, the flag \texttt{GAUGE\_COUPLING} can be switched on to use as inputs the values of the SM gauge couplings encoded in the SLHA block \textsc{GAUGE}, or the flag \texttt{MADGRAPH\_COUPLING} can be turned on to select the electroweak parameters as in~\cite{Frixione:2019fxg}.

\subsection{The command line interface of \resummino}
Several of the configuration settings of the code can be fixed through optional arguments when casting the \resummino executable command in a shell, as in 
\begin{verbatim}
    ./bin/resummino <input-file> [options]
\end{verbatim}
Available options facilitate, for example, iterations over different PDF sets, calculations at different perturbative orders, or the implementation of similar calculations for various processes. Users do not therefore have to manually modify the input file (generically denoted by \verb+<input-file>+ in the above command), as any parameter value passed through the command line supersedes that included in the input file. The full list of options being detailed in \ref{app:cli}, we only briefly introduce them below.

The perturbative order of the calculation can be fixed through the options \texttt{-{}-lo} (or \texttt{-l}, for LO-accurate calculations), \texttt{-{}-nlo} (or \texttt{-n}, for NLO-accurate calculations) and \texttt{-{}-nll} (or \texttt{-s}, for NLO+NLL-accurate calculations), as well as through \texttt{-{}-nnll} (or \texttt{-z}) for aNNLO+NNLL-accurate calculations when available (namely for implemented Drell-Yan-like processes). Casting the \resummino command with the \texttt{-{}-center\_of\_mass\_energy} flag (or \texttt{-e}) followed by a double-precision number allows for a modification of the collider hadronic centre-of-mass energy value (in GeV). Similarly, the $p_T$ and $M$ values (in GeV) relevant for calculations at fixed transverse momentum or fixed invariant mass can be fixed through the two options \texttt{-{}-transverse-momentum} (or \texttt{-t}) and \texttt{-{}-invariant-mass} (or \texttt{-m}), both followed by a double-precision number. Additionally, the PDG identifiers of the final state particles can be modified through the options \texttt{-{}-particle1} (or \texttt{-c}) and \texttt{-{}-particle2} (or \texttt{-d}), both followed by an integer.

The evaluation of the theory uncertainties associated with any given calculation can be easily performed through the modification of the PDF set considered and the chosen unphysical scales. This is achieved through the options \texttt{-{}-pdfset\_lo} (or \texttt{-a}) and \texttt{-{}-pdfset\_nlo} (or \texttt{-b}), both followed by an integer, and \texttt{-{}-mu\_f} (or \texttt{-f}) and \texttt{-{}-mu\_r} (or \texttt{-r}), both followed by a double-precision number. The former two flags allow users to change the identifier of the PDF member set (within a specific \textsc{LHAPDF} collection of parton densities) used for calculations at LO and beyond, whereas the latter two flags provide multiplicative factors to include when fixing the factorisation and renormalisation scales.

In addition, the code can also be started with the options \texttt{-{}-version} (or \texttt{-v}), \texttt{-{}-help} (or \texttt{-h}), \texttt{-{}-parameter-log} (or \texttt{-p}) followed by a string, and \texttt{-{}-output} (or \texttt{-o}) followed by a string. The first of these displays the \resummino release number to the screen, whereas the second of them prints a help message indicating how to run the code. The third possibility makes the code writing all the parameters inherent to the calculation considered in a file (whose path is provided through the included string), and the last one defines the directory (defined through the included string) in which all files created by the code on run time are stored.

\subsection{Docker containers}

We provide convenient `docker’ containers that allow users to employ \resummino without the need to compile it from its source code. This requires to have either \texttt{Docker} or \texttt{Singularity/Apptainer} available on the system, the latter choice being recommended for secured high-performance computing. In this case, \resummino\ can be run from the scripts localised in the \texttt{scripts} folder,
\begin{verbatim}
    # if Docker is available
    source scripts/docker_alias.sh 

    # if Singularity/Apptainer is available
    source scripts/apptainer_alias.sh 
\end{verbatim}

These scripts automatically detect a local \textsc{LHAPDF} installation, and then enable the usage of any PDF set available within the system. If no installed version of \textsc{LHAPDF} is found, then \resummino is restricted to make use of the default PDF sets included within the docker images. New PDFs can always be added via the command `\verb+lhapdf install+’, to be cast within a shell. In addition, when it is run from a script, \resummino only accepts input files located in the current directory. Such a behaviour can however be circumvented through modifications of the \verb+docker run+ command's bound directories, through its usual \verb+-v+ option.
\section{Precision predictions for electroweak SUSY processes in simplified models}
\label{sec:num} \label{sec:models}

In this section we provide state-of-the-art cross section predictions for slepton and electroweakino pair production at the LHC Run~3 operating at an increased centre-of-mass energy $\sqrt{S}$ = \SI{13.6}{TeV}, and we compare our findings with predictions relevant for the LHC Run-2 at $\sqrt{S}$ = \SI{13}{TeV}. We employ the PDF4LHC21\_40 set of parton distribution functions~\cite{PDF4LHCWorkingGroup:2022cjn}, and we provide results together with the associated PDF and scale uncertainties (the latter being obtained with the seven-point method) added in quadrature. Complementary to the figures shown in this section, the complete collection of numerical predictions are shown in the tables of~\ref{app:table}.

We consider simplified SUSY models inspired by the MSSM, and we explore several typical scenarios. In the context of slepton pair production, we focus on a new physics configuration in which all SUSY particles are decoupled by setting their masses at \SI{100}{TeV}, with the exception of a single slepton species that is taken either left-handed ($\tilde e_L$), right-handed ($\tilde e_R$) or maximally mixed ($\tilde\tau_1 = 1/\sqrt{2} \big[\tilde\tau_L + \tilde\tau_R\big]$). This last scenario is representative of models featuring light tau sleptons, as originating from many SUSY scenarios~\cite{Fuks:2013lya}.

Electroweakino pair production rates are estimated in similar scenarios, in which all SUSY particles are decoupled with the exception of the lightest electroweakinos~\cite{Fuks:2012qx}. We begin our study with scenarios in which the three lightest electroweakinos are all higgsinos. In a first setup, all higgsinos are taken mass-degenerate and the lightest states are defined by
\begin{equation}
 \tilde \chi_{1,2}^0 \sim \frac {1}{\sqrt{2}} \left( \tilde H^0_u \pm \tilde H^0_d \right)\,,\qquad 
 \tilde \chi_{1}^\pm \sim  \tilde H_{u,d}^\pm\,.
\label{eq:hino}\end{equation}
Whereas this choice of a real neutralino mixing matrix implies a negative $m(\chi^0_1)$ eigenvalue, it can always be transformed back to positive a mass eigenvalue through a chiral rotation~\cite{Skands:2003cj}. In a second setup, we introduce some mass splitting between the three electroweakinos, such a splitting being typical of next-to-minimal electroweakino simplified models studied at the LHC, and that turn out to be more realistic in the light of concrete MSSM scenarios~\cite{Fuks:2017rio}. In this case, we define the three lightest electroweakinos by
\begin{equation}
 \tilde \chi_{1,2}^0 \sim -\frac {1}{\sqrt{2}} \left( \mp\tilde H^0_u + \tilde H^0_d \right)\,,\qquad 
 \tilde \chi_{1}^\pm \sim  \tilde H_{u,d}^\pm\,.
\label{eq:hino2}\end{equation}
Finally, we consider a scenarios in which all lightest electroweakinos are mass-degenerate gauginos,
\begin{equation}
 \tilde \chi_{1}^0 \sim i \tilde B\,,\qquad 
 \tilde \chi_{2}^0 \sim i \tilde W^3\,,\qquad 
 \tilde \chi_{1}^\pm \sim  \tilde W^\pm\,.
\label{eq:gino}\end{equation}
In this last SUSY configuration, we additionally investigate squark mass effects on gaugino pair production. Here, we consider an eight-fold degeneracy of all first-generation and second-generation squarks, the spectrum featuring thus a large number of states reachable at the LHC.

\begin{figure}\centering
	  \includegraphics[width=\columnwidth]{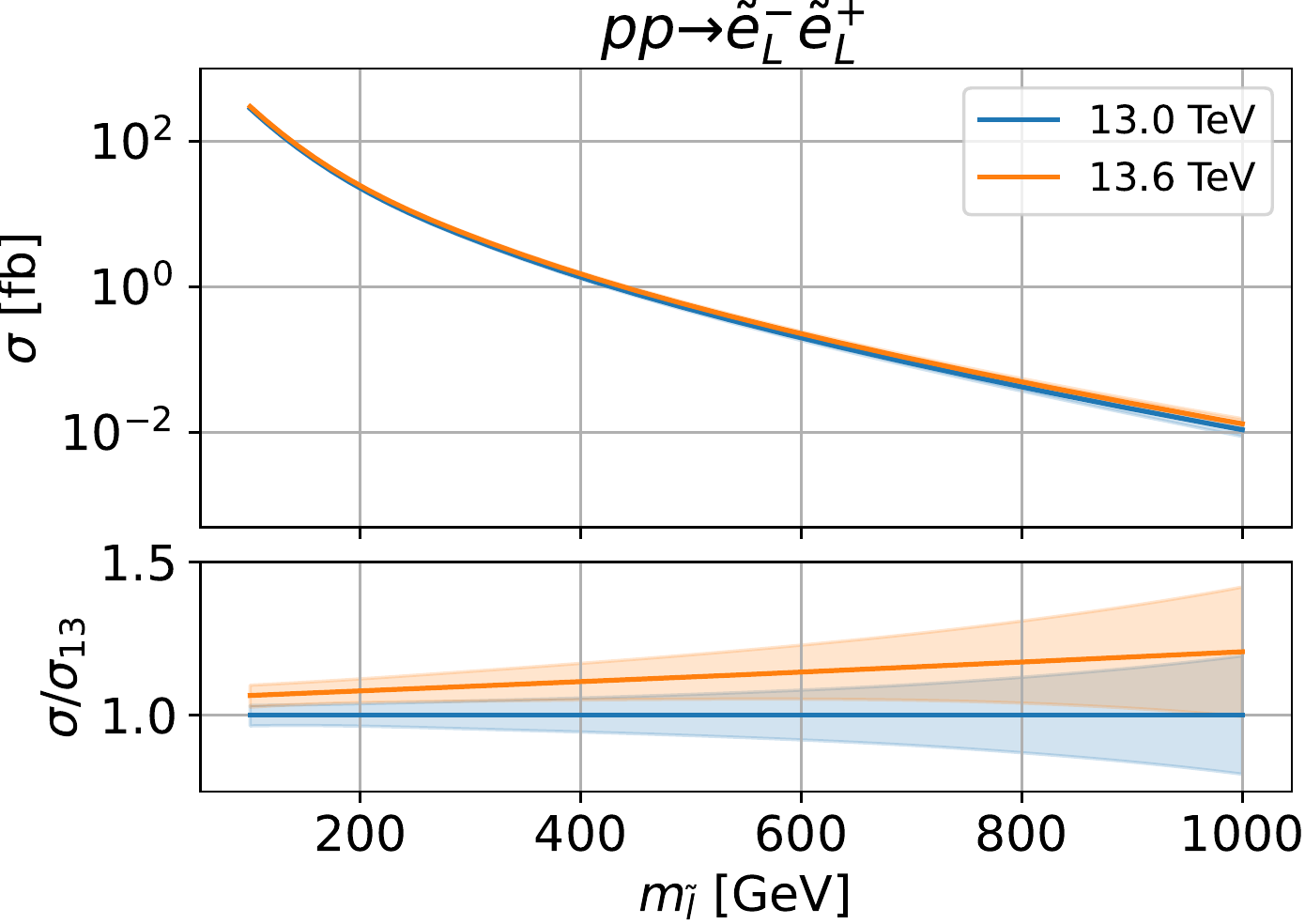}\vspace{.6cm}
    \includegraphics[width=\columnwidth]{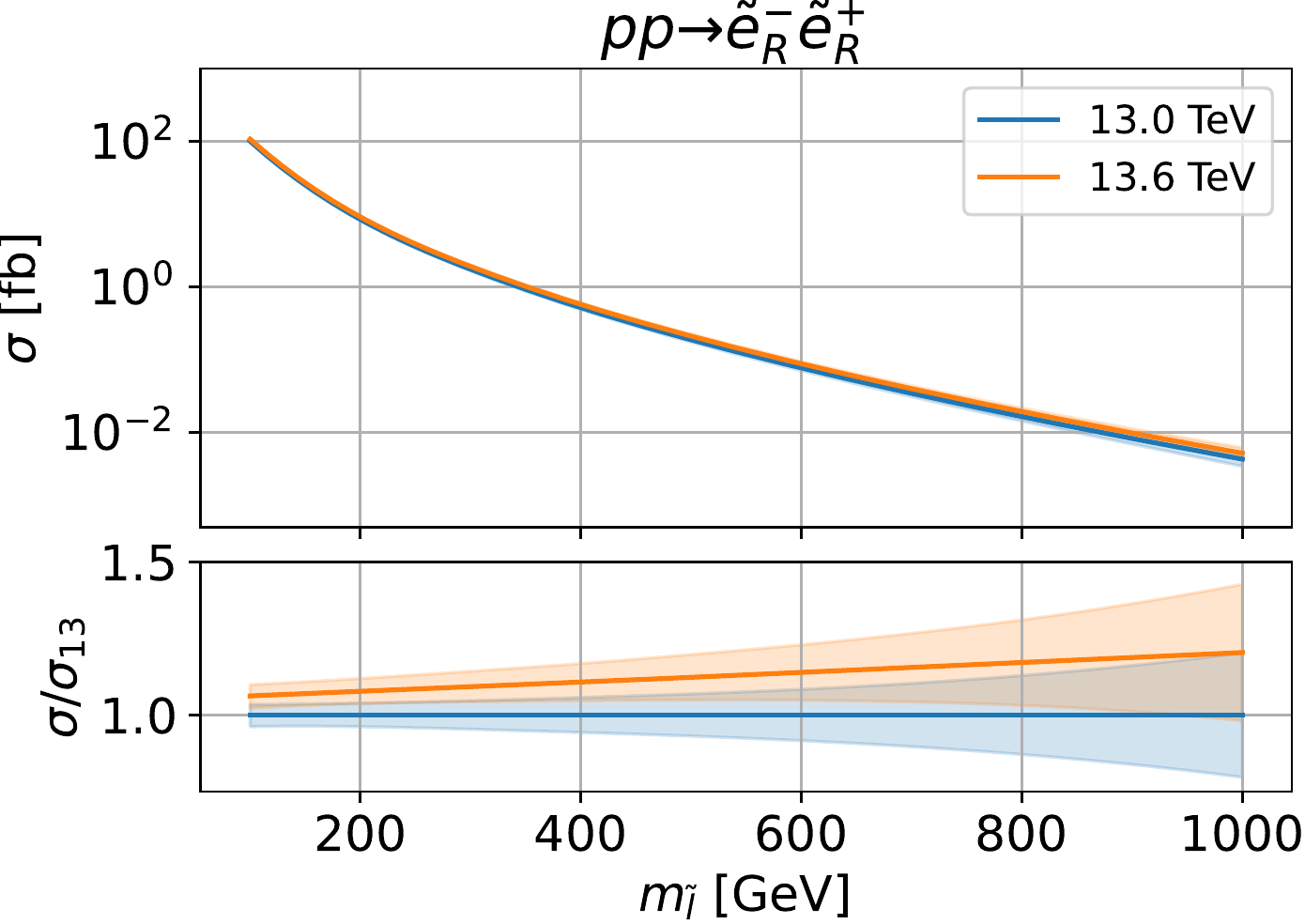}\vspace{.6cm}
    \includegraphics[width=\columnwidth]{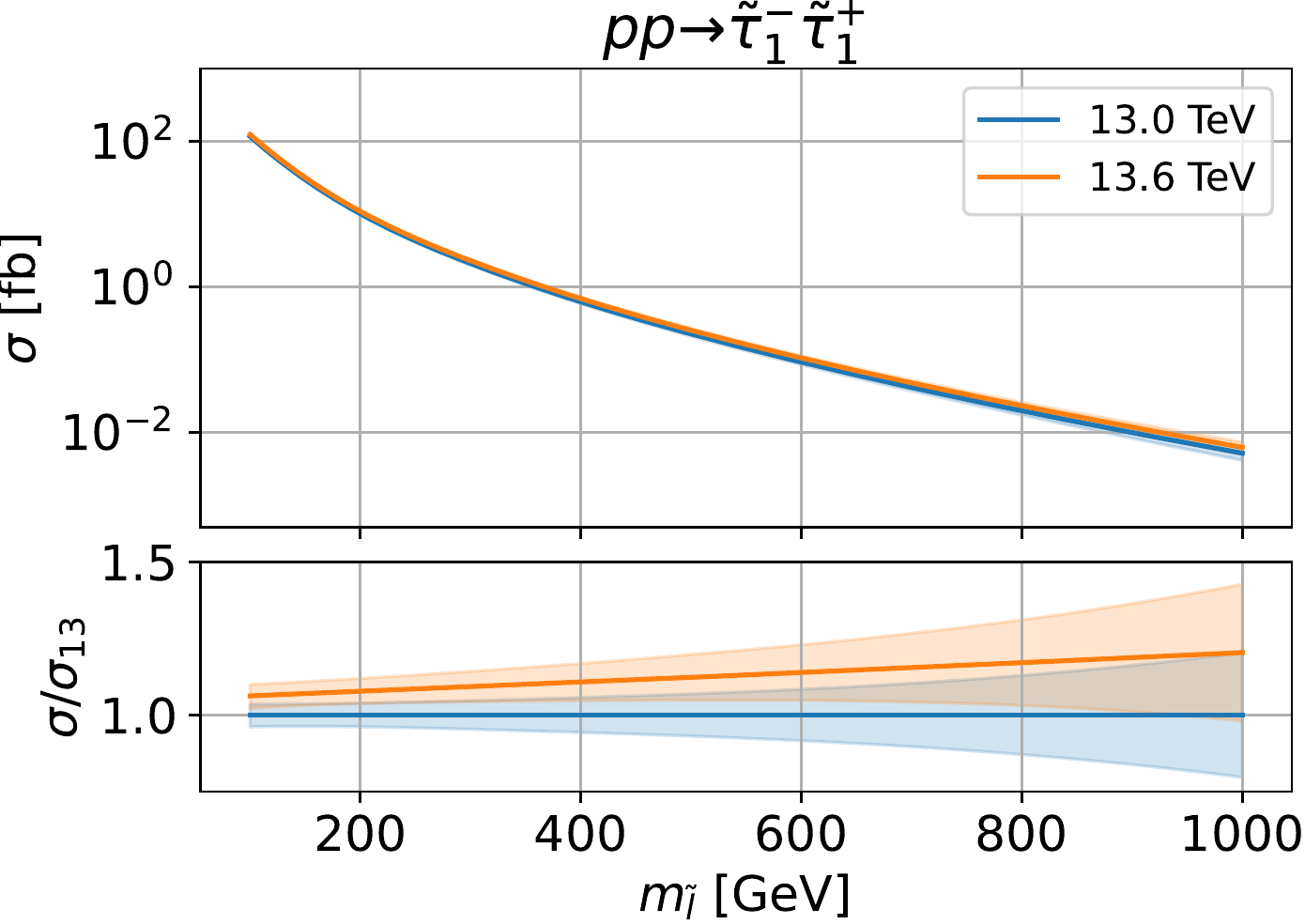}
    \caption{Total cross sections for slepton pair production at the LHC, operating at centre-of-mass energies $\sqrt S$ = \SI{13.6}{TeV} and \SI{13}{TeV} (upper insets), shown together with their ratios to the \SI{13}{TeV} total rates (lower insets) in which combined scale and PDF uncertainties are included. We consider the production of a pair of left-handed sleptons (top), right-handed sleptons (centre) and maximally-mixed sleptons (bottom), and predictions are presented as a function of the slepton mass $m_{\tilde\ell}$.
} \label{fig:sleptons}
\end{figure}

\subsection{Slepton pair production}\label{subsec:sleptons}

In figure~\ref{fig:sleptons} we display aNNLO+NNLL cross section predictions for slepton pair production at the LHC as a function of the slepton mass $m_{\tilde\ell}$. We consider two centre-of-mass energies fixed to $\sqrt{S}$ = \SI{13.6}{TeV} (orange) and \SI{13}{TeV} (blue), and we focus in the upper, central and lower panel of the figure on the respective processes
\begin{equation}
p p \to \tilde e^+_L \tilde e^-_L\,, \qquad \tilde e^+_R \tilde e^-_R\,,\qquad
\tilde \tau^+_1 \tilde \tau^-_1\,.
\end{equation}
In the figures, we restrict the mass range shown to $m_{\tilde\ell} \lesssim \SI{1}{TeV}$. This corresponds to cross section values larger than \SI{0.01}{fb}, to which the LHC Run~3 is in principle sensitive as dozens signal events could populate the signal regions of the relevant ATLAS and CMS analyses.\footnote{The actual numbers of events populating these signal regions depend on the details of the different relevant LHC analyses. However, their precise estimation lies beyond the scope of this article that is solely dedicated to precision predictions for total slepton and electroweakino production rates at the LHC Run~3.} 

Whereas cross sections for $\sqrt{S}$ = \SI{13}{TeV} and \SI{13.6}{TeV} are both shown in the upper insets of the three subfigures, the gain in rate at Run~3 is more visible from the ratio plots presented in their lower insets. This indeed illustrates better how cross section increases ranging up to 20\% can be obtained in the three classes of scenarios considered, especially for large slepton masses. As expected from the structure of the slepton couplings to the $Z$-boson (see {\it e.g.} in~\cite{Bozzi:2004qq}), left-handed sleptons are more easily produced in high-energy hadronic collisions than their right-handed counterparts that only couple through their hypercharge. Consequently, cross sections corresponding to mixed scenarios lie between the two extreme non-mixing cases for a given slepton mass $m_{\tilde\ell}$.

The different ratio plots of the lower insets of the subfigures also show the dependence of the theoretical systematic uncertainty bands on the slepton mass. The collider energy upgrade achieved at Run~3 naturally leads to a reduction of the PDF uncertainties as the gain in centre-of-mass energy yields a smaller relevant Bjorken-$x$ regime in which parton distribution functions are better fitted. As shown in table~\ref{tab:sleptons} (see \ref{app:table}), scale uncertainties contribute to at most 2\% of the combined theoretical uncertainty in the entire mass range probed, regardless of the centre-of-mass energies considered. In contrast, PDF errors vary from 3\% to 18\% at \SI{13.6}{TeV}, which must be compared to a variation ranging from 3\% to 20\% at \SI{13}{TeV}. For a phenomenological study on the reduction of PDF uncertainties in slepton pair production see \cite{Fiaschi:2018xdm}.

\subsection{Higgsino pair production}

\begin{figure*}
     \centering
	  \includegraphics[width=\columnwidth]{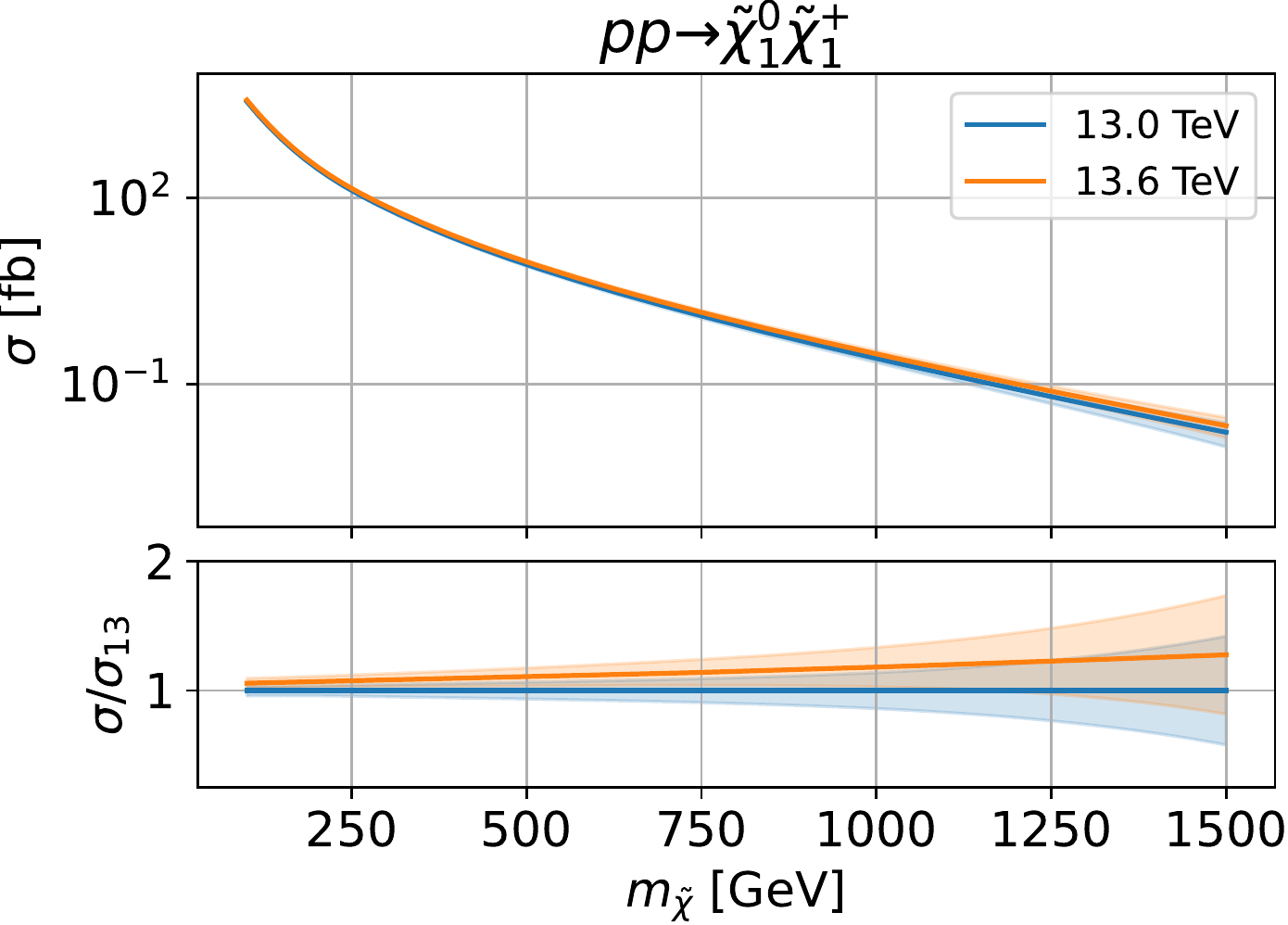}
	  \includegraphics[width=\columnwidth]{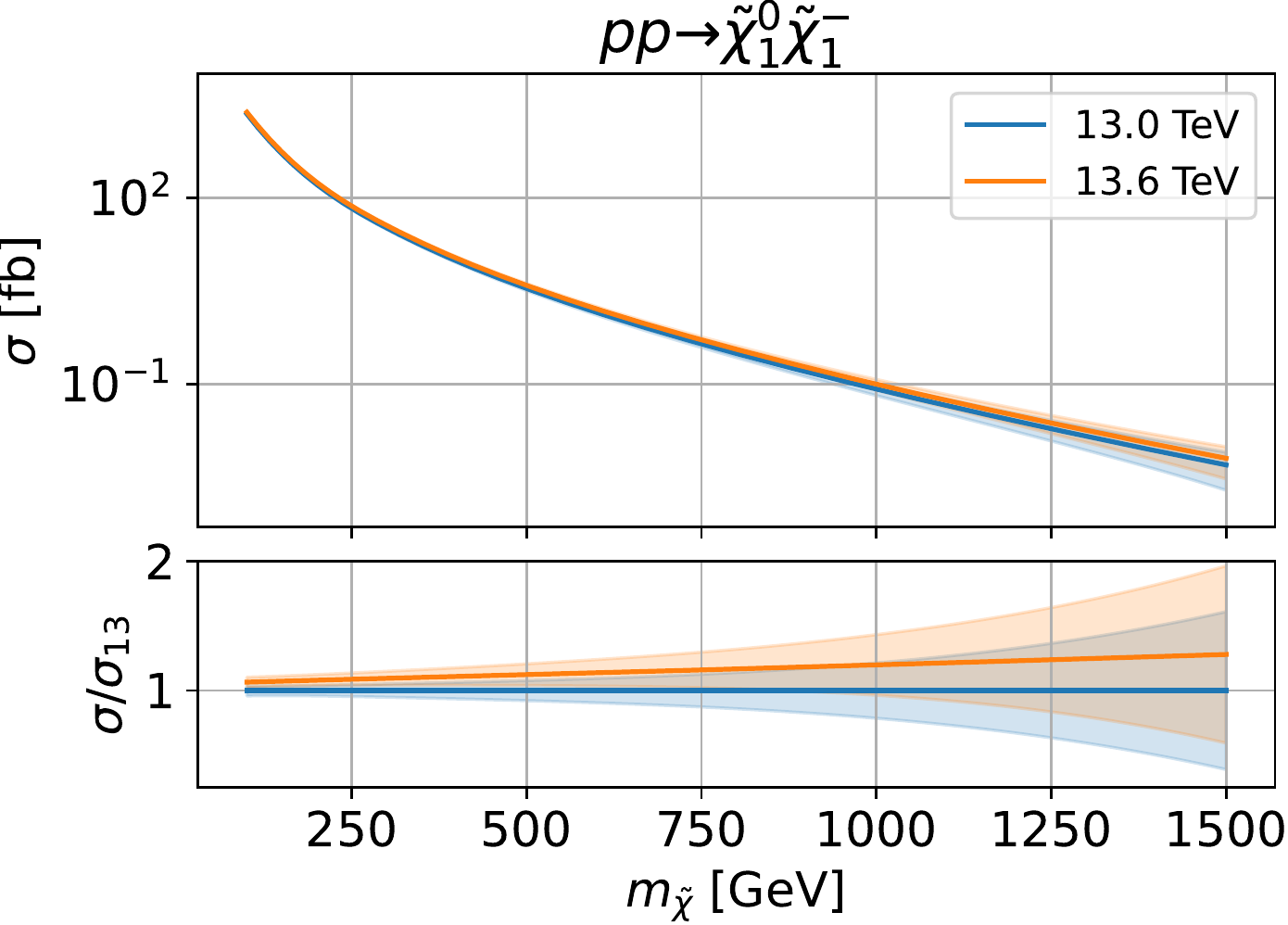}\\ \vspace{.3cm}
	  \includegraphics[width=\columnwidth]{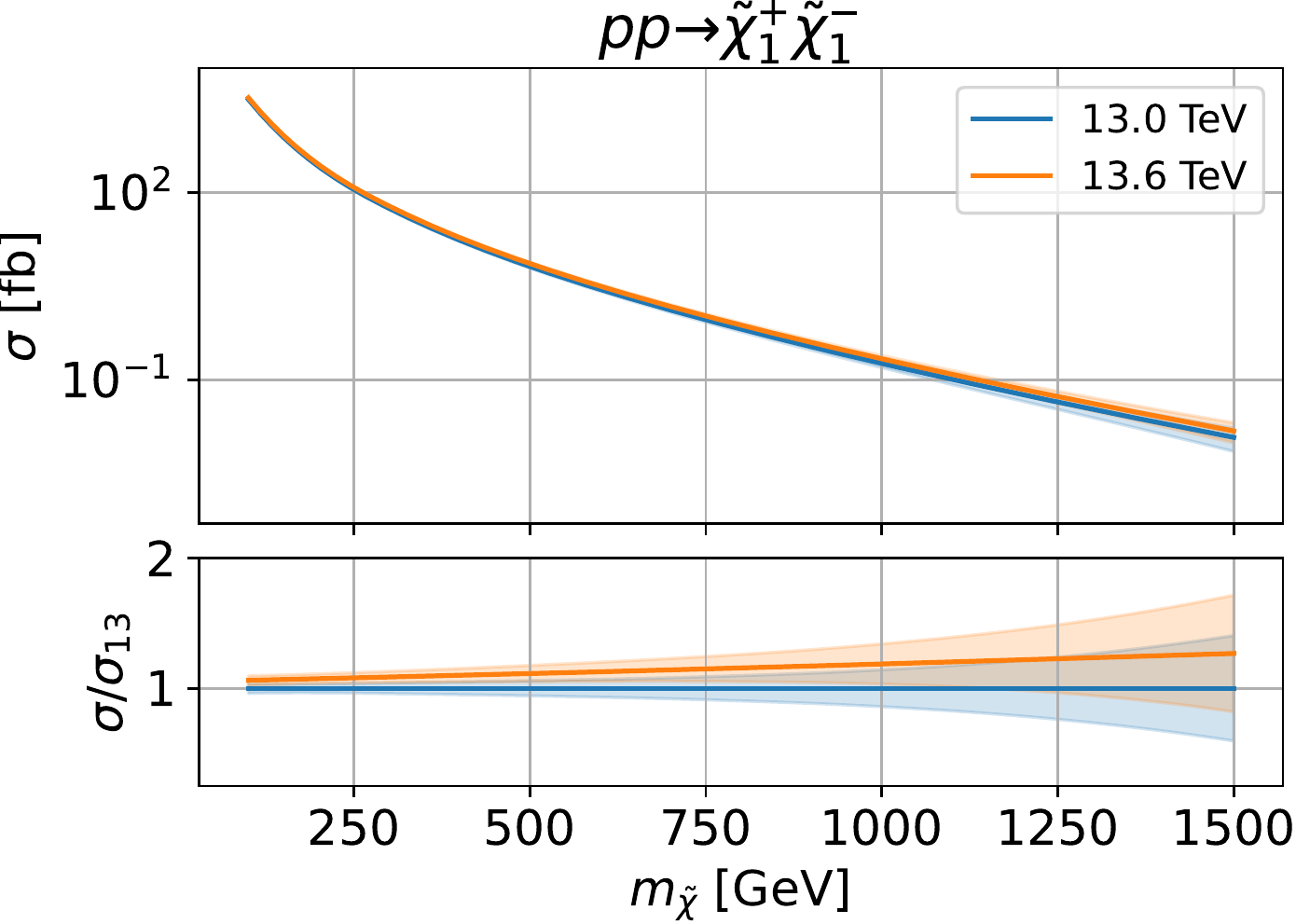}
	  \includegraphics[width=\columnwidth]{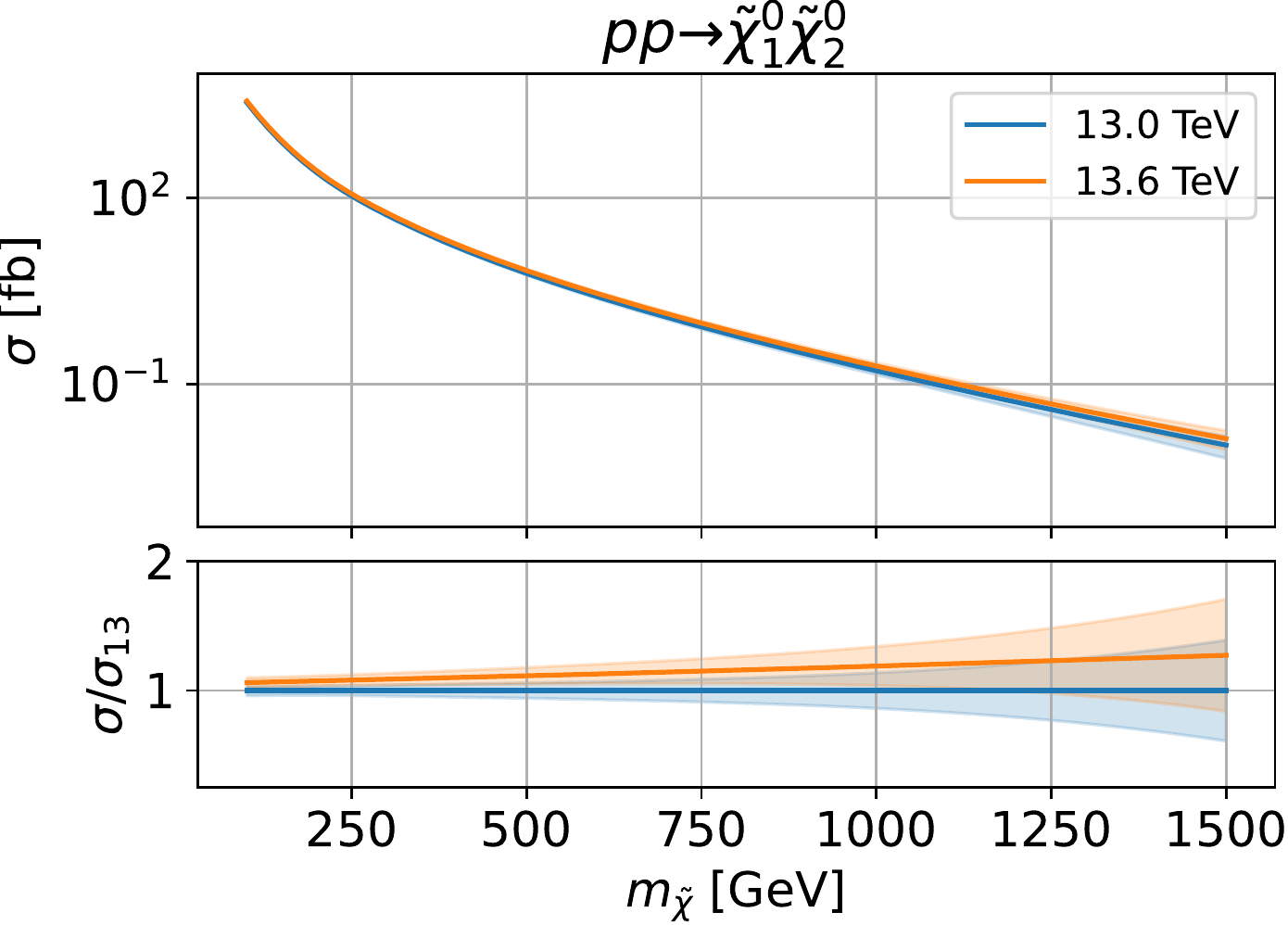}
     \caption{Total cross sections for mass-degenerate higgsino pair production at the LHC, operating at centre-of-mass energies $\sqrt S$ = \SI{13.6}{TeV} and \SI{13}{TeV} (top panels), shown together with their ratios to the \SI{13}{TeV} total rates (bottom panels) in which combined scale and PDF uncertainties are included. We consider the production of associated $\tilde\chi_1^0\tilde\chi_1^+$ (upper left) and  $\tilde\chi_1^0\tilde\chi_1^-$ (upper right) pairs, as well as that of a pair of charginos $\tilde\chi_1^+\tilde\chi_1^-$ (lower left) and neutralinos $\tilde\chi_2^0\tilde\chi_1^0$ (lower right). Predictions are presented as a function of the electroweakino mass $m_{\tilde\chi}$.} \label{fig:hinos_deg}
\end{figure*}

We now turn to scenarios in which all SUSY particles are decoupled by setting their masses at \SI{100}{TeV}, with the exception of all higgsino states. We begin with a calculation of aNNLO+NNLL predictions relevant for higgsino pair production in a scenario in which all higgsinos, defined as in eq.~\eqref{eq:hino}, are mass-degenerate, {\it i.e.}\ in which 
\begin{equation}
  m(\tilde \chi^0_{1})=m(\tilde \chi^0_{2})=m(\tilde \chi^\pm_1) \equiv m(\tilde \chi)\,.
\end{equation}
Our results are shown in figure~\ref{fig:hinos_deg} for the four processes
\begin{equation}
pp \to \tilde\chi_1^0\tilde\chi_1^+\,, \qquad \tilde\chi_1^0\tilde\chi_1^-\,, \qquad \tilde\chi_1^+\tilde\chi_1^-\,, \qquad\tilde\chi_2^0\tilde\chi_1^0\,,
\end{equation}
since in the case of a degenerate spectrum $\sigma(\tilde \chi_1^0 \tilde \chi_1^\pm) = \sigma(\tilde \chi_2^0 \tilde \chi_1^\pm)$. In the results displayed, we restrict the mass range considered to $m_{\tilde\chi}\lesssim \SI{1.5}{TeV}$, which corresponds to production rates at the LHC larger than \SI{0.01}{fb} and therefore potentially reachable at Run~3. The largest cross sections are obtained for the charged-current process $pp\to \tilde \chi_1^0 \tilde \chi_1^+$, such an effect originates from a PDF enhancement related to the ratio of valence and sea quarks in the proton and from the structure of the higgsino gauge couplings. This additionally leads to similar neutral-current higgsino production rates ($pp\to \tilde \chi_1^+ \tilde \chi_1^-$ and $pp\to \tilde \chi_2^0 \tilde \chi_1^0$), and the cross section of the negative charged-current process $pp\to \tilde \chi_1^0 \tilde \chi_1^-$ is then smaller.

As in the slepton case explored in section \ref{subsec:sleptons}, we find an enhancement of total production cross sections at \SI{13.6}{TeV} relative to those at \SI{13}{TeV} thanks to the modest gain in phase space. Rates are indeed found to be 10\% to 30\% larger at $\sqrt{S}$ = \SI{13.6}{TeV} than at $\sqrt{S}$ = \SI{13}{TeV}, for low and high electroweakino masses respectively. 

Still similarly to the slepton case, theoretical uncertainties get reduced with the increase in centre-of-mass energy. As shown in table~\ref{tab:hinos} (see \ref{app:table}), scale uncertainties negligibly contribute to the total theory errors for both centre-of-mass energies, scale variations indeed leading to errors of about 1\% -- 2\% for low higgsino masses and lying in the permille range for $m_{\tilde\chi} \gtrsim \SI{300}{GeV}$. In contrast, total rates at \SI{13}{TeV} are plagued with PDF uncertainties varying from a few percent at low masses to more than 20\% -- 40\% for higgsino masses larger than about \SI{1.2}{TeV}. The reduction of the average Bjorken-$x$ value inherent to the larger centre-of-mass energy of \SI{13.6}{TeV} subsequently leads to smaller PDF errors that are found reduced by about a few permille at low masses, to up to 5\% -- 7\% at large masses. For a phenomenological study on the reduction of PDF uncertainties in higgsino pair production see \cite{Fiaschi:2018hgm}.

We now move on with a second higgsino scenario in which the three lightest higgsino states are defined as in eq.~\eqref{eq:hino2}. Moreover, their spectrum is enforced to feature a significant level of compression, so that all three higgsino states exhibit a mass splitting of a few percent\footnote{The masses of the decoupled squarks are set to \SI{4.5}{TeV} in order to avoid numerical complications.}. In the following, we impose that
\begin{equation}
  m(\chi^\pm_1) = \frac{m(\chi^0_2) - m(\chi^0_1)}{2}\,, 
\end{equation}
with all masses being taken positive.

\begin{figure*}
    \centering
    \includegraphics[width=\columnwidth]{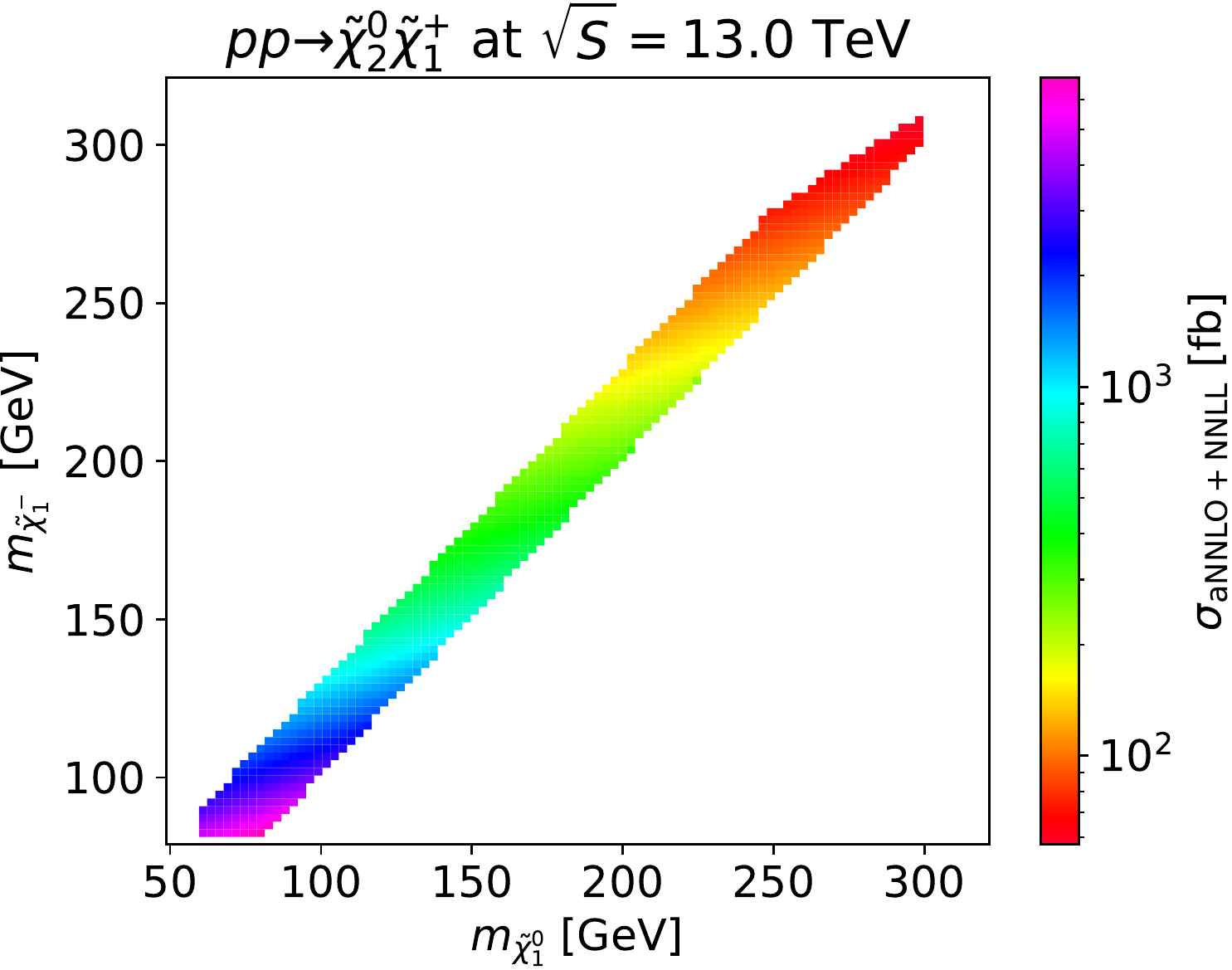}
    \includegraphics[width=\columnwidth]{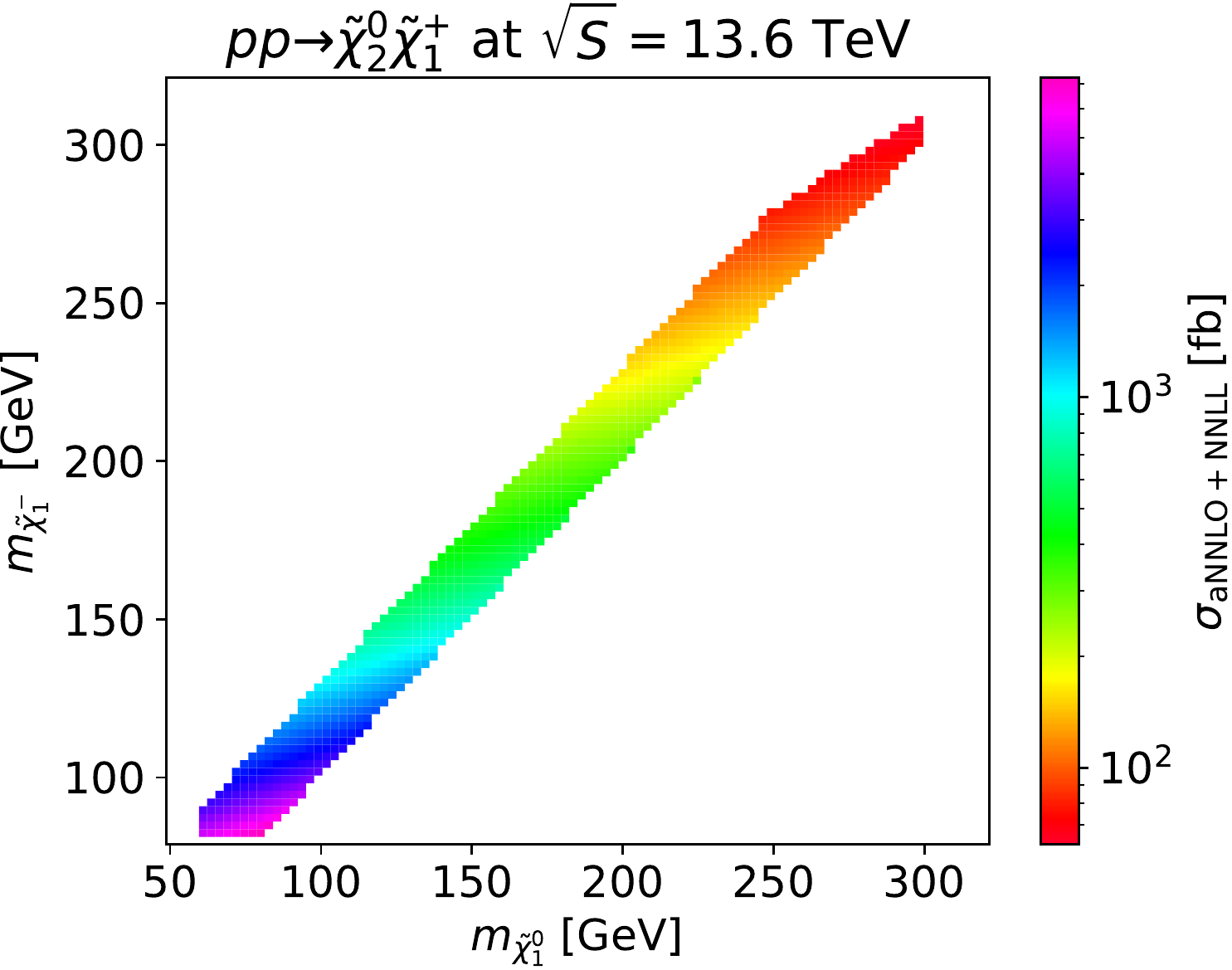}
     \caption{Total cross sections for $\tilde\chi_2^0\tilde\chi_1^+$ production at the LHC, operating at centre-of-mass energies $\sqrt S$ = \SI{13}{TeV} (left) and \SI{13.6}{TeV} (right), in a scenario where all SUSY particles are decoupled with the exception of the non-degenerate produced states. Similar figures can be expected for other higgsino production modes (see table~\ref{tab:hinos_split}).} \label{fig:hinos_nondeg}
\end{figure*}

Given these mass relations, we present in figure~\ref{fig:hinos_nondeg} aNNLO+NNLL total cross sections for the process $pp\to\tilde\chi_1^0\tilde\chi_1^+$ at the LHC, for centre-of-mass energies of $\sqrt S$ = \SI{13}{TeV} (left) and \SI{13.6}{TeV} (right). We consider the mass range in which all higgsinos are lighter than 300~GeV, as this consists of the relevant mass configurations in terms of LHC sensitivity to compressed SUSY higgsino scenarios~\cite{CMS:2018kag, ATLAS:2019lng, ATLAS:2021moa}. 

In the figures, associated rates are shown logarithmically through a colour code. In this scheme, other higgsino production modes yield almost identical figures, that we therefore omit for brevity. Numerical results are nevertheless provided for all processes in table~\ref{tab:hinos_split} (see \ref{app:table}), together with separate scale and PDF uncertainties. For the considered mass range, theoretical systematics are found in very good control, the combined uncertainties being of about 5\% for all mass configurations explored. Whereas scale uncertainties decrease from 2\% -- 3\% in the lightest configurations considered to a few permille for higgsinos of about 200~GeV -- 300~GeV, PDF errors increase from 2\% -- 3\% in the lightest scenarios to 4\% -- 5\% in the heavier cases. The combined theory errors are thus similar in size for all scenarios studied.

As for the previous calculations achieved, a cross section increase at \SI{13.6}{TeV} results from the phase-space enhancement inherent to the increased centre-of-mass energy relative to the \SI{13}{TeV} case, these findings being numerically testified by the results displayed in table~\ref{tab:hinos_split}. Moreover, a rate hierarchy similar to that observed in the mass-degenerate case is obtained, phase-space effects being minimal for compressed scenarios with a non-degenerate spectrum compared to mass-degenerate scenarios in which all higgsinos have exactly the same mass. The process $pp\to\tilde\chi_1^0\tilde\chi_1^+$ hence dominates, followed by the neutral current modes ($pp\to\tilde\chi_1^0\tilde\chi_2^0$ and $pp\to\tilde\chi_1^+\tilde\chi_1^-$) and finally the charged-current channel $pp\to\tilde\chi_1^0\tilde\chi_1^-$. We remind that such a hierarchy is dictated by the structure of the higgsino gauge couplings to the $W$ and $Z$ bosons, and by the PDF ratio of valence and sea quarks in the proton.

\subsection{Gaugino pair production}
\begin{figure}
    \centering
	  \includegraphics[width=\columnwidth]{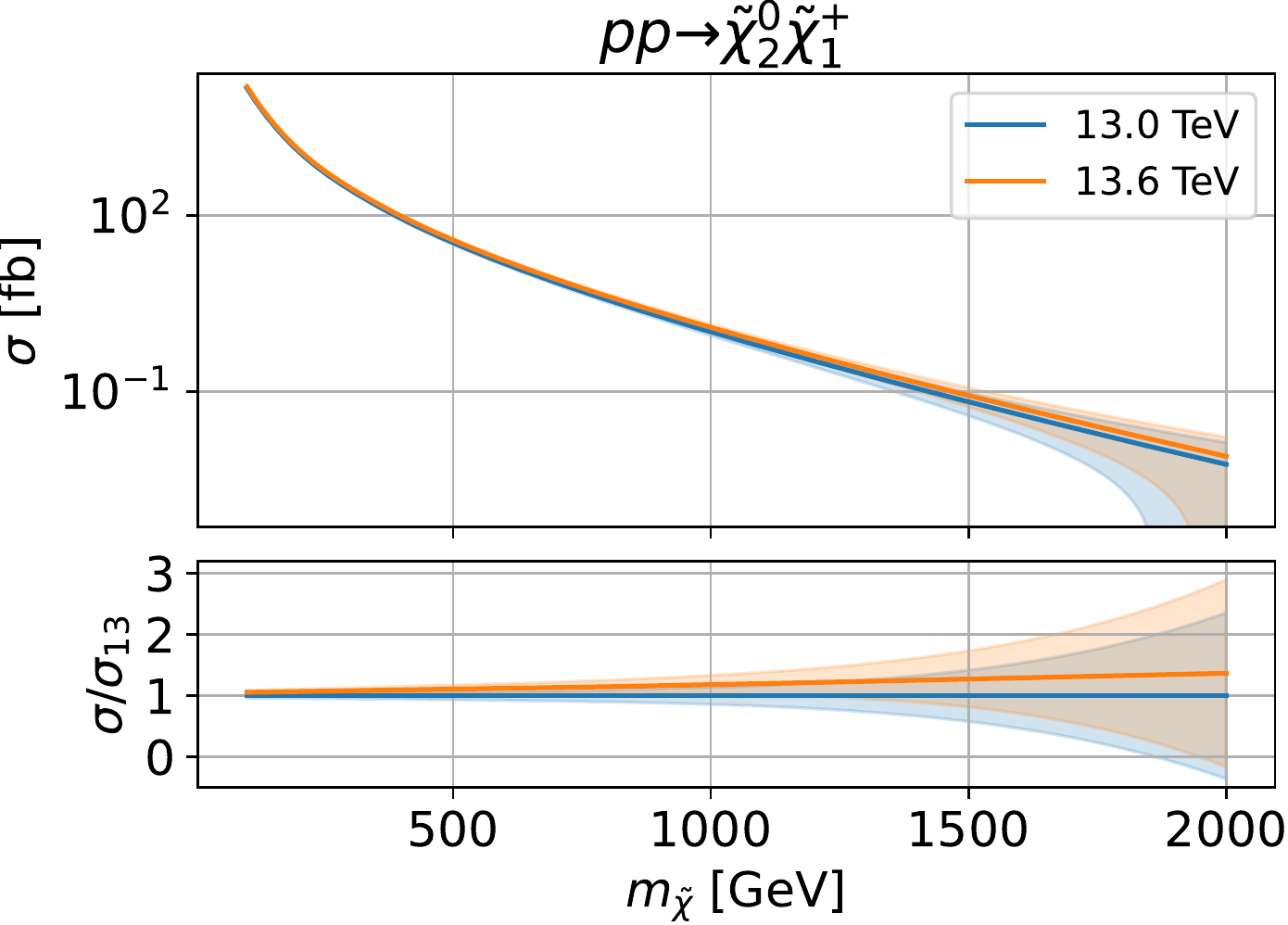}\vspace{.6cm}
	  \includegraphics[width=\columnwidth]{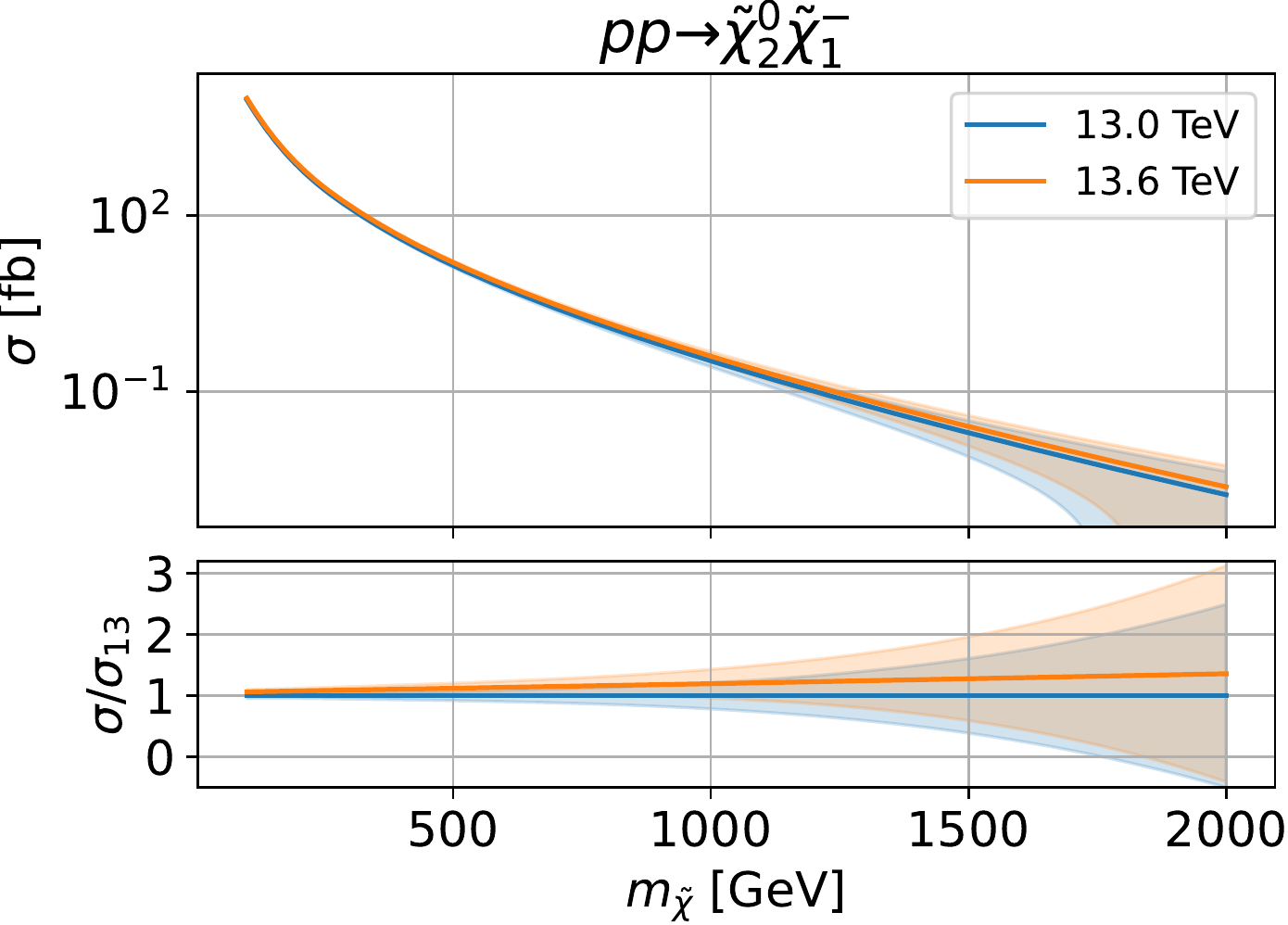}\vspace{.6cm}
	  \includegraphics[width=\columnwidth]{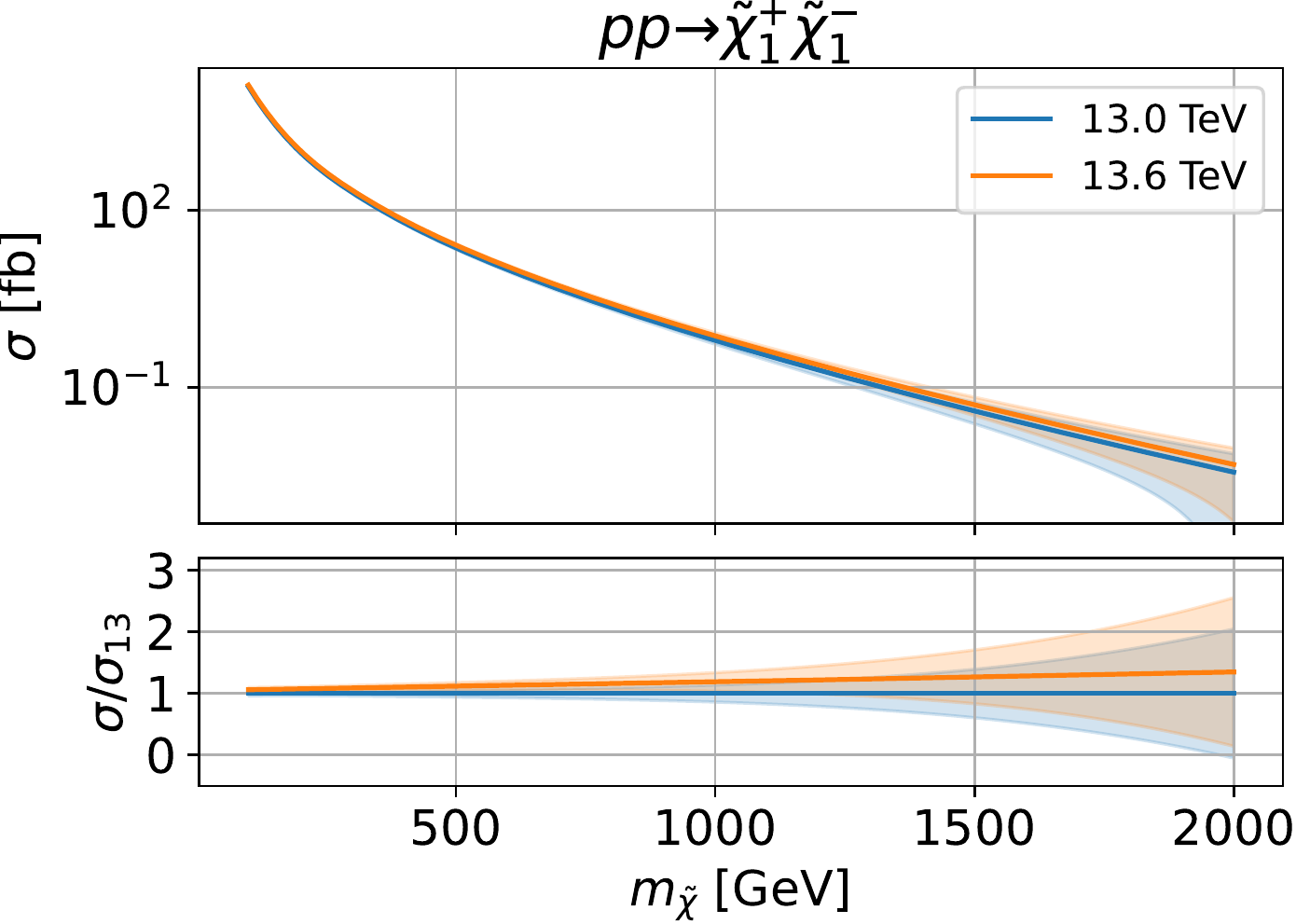}
     \caption{Total cross sections for wino pair production at the LHC, operating at centre-of-mass energies $\sqrt S$ = \SI{13.6}{TeV} and \SI{13}{TeV} (upper insets), shown together with their ratios to the \SI{13}{TeV} total rates (lower insets) in which combined scale and PDF uncertainties are included. We consider the charged-current production of a $\tilde\chi_2^0\tilde\chi_1^+$ (top) and $\tilde\chi_1^0\tilde\chi_1^-$ (centre) wino pair, as well as the neutral-current production of a $\tilde\chi_1^+\tilde\chi_1^-$ pair (bottom). Predictions are presented as a function of the wino mass $m_{\tilde\chi}$.}\label{fig:winos}
\end{figure}

We now turn to the analysis of the gaugino scenarios introduced in eq.~\eqref{eq:gino}, in which the lightest electroweakinos consist of bino and wino eigenstates. In our analysis, we consider that the two wino eigenstates are mass-degenerate, with respective masses satisfying
\begin{equation}
    m(\tilde \chi_2^0) = m(\tilde \chi^\pm_1) \equiv m(\tilde \chi)\,.
\end{equation}
In the following, we then focus on the processes
\begin{equation}
  pp \to \tilde\chi^0_2 \tilde\chi^+_1\,, \qquad \tilde\chi^0_2 \tilde\chi^-_1\,, \qquad \tilde\chi^+_1 \tilde\chi^-_1\,,
\end{equation}
and we present aNNLO+NNLL predictions for the associated total rates in figure~\ref{fig:winos}. Other processes are irrelevant as the corresponding cross sections vanish due to the structure of the bino and wino gauge couplings. As in the previous subsections, we consider results at centre-of-mass energies of \SI{13}{TeV} (blue) and \SI{13.6}{TeV} (orange) in the upper insets of the figures. This time, however, we display predictions for wino masses ranging up to \SI{2}{TeV}, the cross sections being much larger than in the higgsino case by virtue of the weak triplet nature of the winos. Consequently, we can expect a better LHC sensitivity to signatures of wino production and decays, due to the machine being capable to naturally probe a larger mass regime. 

In accord with parton density effects, the charged-current process $pp\to\tilde\chi^0_2 \tilde\chi^+_1$ dominates for a given wino mass, its rate being a factor of 1.5 to 3 larger than that of the other charged-current process $pp\to\tilde\chi^0_2 \tilde\chi^-_1$ in the case of lighter and heavier mass setups respectively. Furthermore, total cross sections for the neutral current process $pp\to\tilde\chi^+_1 \tilde\chi^-_1$, mediated by virtual photon and $Z$-boson exchanges, are usually 1.25 -- 1.5 smaller than rates corresponding to the charged-current mode $pp\to\tilde\chi^0_2 \tilde\chi^+_1$, in which the final state is produced from virtual $W$-boson exchanges. On the other hand, the increase in cross section observed when the hadronic centre-of-mass energy is modified from \SI{13}{TeV} to \SI{13.6}{TeV} can be quite substantial in such mass-degenerate wino scenarios. While for light produced particles the increase is only modest and lies in the 5\% -- 10\% range for all three processes, it increases with the wino mass $m_{\tilde\chi}$ and reaches 35\% -- 40\% for wino masses of about \SI{2}{TeV}.

In the lower insets of the three subfigures, we display the ratio of the total production rates at \SI{13}{TeV} and \SI{13.6}{TeV} to that at \SI{13}{TeV}, including the combined theory systematic error. We remind that numerical values for the cross sections are presented together with separate scale and PDF uncertainties in table~\ref{tab:winos} (see \ref{app:table}). For the entire mass range considered, scale uncertainties are under good control. They are about 1\% -- 2\% at low masses, and then decrease to a few permille for wino masses larger than \SI{300}{GeV}. In contrast, PDF errors are smaller than 10\% for wino masses smaller than about \SI{1}{TeV}, but quickly increase for heavier mass configurations. In this case, typical Bjorken-$x$ values are large and correspond to phase space regimes in which parton densities are poorly constrained, as already pointed out in \cite{Frixione:2019fxg}. This issue will nevertheless be automatically cured with time. Time will indeed allow the LHC collaborations to collect better-quality SM data at large scales, which will consequently help in reducing the PDF errors.

In order to explore the phenomenological consequences of next-to-minimality, we now consider scenarios in which first-generation and second-generation squarks are mass-degenerate, but not decoupled. We introduce the squark mass parameter $m_{\tilde q}$ defined by
\begin{equation}
    m(\tilde u_{L,R}) = m(\tilde d_{L,R})=m(\tilde s_{L,R})=m(\tilde c_{L,R}) \equiv m_{\tilde q}\,,
\end{equation}
that we then vary between \SI{800}{GeV} and \SI{4}{TeV}. For illustrative purposes, we focus on the charged-current process
\begin{equation}\label{eq:procwsq}
    p p \to \tilde\chi^0_2 \tilde\chi^+_1\,,
\end{equation}
that gives rise to the largest wino production cross sections for a specific mass spectrum. The discussion and the results below are however applicable to other wino production modes as well. This is further numerically depicted in table~\ref{tab:wino_sq} (see \ref{app:table}). The latter indeed includes a complete set of numerical predictions for all wino pair production processes in the presence of not too heavy squarks, once again together with separate information on the total rate values, and the associated scale and PDF uncertainties presented separately.

\begin{figure*}
     \centering
	  \includegraphics[width=\columnwidth]{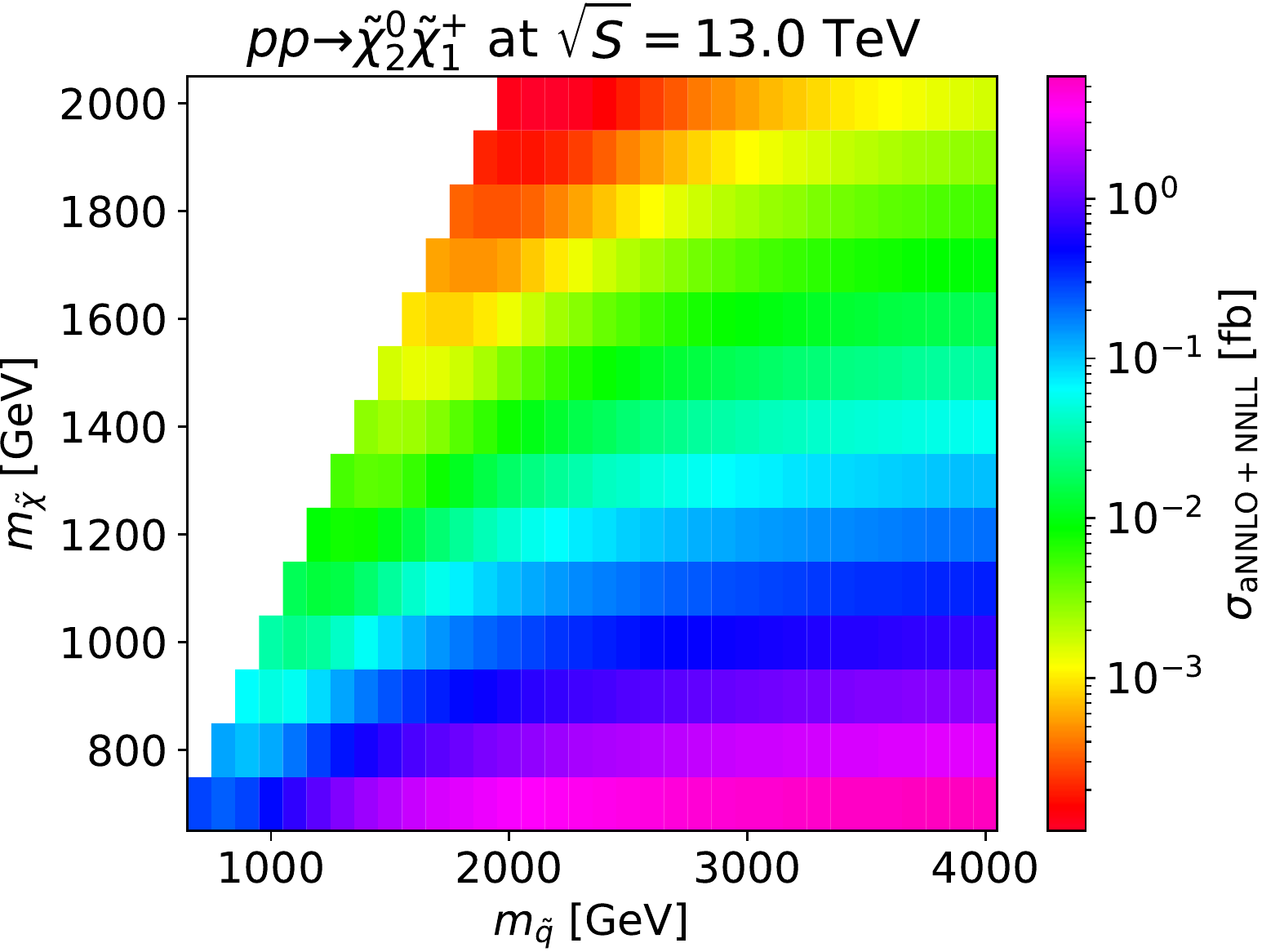}
	  \includegraphics[width=\columnwidth]{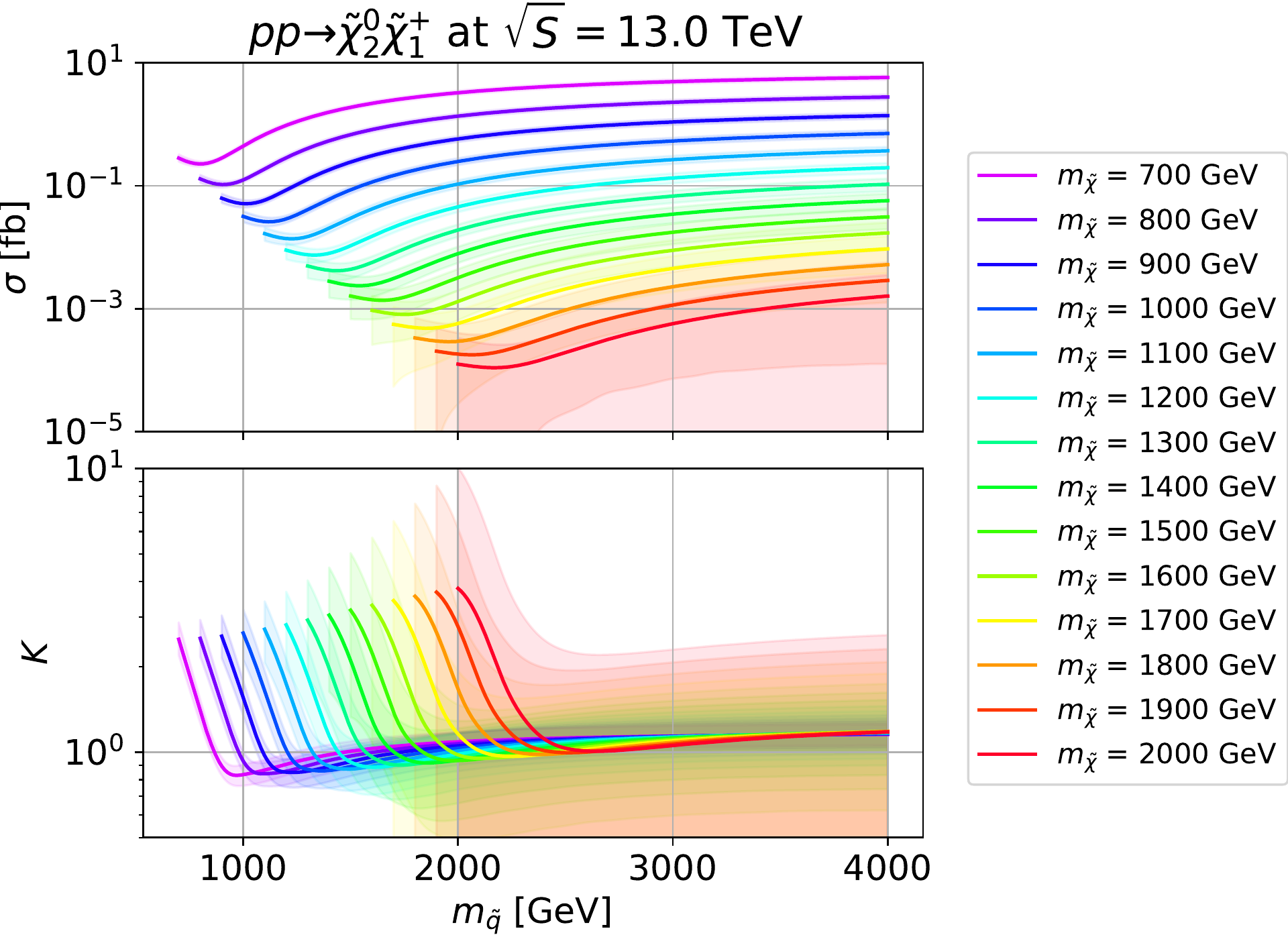}\vspace{.3cm}
	  \includegraphics[width=\columnwidth]{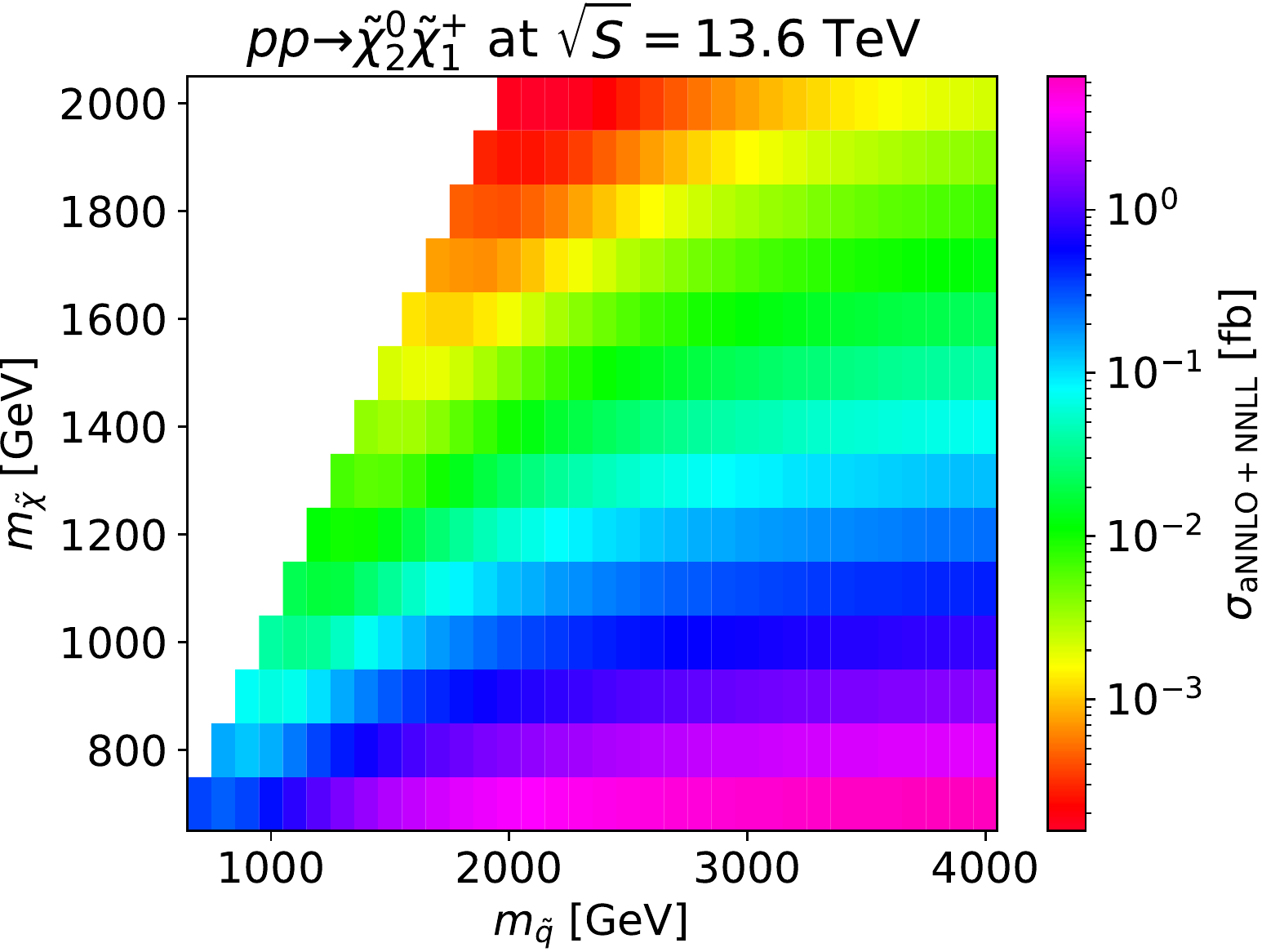}
	  \includegraphics[width=\columnwidth]{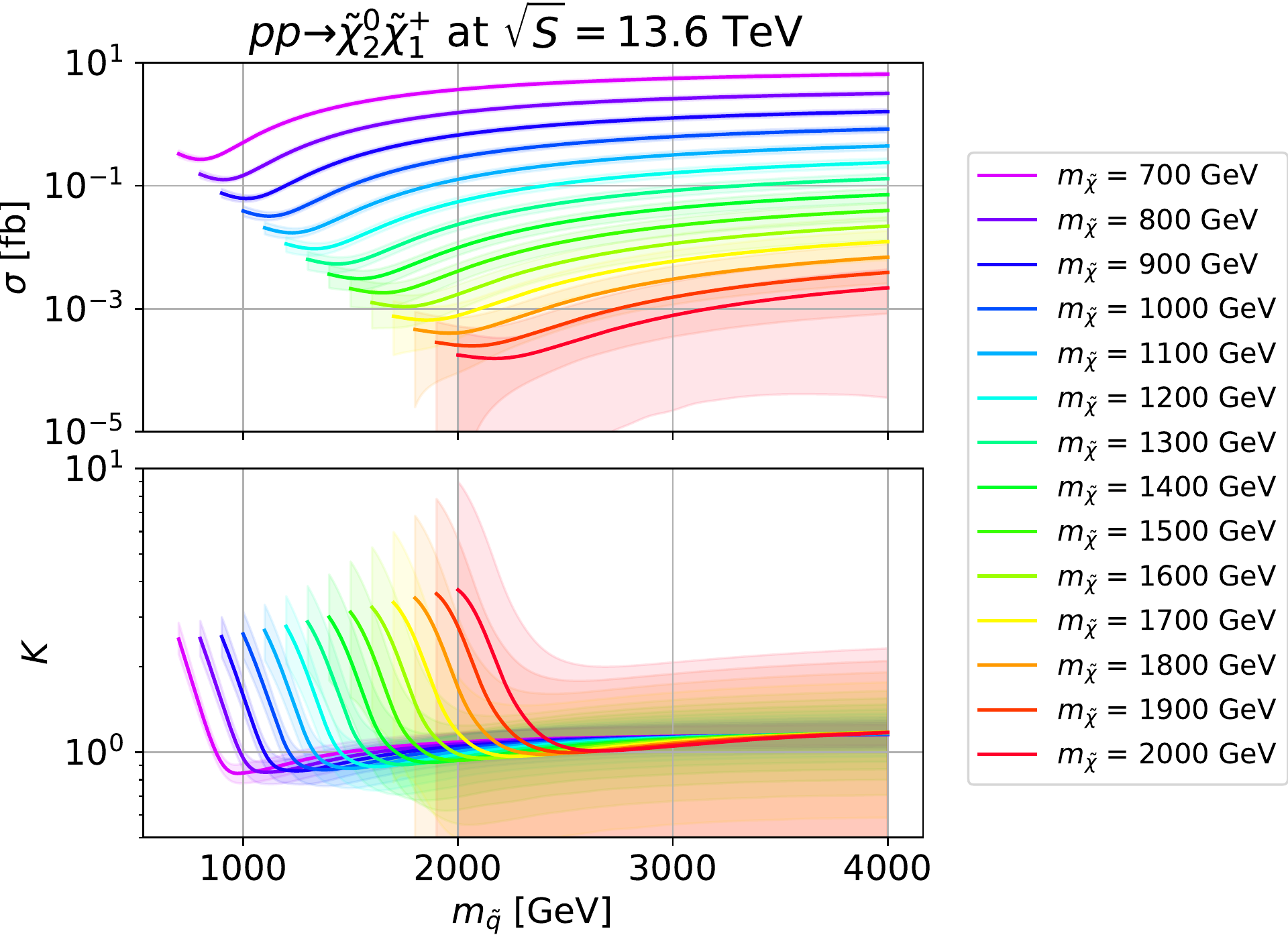}
     \caption{Total cross sections for wino pair production at the LHC, operating at centre-of-mass energies $\sqrt S$ = \SI{13}{TeV} (upper) and \SI{13.6}{TeV} (lower). We consider the charged-current production of a $\tilde\chi_2^0\tilde\chi_1^+$ wino pair, and present predictions as a function of the wino mass $m_{\tilde\chi}$ and the common first-generation and second-generation squark mass $m_\squark$.}\label{fig:wino_sq}
\end{figure*}

In figure~\ref{fig:wino_sq} we present aNNLO+NNLO total cross sections for the process of eq.~\eqref{eq:procwsq} as a function of the wino and squark masses $m_{\tilde\chi}$ and $m_\squark$\footnote{Similarly to the non-fully-degenerate higgsino scenario, here the mass of the decoupled gluino is set between \SI{5}{TeV} and \SI{100}{TeV}.}. Results are presented for collider energies of $\sqrt{S}$ = \SI{13}{TeV} and \SI{13.6}{TeV} in the top and bottom row of the figure respectively. In the left subfigures, the rates are represented through a logarithmic colour code, which shows that they exhibit a non-trivial dependence on the squark mass. Due to the destructive interference of $s$-channel gauge boson exchange diagrams with $t/u$-channel squark exchange diagrams, the cross section starts by decreasing when the SUSY spectrum is varied from a configuration in which $m_{\tilde\chi}=m_\squark$ to one with a larger squark mass value still in the vicinity of the same wino mass $m_{\tilde\chi}$. The rate then gets larger and larger with increasing squark masses (the wino mass being constant), and it finally saturates when squarks decouple. 

This feature is further illustrated in the two upper insets of the right subfigures, that display aNNLO+NNLL production cross sections for $\tilde\chi^0_2 \tilde\chi^+_1$ production at the LHC as a function of the squark mass for several choices of wino masses. For a given wino mass $m_{\tilde\chi}$, we observe that the cross section always begins by decreasing before quickly reaching a minimum, and then increases for larger and larger squark masses. In the limit of very heavy squarks, the latter decouple and the rates become independent of the squark properties. They hence solely depend on the wino mass.

In the lower inset of these two right subfigures, we present $K$-factors defined as the ratio of the most precise aNNLO+NNLL rates to the LO ones for a given mass configuration,
\begin{equation}
  K = \frac{\sigma_\text{aNNLO+NNLL}}{\sigma_\text{LO}}\,.
\end{equation}
This further illustrates the squark decoupling at large $m_\squark$ values, the $K$-factor becoming constant. Moreover, the dependence of the $K$-factors on the squark mass additionally shows how the minimum of the cross section, for a given wino mass, is shifted by tens of GeV by virtue of the higher-order corrections. 

In all the insets included in the right subfigures, we additionally include combined theory uncertainties. In general, those uncertainties, that are numerically given in table~\ref{tab:wino_sq}, are under good control, with the exception of the heaviest scenarios in which the PDF errors get very large, as already above-mentioned. On the contrary, scale uncertainties always lie in the percent or permille level for all scenarios considered, the error bars being in fact drastically impacted (and reduced) by QCD resummation.

\section{Conclusion}\label{sec:summary}

We have updated the documentation of the \resummino package dedicated to precision calculations of total rates and invariant-mass and transverse-momentum spectra in the context of slepton and electroweakino pair production, the production of a pair of leptons in the presence of additional gauge bosons, and that of an associated electroweakino-squark or electroweakino-gluino pair. This update of the documentation is motivated by the significant extensions that have been implemented since the initial release of the code a decade ago, and that have never been collected in a single document. Whereas these new features of \resummino have already been used for various phenomenological studies and experimental searches for particles beyond the Standard Model at the LHC, we have taken the opportunity of this paper to provide a new useful illustration of the capabilities of the program. We have computed and tabulated for the first time precision predictions of total rates relevant to searches for sleptons and electroweakinos at the ongoing Run~3 of the LHC, operating at a centre-of-mass energy of $\sqrt{S}=\SI{13.6}{TeV}$.

\renewcommand{\arraystretch}{1.5}
\setlength{\tabcolsep}{5pt}
\begin{table}
\begin{center}
	\begin{tabular}{ cccc }
	 Final-state particle mass  & 1 TeV & 1.5 TeV & 2 TeV \\ \hline 
	 $\sigma_{\SI{13.6}{TeV}}/\sigma_{\SI{13}{TeV}}$ & $\sim$1.2 & $\sim$1.3 & $\sim$1.6 \\ 
	PDF4LHC21 unc. & 13\% -- 20\% & 35\% -- 40\% & $> $100\% \\ 
	Scale unc. &$<$0.3\%& $<$0.2\% & $<$0.1\% \\
	\end{tabular}
\end{center}
\caption{Behaviour of aNNLO+NNLL total cross sections for slepton and electroweakino pair production at the LHC, for a centre-of-mass energy of $\sqrt{S}=\SI{13.6}{TeV}$. The results are given as a function of the average mass of the final-state particles, and we provide the typical increase in cross section related to predictions at $\sqrt{S}=\SI{13}{TeV}$, and information about the size of the theoretical PDF and scale uncertainties inherent to the calculations achieved.}\label{tab:summary}
\end{table}

We have considered several simplified models inspired by the MSSM in which all superpartners are decoupled, with the exception of a few states whose production at the LHC has been explored. We have presented predictions matching fixed-order calculations at approximate NNLO in QCD (in which SUSY two-loop contributions have been neglected) with predictions including the resummation of soft-gluon radiation at NNLL. In our results, we have put emphasis on the gain obtained from updating the centre-of-mass energy from \SI{13}{TeV} to \SI{13.6}{TeV}, and on the theory uncertainties inherent to all calculations achieved. A brief summary of our findings is given in table~\ref{tab:summary}.

As a rule of thumb valid for all investigated processes, QCD threshold resummation reduces the typical scale dependence to less than 1\% at aNNLO+NNLL. These contributions to the combined theory error hence become generally negligible, or at least subleading. For the production of low-mass SUSY states, PDF and scale uncertainties are of the same order of magnitude so that theory systematics are under good control. 
In our predictions, we have employed the PDF4LHC21 set of parton densities that originates from a combination of variants of the CT18~\cite{Hou:2019efy}, MSHT20~\cite{Bailey:2020ooq} and NNPDF3.1~\cite{NNPDF:2017mvq} global PDF fits. Such a set generally leads to larger PDF uncertainties compared to the individual global sets, particularly at large Bjorken $x \gtrsim 0.4$~\cite{PDF4LHCWorkingGroup:2022cjn}. In this region, yet poorly constrained by experimental data, discrepancies between predictions relying on the different global sets can be significant as the parton density behaviour is mainly driven by the theoretical assumptions made. For instance, recent studies have shown how the recent NNPDF4.0~\cite{NNPDF:2021njg} set predicts (anti)quark distributions falling much faster (slower) with respect to the other sets~\cite{Ball:2022qtp}. Moreover, the sea over valence quark ratio is a determining factor for Drell-Yan-like processes~\cite{Fiaschi:2022wgl}, and in turn it also has a strong impact on the determination of the gluon density at high $x$ values and large scale~\cite{Alekhin:2017kpj}. Predictions for (differential) cross section can therefore vary substantially with the choice of the PDF set, especially in scenarios featuring heavy SUSY particles.
This issue will however be naturally fixed with time, as more data gets collected at high scale by the LHC collaborations.

We have observed that the modest increase in centre-of-mass energy at the LHC Run~3 brings a typical increase in the rates associated with colourless SUSY processes of 20\% for light spectra to 40\% in the heavier cases considered. In addition, we have explored the impact of next-to-minimality following two aspects. First, we have considered the production of a pair of non-degenerate compressed higgsinos, and next we have studied the impact of squark decoupling on wino pair production. Both setups have led to visible effects, so that their impact will have to be considered for non-minimal excursions in the interpretation of future search results. 

These findings additionally pave the way to future work. In particular, it will be important to assess the phenomenology of non-minimal model spectra exhibiting several light strongly-interacting and non-strongly-interacting SUSY states. In this case, rarely-studied semi-strong processes (such as squark-electroweakino associated production) could be relevant, and worth to be explored.

The package \resummino and its current release 3.1.2 is ready to achieve related calculations. The source code is publicly available from \texttt{HEPForge} and can be used for aNNLO+NNLL calculations for slepton and electroweakino pair production (as studied in this article), as well as for NLO+NLL calculations for the associated production of a squark-electroweakino and gluino-electroweakino pair. Moreover, non-supersymmetric models featuring additional $W'$ and $Z'$ boson could also be investigated at aNNLO+NNLL accuracy, and in addition to total rates \resummino can be used to calculate invariant-mass and transverse-momentum distributions for all implemented processes (with an NLO+NLL precision).


\begin{acknowledgements}
This work has been supported by the BMBF under contract 05P21PMCAA, by the DFG through the Research Training Network 2149 “Strong and Weak Interactions - from Hadrons to Dark Matter” and by the French ANR (grant ANR-21-CE31-0013 “DMwithLLPatLHC”). The work of JF has been supported by STFC under the Consolidated Grant ST/T000988/1. 
\end{acknowledgements}

\newpage 
\appendix
\section{Details on the threshold resummation formulas implemented in \resummino}
\label{app:Resummation}

In this appendix we collect analytic expressions for the exponentiated logarithmic contributions and the hard matching coefficients relevant for all processes supported in \emph{Resummino}. 

The logarithmic coefficients of~\eqref{eq:HtimesG} are given, for a Born process initiated by the initial-state partons $a$ and $b$ by~\cite{Sterman:1986aj, Catani:1989ne, Catani:1990rp, Kramer:1996iq, Vogt:2000ci}
\begin{equation}\begin{split}
&G^{(1)}_{ab}(\lambda) \!=\! \sum_{c=a,b} g_c^{(1)}(\lambda) \,,\\
&G^{(2)}_{ab \to ij}(\lambda,M^2,\mu_F^2,\mu_R^2) \!=\!\!\! \sum_{c=a,b}  g_c^{(2)}(\lambda,M^2,\mu_F^2,\mu_R^2)\\ &\qquad\qquad + h^{(2)}_{ab\to ij}(\lambda)\,,\\
 &G^{(3)}_{ab \to ij}(\lambda,M^2,\mu_F^2,\mu_R^2) \!=\!\!\! \sum_{c=a,b} g_c^{(3)}(\lambda,M^2,\mu_F^2,\mu_R^2) \\ &\qquad\qquad  + h^{(3)}_{ab\to ij}(\lambda)\,,
\end{split}\label{eq:Gab}\end{equation}
where $\lambda = \alpha_s b_0 L$ with $L = \ln(\bar N)$ and $\bar N = Ne^{\gamma_E}$, and where $b_0$ is defined from the first coefficient of the QCD beta function. We provide its expression together with that of the next two coefficients of the QCD beta function, that are relevant for the formulas given below. They are normalised according to $b_n=\beta_n/(2\pi)^{n+1}$~\cite{Tarasov:1980au,Larin:1993tp}, which gives
\begin{equation}\begin{split}
 &b_0 = \frac{1}{12 \pi} (11 C_A - 2 n_f)\,, \\
 &b_1 = \frac{1}{24 \pi^2} (17 C_A^2 - 5 C_A n_f - 3 C_F n_f)\,, \\
 &b_2 = \frac{1}{64 \pi^3} \bigg(\frac{2857}{54} C_A^3 - \frac{1415}{54} C_A^2 n_f + C_F^2 n_f\\ &\qquad \quad - \frac{205}{18} C_A C_F n_f + \frac{79}{54} C_A n_f^2 + \frac{11}{9} C_F n_f^2 \bigg)\,,
\end{split}\end{equation}
with $C_A=N_C=3$, $C_F=(N^2-1)/(2N_C)=4/3$ and the number of active quark
flavours $n_f=5$. 

The individual process-independent resummation coefficients $g_c^{(n)}$ for quarks ($c=q$) and gluons ($c=g$) only depend on the nature of the initial state. They are given as compact functions of the universal process-independent resummation coefficients $A_c^{(n)}$, the first two of these being written as
\begin{equation}\begin{split}
A_c^{(1)} =&\ 2 C_c \,,\\
A_c^{(2)} =&\ 2 C_c \left( C_A\frac{67}{18} - \frac{\pi^2}{6} C_A - \frac{5}{9} n_f \right)\,,
\end{split}\end{equation}
with $C_c=C_F$ for quarks and $C_c=C_A$ for gluons. As in \resummino NNLL precision is only achieved for slepton and electroweakino pair production, the implemented formulas only require the knowledge of the third universal quark coefficient $A_q^{(3)}$,
\begin{equation}\begin{split}
A_q^{(3)} =& \frac{1}{2} C_F \Bigg[C_A^2 \left(\frac{245}{24} - \frac{67}{9}\zeta_2 + \frac{11}{6}\zeta_3 + \frac{11}{5}\zeta_2^2 \right) \\ &\ + C_A n_f \left(\frac{10}{9}\zeta_2 - \frac{7}{3} \zeta_3 - \frac{209}{108} \right) - \frac{n_f^2}{27}\\ &\ +C_F n_f \left(2\zeta_3 - \frac{55}{24} \right)\Bigg] \,.
\end{split}\end{equation}
The first two resummation coefficients $g_c^{(n)}$ read
\begin{equation}\begin{split}
  & g_c^{(1)} = \frac{A_c^{(1)}}{4\pi b_0 \lambda} \left[ 2 \lambda + (1- 2 \lambda) \ln (1- 2\lambda)\right]\,,\\
  &g_c^{(2)} = \frac{A_c^{(1)} b_1}{4\pi b_0^3 } \left[ 2 \lambda + \ln(1- 2 \lambda) + \frac 1 2 \ln^2 (1- 2\lambda)\right] \\
  & \quad - \frac{A_c^{(2)}}{2(2\pi)^2 b_0^2} \left[ 2 \lambda + \ln (1- 2 \lambda)\right] \\
  & \quad + \frac{A_c^{(1)}}{4\pi b_0} \left[ \ln (1 \!-\! 2 \lambda ) \ln\left(\frac{M^2}{\mu_R^2}\right) \!+\! 2 \lambda \ln\left( \frac{\mu_F^2}{\mu_R^2}\right)\right]\,,
\end{split}\end{equation}
whereas the third quark coefficient (the only coefficient needed for the processes implemented at aNNLL+NNLL in \resummino) is given by
\begin{equation}\begin{split}
  & g_q^{(3)} = \frac{A_q^{(3)}}{\pi^3 b_0^2} \frac{\lambda^2}{1 - 2 \lambda} + \frac{2A_q^{(1)}}{\pi} \zeta_2 \frac{\lambda}{1 - 2 \lambda} \\
  & \quad - \frac{A_q^{(2)} b_1}{(2 \pi)^2 b_0^3}\ \frac{2 \lambda^2 + 2 \lambda + \ln(1-2\lambda)}{1 - 2 \lambda}\\
  & \quad - \frac{A_q^{(2)}}{2 \pi^2 b_0} \left[\frac{\lambda}{1 - 2 \lambda}\ln\left(\frac{M^2}{\mu_R^2}\right)- \lambda\ln\left(\frac{\mu_F^2}{\mu_R^2}\right)\right] \\
  & \quad + \frac{A_q^{(1)} b_1^2}{2 \pi b_0^4}\ \frac{2 \lambda^2 + 2 \lambda \ln(1-2\lambda) + \frac{1}{2}\ln^2(1-2\lambda)}{1 - 2 \lambda}  \\
  & \quad + \frac{A_q^{(1)} b_2}{2 \pi b_0^3} \left[2 \lambda + \ln(1-2\lambda) + \frac{2\lambda^2}{1 - 2 \lambda}\right]\\
  & \quad + \frac{A_q^{(1)} b_1}{2 \pi b_0^2} \frac{2 \lambda + \ln(1-2\lambda)}{1 - 2 \lambda}\ \ln\left(\frac{M^2}{\mu_R^2}\right) \\
  & \quad +\frac{A_q^{(1)}}{2 \pi} \left[\frac{\lambda}{1 - 2 \lambda}\ln^2\left(\frac{M^2}{\mu_R^2}\right) - \lambda\ln^2\left(\frac{\mu_F^2}{\mu_R^2}\right)\right]\,.
\end{split}\end{equation}

The process dependent term $h^{(2)}_{ab\rightarrow ij}$ appearing in \eqref{eq:Gab} can be expressed in terms of the soft anomalous dimension $\Gamma_{ab\rightarrow ij}$ associated with the partonic process $ab\rightarrow ij$, from which the Drell-Yan soft anomalous dimension $\Gamma^{\rm DY}_{ab}$ has been subtracted. The coefficient $h^{(2)}_{ab\rightarrow ij}$ hence reads\footnote{The coefficient $h^{(2)}_{ab\rightarrow ij}$ therefore vanishes for all Drell-Yan-like processes.}
\begin{equation}\begin{split}
    h^{(2)}_{ab\rightarrow ij}(\lambda) =&\  \frac{\ln(1-2\lambda)}{b_0} \ \frac{\Re (\Gamma_{ab\rightarrow ij} - \Gamma^{\rm DY}_{ab})}{\alpha_s}\\
    \equiv &\  \frac{\ln(1-2\lambda)}{b_0} \ \frac{\Re (\bar \Gamma_{ab\rightarrow ij})}{\alpha_s}\,,
\end{split}\label{h2ab}\end{equation}
with
\begin{equation}\begin{split}
 \Gamma^{\rm DY}_{ab} = \ \frac{\alpha_s}{2\pi} \sum_{k=\{a,b\}} C_k &\bigg[1 - \ln(2) - i\pi   \\
 &- \ln\left(\frac{(v_k \cdot n)^2}{|n|^2}\right)  \bigg].
\end{split}\end{equation}
For Drell-Yan-like processes computed at the aNNLO+NNLL accuracy (namely electroweakino and slepton pair production), the coefficient $h^{(3)}_{ab\rightarrow ij}$ must be included too. It is in this case universal, and it is given by
\begin{equation}\begin{split}
  & h^{(3)}_{ab\rightarrow ij}(\lambda) = - \frac{2C_F}{2 \pi^2 b_0}\ \frac{\lambda}{1 - 2 \lambda} \bigg[n_f\left(\frac{14}{27}-\frac23\zeta_2\right)\\
  & \qquad \ + C_A\left(-\frac{101}{27}+\frac{11}{3}\zeta_2+\frac{7}{2}\zeta_3\right)\bigg]\,.
\end{split}\end{equation}

In the next subsections, we collect the process-dependent hard matching coefficients appearing in~\eqref{eq:H} and~\eqref{eq:Cab}, expressions for the modified soft anomalous dimension $\bar \Gamma_{ab\rightarrow ij}$, as well as the ingredients allowing for the computation of the resummed cross section at a given order in $\alpha_s$, the latter being necessary for the matching of the resummed and fixed-order predictions shown in~\eqref{eq:match}.

\subsection{Supersymmetric Drell-Yan-like processes}
This subsection collects the formulas relevant for the purely electroweak production of pairs of sleptons and electroweakinos.

These processes do not feature any coloured final-state particle, and they only involve a quark-antiquark initial state at Born level. The associated hard matching coefficients $C^{(n)}_{q\bar{q}}$ are hence identical for all Dell-Yan-like processes, the process dependence being factorised in the Born cross section in~\eqref{eq:H}. The first two $C^{(n)}_{q\bar{q}}$ coefficients can be written in a compact form in Mellin space,
\begin{equation}\begin{split}
    & C^{(0)}_{q\bar{q}} = 1 \,,\\
    & C^{(1)}_{q\bar{q}} = C_F \left[\frac{4}{3}(\pi^2 - 6) - 3 \ln\left(\frac{\mu_F^2}{M^2}\right)\right] \,,\\
\end{split}\end{equation}
whereas the third coefficient is
\begin{equation}\begin{split}
    & C^{(2)}_{q\bar{q}} = \frac{C_F}{720} \bigg\{ 5(762~n_f - 4605~C_A + 4599~C_F) \\ 
    & \quad + 80 (151~C_A - 135~C_F + 2~n_f) \zeta_3\\
    &\quad + 20 \pi^2 (188~C_A - 297~C_F - 32~n_f) \\ 
    & \quad - 92 \pi^4 (C_A - 6~C_F) \\ 
    & \quad + 180 (11~C_A + 18~C_F - 2~n_f) \ln^2\left(\frac{\mu_F^2}{M^2}\right) \\ 
    & \quad - 160 (11~C_A - 2~n_f)(6 - \pi^2)\ln\left(\frac{\mu_R^2}{M^2}\right) \\ 
    & \quad + 20 \ln\left(\frac{\mu_F^2}{M^2}\right) \bigg[-51~C_A + 837~C_F + 6~n_f\\
    &\qquad - 4 \pi^2 (11~C_A + 27~C_F - 2~n_f) +\\
    &\qquad (-198~C_A + 36~n_f)\ln\left(\frac{\mu_R^2}{M^2}\right) \\
    &\qquad + 216 (C_A - 2~C_F)\zeta_3\bigg]  \bigg\}\,.
\end{split}\end{equation}

For SUSY Drell-Yan-like processes, fixed-order predictions are available at aNNLO in QCD. Calculations hence include SM NNLO contributions, but neglect NNLO SUSY-QCD corrections. The latter are however expected to be small, given current bounds on SUSY particles. The matching of the resummed cross sections at NNLL with the aNNLO fixed-order ones is performed by expanding the resummed cross section up to ${\cal O}(\alpha_s^2)$. Starting from the expressions given in \eqref{eq:G}, \eqref{eq:H} and \eqref{eq:Cab}, we get
\begin{equation}\begin{split}
   & \sigma^{\rm (exp.)}_{ab}(N,M^2,\mu^2_F, \mu_R^2) =\sigma^{(0)}_{ab\to ij}\  C_{ab\to ij}(M^2,\mu^2_F, \mu_R^2)\\
   &\qquad \quad\times \exp[G_{ab\to ij}(N,M^2,\mu^2_F, \mu_R^2)] \\
   &\quad  = \sigma^{(0)}_{ab\to ij} \bigg[1 + \left(\frac{\alpha_s}{2\pi}\right) C_{ab}^{(1)} + \left(\frac{\alpha_s}{2\pi}\right)^2 C_{ab}^{(2)} +\dots\bigg]\\
   &\qquad\quad \times \left[1 + \left(\frac{\alpha_s}{2\pi}\right) K^{(1)} + \left(\frac{\alpha_s}{2\pi}\right)^2 K^{(2)} +\dots\right] \\
   &\quad= \sigma^{(0)}_{ab}\bigg[1 + \left(\frac{\alpha_s}{2\pi}\right) \left(C_{ab}^{(1)} + K^{(1)}\right) \\
   &\qquad\quad + \left(\frac{\alpha_s}{2\pi}\right)^2 \left(C_{ab}^{(2)} + K^{(2)} + C_{ab}^{(1)} K^{(1)}\right) + \dots\bigg].
\end{split}\end{equation}
In the above expressions, we have omitted the arguments of the various functions from the second equality onward, in order to simplify the notation. The coefficients $K^{(n)}$ of the expanded exponential factor can be organised in powers of the logarithm $L = \ln(Ne^{\gamma_E})$, 
\begin{equation}\begin{split}    
 K^{(1)} = K^{(1,1)} L + K^{(1,2)} L^2\,, \\
 K^{(2)} \!=\! K^{(2,1)} L \!+\! K^{(2,2)} L^2 \!+\! K^{(2,3)} L^3 \!+\! K^{(2,4)} L^4,
\end{split}\end{equation}
with the various $N$-independent contributions being given by 
\begin{equation}\begin{split}  
 &K^{(1,1)} = 4 C_F \ln\left(\frac{\mu_F^2}{s}\right)\,, \\
 &K^{(1,2)} = 4 C_F\,, \\
 &K^{(2,1)} = -\frac{C_F}{27} \bigg\{56~n_f - 404~C_A + 378~C_A~\zeta_3 + \\ 
 & \quad + 3\ln\left(\frac{\mu_F^2}{s}\right) \Bigg[20~n_f + 2~C_A (-67 + 3\pi^2) \\
 & \quad + 3 (11~C_A - 2~n_f) \left(\ln\left(\frac{\mu_F^2}{\mu_R^2}\right) - \ln\left(\frac{\mu_R^2}{s}\right)\right)\Bigg] \bigg\}\,, \\
 &K^{(2,2)} = \frac{2}{9} C_F \Bigg[-10~n_f + 67~C_A - 3~C_A \pi^2\\
 &\quad +\! 36~C_F \ln^2\left(\frac{\mu_F^2}{s}\right) \!+\! (33~C_A \!-\! 6~n_f) \ln\left(\frac{\mu_R^2}{s}\right) \Bigg]\,, \\
 &K^{(2,3)} = \frac{4}{9} C_F \left[11~C_A - 2~n_f + 36~C_F \ln\left(\frac{\mu_F^2}{s}\right)\right]\,, \\
 &K^{(2,4)} = 8 C_F^2\,.
\end{split}\end{equation}

\subsection{Gluino-electroweakino associated production}\label{app:gone}

The soft anomalous dimension ${\Gamma}_{q\bar q\rightarrow \tilde{g}\tilde{\chi}}$ associated with the process $q\bar q\to \tilde{g}\tilde{\chi}$ can be calculated following standard techniques, from eikonal Feynman rules. The corresponding modified soft anomalous dimension $\bar{\Gamma}_{q\bar q\rightarrow \tilde{g}\tilde{\chi}}$ is given, after that the Drell-Yan contribution has been subtracted, by
\begin{equation}\begin{split}
  \bar{\Gamma}_{q\bar q\rightarrow \tilde{g}\tilde{\chi}} =  \frac{\alpha_s}{2\pi} C_A \Bigg[ &\ln\left(\frac{m_{\tilde{g}}^2 \!-\! t}{m_{\tilde{g}} \sqrt{s}}\right) \!+\! \\
  &\ln\left(\frac{m_{\tilde{g}}^2 \!-\! u}{m_{\tilde{g}} \sqrt{s}}\right) -1 + i\pi \Bigg]\,.
\end{split}\end{equation}

\noindent
For such a process and predictions at NLO+NLL, the hard matching coefficients can be derived from the collinear remainder originating from the ${\bf P}$ and ${\bf K}$ insertion operators defined in the language of dipole subtraction~\cite{Catani:1996vz, Catani:2002hc, Catani:2000ef}. This is achieved by restricting these operators to their $N$-independent pieces, and by adding to the result the finite part of the virtual corrections,
\begin{equation}\begin{split}
  & \mathcal C^{(0)}_{q\bar q}  =  1\,,\\
  & \frac{\alpha_s}{2\pi} \mathcal C^{(1)}_{q\bar q} \!=\!
   2 \Big\langle{\bf P} \!+\! {\bf K}\Big\rangle_{q, N\text{-ind.}}
   \!+\! \Big(\sigma^\text{V} \!+\! \int_1 \dd\sigma^\text{A} \Big) \,.
\end{split}\label{eq:PKgone}\end{equation}
In this expression, the arguments of the various functions are omitted for brevity. Moreover, $\sigma^V$ stands for the virtual corrections  contributing at NLO, and $\sigma^A$ consists of the auxiliary cross section shifting the divergences so that integration over phase space can be numerically achieved. Explicit expressions for the insertion operators ${\bf P}$ and ${\bf K}$ for an initial quark are
\begin{equation}\begin{split}
  & \langle {\bf P}\rangle_q = \frac{\alpha_s}{2\pi}\ \left(\ln\bar{N} - \frac{3}{4} \right)\\
  &\qquad \times  \left[(2C_F \!-\! C_A)\ln\frac{\mu_F^2}{s} \!+\! C_A \ln\frac{\mu_F^2}{m_{\tilde{g}}^2 \!-\! t}\right]\,,\\
 & \langle {\bf K}\rangle_q =  \frac{\alpha_s}{2\pi} \Bigg\{ 2 C_F \ln^2\bar{N}
     + C_F \left( \frac{2}{3} \pi^2 - 5\right)\\
  &\quad +  C_A \ln\bar{N}\left(1+\ln\frac{m_{\tilde{g}}^2}{m_{\tilde{g}}^2-t}\right)  \\
  &\quad + \frac{C_A}{4}\Bigg[1 + 4{\rm Li}_2\left(\frac{2m_{\tilde{g}}^2 - t}{m_{\tilde{g}}^2}\right)\\
  &\qquad + \left(1 \!+\! 2\frac{m_{\tilde{g}}^2}{m_{\tilde{g}}^2-t} \!+\! 4\ln\frac{m_{\tilde{g}}^2}{m_{\tilde{g}}^2-t} \right) \ln\frac{m_{\tilde{g}}^2}{2m_{\tilde{g}}^2-t}\\
  &\qquad  + 3\ln\frac{3m_{\tilde{g}}^2-t-2 m_{\tilde{g}} \sqrt{2m_{\tilde{g}}^2-t}}{m_{\tilde{g}}^2-t} \\
  &\qquad + 6\frac{m_{\tilde{g}}}{\sqrt{2m_{\tilde{g}}^2-t}+m_{\tilde{g}}} - 3 \bigg]
  \bigg\}\,,
\end{split}\end{equation}
where the Mandelstam variables are defined according to the ordering of the two-to-two partonic process $q\bar q\rightarrow \tilde{g}\tilde{\chi}$. Similar equations hold for $u$-channel contributions, that must be included as well when computing the $\langle{\bf P} \!+\! {\bf K}\rangle_q$ term in \eqref{eq:PKgone},
\begin{equation}
  \Big\langle{\bf P} \!+\! {\bf K}\Big\rangle_q = 
      \langle{\bf P}\rangle_q + \langle{\bf K}\rangle_q
    + \langle{\bf P}\rangle_q|_{t\leftrightarrow u} + \langle{\bf K}\rangle_q|_{t\leftrightarrow u}\,.
\end{equation}

\subsection{Squark-electroweakino associated production}

Similarly to \ref{app:gone}, the modified soft anomalous dimension $\bar{\Gamma}_{qg\rightarrow \tilde{q}\tilde{\chi}}$ associated with the process $qg\to \tilde{q}\tilde{\chi}$ is obtained from eikonal Feynman rules, and it is given by 
\begin{equation}\begin{split}
    \bar{\Gamma}_{qg\rightarrow \tilde{q}\tilde{\chi}} &= \frac{\alpha_s}{2\pi} C_F \left[ 2\ln\left(\frac{m_{\tilde{q}}^2 - t}{\sqrt{s} m_{\tilde{q}}}\right) - 1 + i\pi \right] \\ 
    &+ \frac{\alpha_s}{2\pi}  C_A  \ln\left(\frac{m_{\tilde{q}}^2 - u}{m_{\tilde{q}}^2 - t}\right)\,.
\end{split}\end{equation}
The hard-matching coefficients are similarly derived from the $N$-independent parts of the insertion operators ${\bf P}$ and ${\bf K}$,
\begin{equation}\begin{split}
  \mathcal C^{(0)}_{qg}  = &\ 1\,,\\
  \frac{\alpha_s}{2\pi} \mathcal C^{(1)}_{qg} &= 
   \Big\langle{\bf P} \!+\! {\bf K}\Big\rangle_{q, N\text{-ind.}} \!+\! \Big\langle{\bf P} \!+\! {\bf K}\Big\rangle_{g, N\text{-ind.}}\\
   &\!+\! \Big(\sigma^\text{V} \!+\! \int_1 \dd\sigma^\text{A} \Big) \,,
\end{split}\end{equation}
in which we use the explicit expressions
\begin{equation}\begin{split}
  &\langle{\bf P}\rangle_q = \frac{\alpha_s}{2\pi}\left(\ln\bar{N} - \frac{3}{4}\right)\\
    &\quad \times \left(2 C_F \ln\frac{\mu_F^2}{m_{\tilde{q}}^2 - t} - C_A \ln\frac{s}{{m_{\tilde{q}}^2 - t}}\right)\,, \\
  & \langle{\bf P}\rangle_{g} = \frac{\alpha_s}{2\pi}\left(C_A \ln\bar{N} \!-\! \frac{11}{4} \!+\! \frac{n_f}{6} \right)\ln\frac{\mu_F^4}{s(m_{\tilde{q}}^2 \!-\! u)}\,,\\
  & \langle{\bf K}\rangle_q =  \frac{\alpha_s}{2\pi}\Bigg\{ C_F \left(2\ln^2\bar{N} + \frac{2}{3}\pi^2 - 5 \right) \\
    & \   +\! \left(C_F \!-\! \frac{C_A}{2}\right)\! \left[2\ln\bar{N}\left(\!1 \!+\! \log\frac{m_{\tilde q}^2}{m_{\tilde q}^2\!-\!t}\!\right) \!+\! \mathcal{Q}\right]\! \Bigg\}\,,\\
  & \langle{\bf K}\rangle_g = \frac{\alpha_s}{2\pi}\frac{C_A}{2}\Bigg[ 4\ln^2\bar{N} \!+\! 2\ln\bar{N}\left(\!1 \!+\! \ln\frac{m_{\tilde q}^2}{m_{\tilde q}^2\!-\!u}\!\right) \\
    &\quad -\frac{100}{9} + \frac{4}{3}\pi^2 + \frac{16}{9}\frac{n_f}{C_A} + \mathcal{G}\Bigg]\,.
\end{split}\end{equation}
The two quantities $\mathcal{Q}$ and $\mathcal{G}$ appearing in these formulas are given by
\begin{equation}\begin{split}
 & \mathcal{Q} = 
     \frac{m_{\tilde q}^2 - t}{2 m_{\tilde q}^2-t} 
   + \frac{3 m_{\tilde q}}{m_{\tilde q} + \sqrt{2 m_{\tilde q}^2-t}} - 2
   \\ &\
   + \ln\frac{m_{\tilde q}^2}{2 m_{\tilde q}^2\!-\!t} \left(1 \!+\! 2\ln\frac{m_{\tilde q}^2}{m_{\tilde q}^2\!-\!t}\right) 
   + 2{\rm Li}_2 \frac{2 m_{\tilde q}^2\!-\!t}{m_{\tilde q}^2}
   \\ &\ 
   - \frac{3}{2}\ln\frac{3m_{\tilde q}^2 -t - 2m_{\tilde q} \sqrt{2 m_{\tilde q}^2-t}}{m_{\tilde q}^2-t} \,,\\
 & \mathcal{G} =
     \frac{m_{\tilde q}^2\!-\!u}{2m_{\tilde q}^2\!-\!u} 
     \!+\!
     \ln\frac{m_{\tilde q}^2}{2 m_{\tilde q}^2 \!-\!u}\left(\!1 \!+\! 2 \ln\frac{m_{\tilde q}^2}{m_{\tilde q}^2-u}\!\right) \!-\! 2 
   \\ &\
   +\! 2{\rm Li}_2 \frac{2m_{\tilde q}^2\!-\!u}{m_{\tilde q}^2}  
   \!+\! \left(\! \frac{11}{6} \!-\! \frac{n_f}{3 C_A}\right)  \frac{2m_{\tilde q}}{\sqrt{2m_{\tilde q}^2\!-\!u} \!+\! m_{\tilde q}}
   \\ &\ +\! \left(\frac{11}{6} \!-\! \frac{n_f}{3 C_A}\right) \ln\frac{3m_{\tilde q}^2\!-\!u \!-\! 2m_{\tilde q}\sqrt{2m_{\tilde q}^2\!-\!u}}{m_{\tilde q}^2\!-\!u}\,,
\end{split}\end{equation}
where the Mandelstam variables are defined from the two-to-two partonic process $qg\to \tilde{q}\tilde{\chi}$.

\section{Architecture and structure of the \resummino sources files} \label{app:code}

The source code of \resummino is available from the folder \verb+src+ of any local installation of the code. It includes the following set of files, that we briefly describe.
\begin{itemize}

  \item \texttt{constants.h}: Constants useds throughout the code, such as colour factors and the regulator width used for on-shell subtraction~\cite{Gavin:2013kga}.\vspace{.1cm}

  \item \texttt{dipoles.cc}: NLO dipoles in the Catani-Seymour formalism for the production of heavy SUSY particles \cite{Catani:1996vz, Catani:2002hc}.\vspace{.1cm}

  \item \texttt{hxs.cc}: Phase space implementations and integration routines relevant for resummed cross sections matched to NLO fixed-order predictions evaluated within the dipole formalism. It thus includes separate components for the partonic LO cross section, the associated virtual corrections (to which integrated dipoles are added), the collinear remainders, the real emission corrections (from which dipole contributions are subtracted), and threshold resummation components. The integration relevant for total rate calculations is performed through the function \texttt{hadronic\_xs}. Integrations associated with transverse momentum and invariant mass differential cross sections are similarly implemented in the files \texttt{hxs\_dpt2.cc} and \texttt{hxs\_dlnm2.cc} respectively.\vspace{.1cm}
  
  \item \texttt{kinematics.cc}: Calculation of the kinematics inherent to the different partonic configurations relevant for a given process, and momentum reshuffling as required by the regularisation of infrared divergences in the dipole subtraction formalism. \vspace{.1cm}

  \item \texttt{integration\_method.cc}: General-purpose wrapper for Monte Carlo integration through the \verb+GSL+ routine \verb+VEGAS+.\vspace{.1cm}

  \item \texttt{main.cc}: Initialisation of the code within its main function.\vspace{.1cm}

  \item \texttt{maths.h}: Useful mathematical functions.\vspace{.1cm}

  \item \texttt{options.h}: Different options for the implementation of on-shell resonance subtraction from the real emission corrections to squark-electroweakino associated production.\vspace{.1cm}
  
  \item \texttt{params.cc}: Values of all SUSY and SM couplings and masses (see also details in section~\ref{sec:params}).\vspace{.1cm}

  \item \texttt{pdf.cc}: Initialisation of the PDF functions, interface with \verb+LHAPDF+, and fit of the PDFs in Mellin space.\vspace{.1cm}

  \item \texttt{resummation.cc}: Implementation of the resummed exponent defined in~\eqref{eq:HtimesG}, and the hard-matching coefficients defined in \eqref{eq:C}.\vspace{.1cm}

  \item \texttt{resummino.cc}: Parsing of the input file and calls to the functions appropriate for the process and calculation chosen (total rate, $p_T$ distribution, {\it etc.}).\vspace{.1cm}

  \item \texttt{utils.h}: Additional functions allowing for internal checks.
\end{itemize}

Specific functions relevant for each of the supported processes, and including Born, NLO-real and NLO-virtual contributions, are collected in the sub-folders \texttt{gaugino\_gluino} ($\gaugino\gluino$ production), \texttt{gaugino\_squark} ($\gaugino\squark$ production), \texttt{gauginos} ($\gaugino\gaugino$ production), \texttt{sleptons} ($\slepton\slepton$ production) and \texttt{leptons} (dilepton production including $Z^\prime$ and $W^\prime$ exchange contributions).
\section{Configuration of the \resummino run}
\label{app:input}

In this appendix, we describe how to write a configuration file for \resummino. Such a file includes a set of equalities allowing to set keywords to user-defined values. A first ensemble of keywords is dedicated to the definition of the hadronic process and the type of calculation to be performed. They are given by:
\begin{itemize} 
  \item \texttt{collider\_type = <string>}: information on the initial-state hadrons. Allowed values are \texttt{proton-proton} or \texttt{proton-antiproton}.\vspace{.1cm}
  
  \item \texttt{center\_of\_mass\_energy = <double>}: hadronic centre-of-mass energy in GeV.\vspace{.1cm}
  
  \item \texttt{particle1 = <int>}  and \verb+particle2 = <int>+: nature of the first and second outgoing particles, given through their PDG identifiers~\cite{ParticleDataGroup:2022pth}.\vspace{.1cm}

  \item \verb+result = <string>+: defines the cross section to compute. Allowed values are \texttt{total} (total cross section using the threshold resummation formalism), \texttt{pt} (differential cross section $\dd \sigma/\dd p_T$ using the $p_T$ resummation formalism), \texttt{ptj} (differential cross section $\dd \sigma/\dd p_T$ using the joint resummation formalism) and \texttt{m} (differential cross section $\dd \sigma/\dd M$ using the threshold resummation formalism). \vspace{.1cm}

  \item \texttt{pt = <double>}: transverse momentum in GeV, used for differential cross section calculations at fixed $p_T$. This variable is only relevant when the \texttt{result} parameter is set to \texttt{pt} or \texttt{ptj}. \vspace{.1cm}

  \item \texttt{M = <double>}: invariant mass in GeV, used for differential cross section calculations at fixed $M$. The invariant mass of the final-state system is used by enforcing the specific value \texttt{auto}.\vspace{.1cm}

  \item \texttt{Minv\_min = <double>} and \texttt{Minv\_max = <double>}: invariant-mass window, given in GeV, that is used for total cross section computations relevant for the production of a Drell-Yan pair of leptons. Such parameters can also be set to \texttt{auto}, which corresponds to $M_{\rm min} = (3/4) M_{Z^\prime}$ and $M_{\rm max} = (5/4) M_{Z^\prime}$.
\end{itemize}

Information on the SUSY properties or on the extra gauge boson properties is provided through files encoded in an SLHA(-like) format. Two variables allow users to provide information about the path to these files:
\begin{itemize}
  \item \texttt{slha = <string>}: path to the SLHA input file containing information on the SUSY scenario considered (masses, widths, parameters, {\it etc.}).\vspace{.1cm}
  \item \texttt{zpwp = <string>}: path to the SLHA-like input file containing information on the properties of extra $Z^\prime$ and $W^\prime$ bosons.
\end{itemize}

A third ensemble of keywords are related to the choice of the parton density sets to be used in given calculations, and how their Mellin transform is handled. They are:
\begin{itemize} 
  \item \texttt{pdf\_format = <string>}: format according to which the \textsc{LHAPDF} grids are encoded. Allowed values include \textsc{LHpdf} (default) and \textsc{LHgrid} (deprecated option relevant for \textsc{LHAPDF} versions anterior to 6).\vspace{.1cm}
  
  \item \texttt{pdf\_lo = <string>} and \texttt{pdf\_nlo = <string>}: name of the PDF set to be employed for LO and higher-order calculations respectively, according to the naming scheme of the webpage \url{https://lhapdf.hepforge.org/pdfsets}.\vspace{.1cm}
  
  \item \texttt{pdfset\_lo = <int>} and \texttt{pdfset\_nlo = <int>}: specific PDF set member to be used in LO and higher-order calculations respectively. These integers allow for the selection of the best-fit set or of one of the next-to-best-fit sets provided within a given \textsc{LHAPDF} set, in the context of PDF error estimates.\vspace{.1cm}
  
  \item \texttt{weight\_gluon = <double>}, \texttt{weight\_sea = <double>} and \texttt{weight\_valence = <double>}: weights used in the respective fits of the gluon, sea quark and valence quark densities according to the Levenberg-Marquard algorithm. The (recommended) default values are \texttt{-1.6}. \vspace{.1cm}
  
  \item \texttt{xmin = <double>}: minimum Bjorken-$x$ value to consider in the PDF fitting procedure. When set to \texttt{auto} (default), the code uses $x_{\textrm min}=M^2/S$ where $M$ is the invariant mass of the pair of produced particles.
\end{itemize} 

Renormalisation and factorisation scales are automatically set to specific values for any calculation achieved by \resummino (see section~\ref{subsec:ctype}). Those values can however be multiplied by user-defined factors given through two self-explanatory variables:
\begin{itemize} 
\item \texttt{mu\_f = <double>} and \texttt{mu\_r = <double>}: factors multiplying the factorisation and renormalisation scale values specific to the calculation considered. Default values are \texttt{1.0}.
\end{itemize} 

Finally, the numerical precision of the Monte Carlo integration achieved with the \texttt{VEGAS} algorithm is controlled through two input variables:
\begin{itemize}
	\item \texttt{precision = <double>}: relative precision of the Monte Carlo integration (recommended value: \texttt{0.001} $\equiv 0.1\%$). The integration routine stops when the relative numerical error of the result matches the requested precision. For contributions beyond LO, this error is estimated with respect to the tree-level result, which ensures that no computing power is wasted on small contributions.\vspace{.1cm}
 
	\item \texttt{max\_iters = <int>}: maximum number of integration iterations, excluding the warm-up phase (recommended value: \texttt{50}). After the maximum number of iterations, Monte Carlo integration is stopped, even if the desired precision is not reached.
\end{itemize}

\section{Resummino output} \label{app:output} 

In this appendix, we detail the (screen) output generated by \resummino in the context of the calculations of the total cross section associated with slepton pair-production at the LHC, operating at a centre-of-mass energy of \SI{13}{TeV}. We choose to make use of the LO and NLO sets of CT14 parton densities~\cite{Dulat:2015mca}, and to rely on one of the SUSY scenarios whose SLHA input file is provided in the \texttt{input} folder of the \resummino source files. 

The used input file including all computational settings is available from the folder \texttt{input/resummino.in}. It reads:
\begin{verbatim}
   collider_type         =  proton-proton 
   center_of_mass_energy =  13000 # in GeV
   particle1             =  1000011 # slepton
   particle2             = -1000011 # slepton
 
   result = total   # total, pt, ptj or m
   M      = auto    # auto (M^2 = (p1 + p2)^2)
   pt     = auto

   slha = slha.in   # SLHA SUSY spectrum
   zpwp = ssm.in    # Z'/W' model parameters

   pdf_format = lhgrid  # lhgrid or lhpdf
   pdf_lo     = CT14lo
   pdfset_lo  = 0
   pdf_nlo    = CT14nlo
   pdfset_nlo = 0
 
   mu_f = 1.0   # central scale
   mu_r = 1.0   # central scale
 
   precision = 0.00001  # desired precision
   max_iters = 100  # maximum iterations
 
   Minv_min = auto
   Minv_max = auto 

   weight_valence = -1.6
   weight_sea     = -1.6
   weight_gluon   = -1.6
   xmin           = auto
\end{verbatim}
We remind that several entries (like \verb+zpwp+ or \verb+Min_min+ and \verb+Minv_max+) are irrelevant for the SUSY process considered. The code is started by typing in a shell
\begin{verbatim}
    ./bin/resummino input/resummino.in
\end{verbatim}

\begin{figure*}\centering
\begin{verbbox}
           single.......iteration                   accumulated......results   
    iteration     integral    sigma             integral         sigma     chi-sq/it

       1        2.3648e-04   9.11e-06          2.3648e-04      9.11e-06    0.00e+00
       2        2.5402e-04   3.27e-06          2.5201e-04      3.08e-06    3.28e+00
       3        2.5908e-04   2.01e-06          2.5697e-04      1.68e-06    3.49e+00
       4        2.5453e-04   1.95e-06          2.5593e-04      1.27e-06    2.63e+00
       5        2.5528e-04   1.98e-06          2.5574e-04      1.07e-06    1.99e+00

    num_dim=3, calls=20000, it_num=1, max_it_num=5 verb=0, alph=1.50,
    mode=0, bins=50, boxes=1

           single.......iteration                   accumulated......results   
    iteration     integral    sigma             integral         sigma     chi-sq/it

       1        2.5710e-04   6.19e-07          2.5710e-04      6.19e-07    0.00e+00
       2        2.5659e-04   6.05e-07          2.5684e-04      4.33e-07    3.39e-01
       3        2.5642e-04   5.95e-07          2.5669e-04      3.50e-07    3.31e-01
       4        2.5676e-04   6.17e-07          2.5671e-04      3.04e-07    2.24e-01
       5        2.5605e-04   6.26e-07          2.5658e-04      2.74e-07    3.96e-01
    LO = (2.5658342e-04 +- 2.7361523e-07) pb
\end{verbbox}
\theverbbox
\caption{\resummino screen output relevant for the calculation of the LO total rate associated with slepton pair production at the LHC, using the SUSY spectrum encoded in the file \texttt{input/slha.in} and the collider settings provided in the file \texttt{input/resummino.in}, both files being shipped with the program. \label{fig:screenoutput}}\vspace{.5cm}
\begin{verbbox}
    Performing PDF fit with 5 flavors with M^2/S = 0.0075293, Q^2 = 318113
      and weights: valence: x^-1.6, sea: x^-1.6, gluon: x^-1.6 and xmin = 0.0075293 
      Fit function: f = A0 * x^A1 * (1 - x)^A2 * 
      ( 1 + A3 * x^(1/2) + A4 * x + A5 * x^(3/2) + A6 * x^2 + A7 * x^(5/2) )
    Fitting gluon PDF...#chisq/dof = 9.05796e-09
    #status = success
\end{verbbox}
\theverbbox
\caption{Screen output relevant for the fit of the gluon density in the case of slepton pair production at the LHC operating at $\sqrt{S}=\SI{13}{TeV}$, the details on the calculation being encoded in the file \texttt{input/resummino.in} and the SUSY spectrum in the SLHA file \texttt{input/slha.in}. \label{fig:screenoutput2}}
\end{figure*}

When executed, \resummino begins by printing to the screen its banner (that includes information on the articles to cite when using the code), as well as \textsc{LHAPDF} settings (path to the PDF grids, information on the chosen PDF set, value of the strong coupling, {\it etc}.) used for the first calculation achieved by the code, namely the LO rate associated with the process considered. This gives in our case
\begin{verbatim}
 LHAPDF 6.3.0 loading <lhapth>/CT14lo_0000.dat
 CT14lo PDF set, member #0, version 1; \
     LHAPDF ID = 13200
 alpha_s(M_Z^2) = 0.1180   for LO PDF set 
 alpha_s(564.015^2) = 0.0927  for NLO PDF set 
\end{verbatim}
where \verb+<lhapth>+ corresponds to the path to the folder containing the PDF grid \verb+CT14lo_0000.dat+, that is used for the evaluation of the LO cross section.

Subsequently, \resummino displays to the screen the results of each \texttt{VEGAS} iteration, together with its numerical error, as well as the accumulated result obtained by combining all \texttt{VEGAS} iterations. The input file provided above leads to the screen output shown in figure~\ref{fig:screenoutput}, in which the first five iterations are related to the warm-up stage of the numerical integration process, and the last five to the actual calculation. The latter stops when the required precision is reached. 

Next, \textsc{LHAPDF} information on the NLO PDF set used is printed to the screen, followed by a similar output as in figure~\ref{fig:screenoutput} for each component of the full fixed-order calculation. These consist of a recalculation of the LO rate with NLO parton densities, the sum of the virtual corrections and the integrated dipoles (in the language of the dipole subtraction formalism~\cite{Catani:1996vz, Catani:2002hc, Catani:2000ef}), the collinear remainder stemming from the $\mathbf{P}$ and $\mathbf{K}$ insertion operators (still in the language of the dipole subtraction formalism), and the difference between the real gluon emission corrections and the associated dipoles, and that between the real quark emission corrections and the associated dipoles. In the case in which on-shell resonant contributions appear in the real emission contributions, their integration is handled separately for improved convergence.

The code then moves with the calculation of the resummed components of the cross section, which starts with a fit of the PDFs so that they could be transformed to Mellin space (see section~\ref{sec:resum}). Information on this fit is printed to the screen, as shown in figure~\ref{fig:screenoutput2} for the case of the gluon density. The cross section is then evaluated at NLO+NLL using the collinear-improved resummation formalism of~\cite{Kramer:1996iq, Catani:2001ic, Kulesza:2002rh, Almeida:2009jt}, or at aNNLO+NNLL using the standard threshold resummation formalism of~\cite{Sterman:1986aj, Catani:1989ne, Catani:1990rp, Kidonakis:1997gm, Kidonakis:1998bk, Vogt:2000ci} if the code is run with its \verb|--nnll| option (see \ref{app:cli}). 

The matched results are finally displayed to the screen,\footnote{The number of digits in the results has been truncated to improve readability.}
\begin{verbatim}   
Results:
 LO = (2.56583e-04 +- 2.73615e-07) pb
 NLO = (2.70544e-04 +- 2.72217e-07) pb
 aNNLO+NNLL = (2.71317e-04 +- 2.72777e-07) pb
\end{verbatim}
the last line being replaced by an NLO+NLL prediction if the aNNLO+NNLL one is not required or unavailable.
\section{Options available from the command line interface of \resummino}
\label{app:cli}

The \resummino program can be executed with several options, by typing in a shell
\begin{verbatim}
    ./bin/resummino <input-file> [options]
\end{verbatim}
The keyword \verb+<input-file>+ provides the path to a configuration file detailing the calculation to be achieved, and that must be encoded following the guidelines sketched in section~\ref{sec:input} and \ref{app:input}. All allowed options are listed below. 
\begin{itemize} 
    \item \texttt{-{}-center\_of\_mass\_energy} (or \texttt{-e}), followed by a double-precision number: this sets the hadronic centre-of-mass energy, given in GeV.\vspace{.1cm}

    \item \texttt{-{}-help} (or \texttt{-h}): this displays a help message to the screen, indicating how to run the code.\vspace{.1cm}

    \item \texttt{-{}-invariant-mass} (or \texttt{-m}), followed by a double-precision number: this sets the value, in GeV, of the final-state invariant mass for calculations at fixed invariant mass.\vspace{.1cm}

    \item \texttt{-{}-lo} (or \texttt{-l}): this enforces a calculation at LO.\vspace{.1cm}

    \item \texttt{-{}-mu\_f} (or \texttt{-f}) and \texttt{-{}-mu\_r} (or \texttt{-r}), both followed by a double-precision number: this determines the factors multiplying the factorisation and renormalisation scales, relatively to central scale choices.\vspace{.1cm}

    \item \texttt{-{}-nlo} (or \texttt{-n}): this enforces a calculation at NLO.\vspace{.1cm}

    \item \texttt{-{}-nll} (or \texttt{-s}): this enforces a calculation at NLO+NLL.\vspace{.1cm}

    \item \texttt{-{}-nnll} (or \texttt{-z}): this enforces a calculation at aNNLO+NNLL.\vspace{.1cm}

    \item \texttt{-{}-output} (or \texttt{-o}), followed by a string referring to the path to a specific folder: this defines the folder in which all output files created by \resummino will be stored.\vspace{.1cm}

    \item \texttt{-{}-parameter-log} (or \texttt{-p}), followed by a string representing the path to a file: this makes the code writing all the parameters inherent to the calculation considered in the provided file.\vspace{0.1cm}

    \item \texttt{-{}-particle1} (or \texttt{-c}) and \texttt{-{}-particle2} (or \texttt{-d}), followed by integer numbers: modifications of the nature of the first and second final-state particles through a change in their PDG identifiers.\vspace{.1cm}

    \item \texttt{-{}-pdfset\_lo} (or \texttt{-a}) and \texttt{-{}-pdfset\_nlo} (or \texttt{-b}), both followed by an integer: this determines the exact PDF set, within a given collection of {\tt LHAPDF} parton densities, to employ for calculations at LO and beyond respectively.\vspace{.1cm}

    \item \texttt{-{}-transverse-momentum} (or \texttt{-t}), followed by a double-precision number: this sets the value, in GeV, of the final-state transverse momentum, for calculations at fixed $p_T$.\vspace{.1cm}

    \item \texttt{-{}-version} (or \texttt{-v}): this displays the \emph{Resummino} version number to the screen.

\end{itemize} 

\section{Lists of total cross sections}\label{app:table}

\makeatletter\let\expandableinput\@@input\makeatother

In this appendix we collect total cross section predictions for the different SUSY processes studied in this article, and present them in a tabulated form in which scale and PDF uncertainties are separate. Their digitised version can be found on the webpage \url{https://github.com/APN-Pucky/HEPi/tree/master/hepi/data/json}. 

\renewcommand{\arraystretch}{1.60}
\begin{table*}
\begin{center}
	\begin{tabular*}{\textwidth}{@{\extracolsep{\fill}}cccc@{}}
		$m_{\tilde l}$ [GeV]  &  \multicolumn{3}{c}{aNNLO+NNLL$_{-\text{scale}-\text{PDF}}^{+\text{scale}+\text{PDF}}$ [fb]}  \\
		  &  $pp\to\tilde e_L^- \tilde e_L^+$ & $pp\to\tilde e_R^- \tilde e_R^+$& $pp\to\tilde \tau_1^- \tilde \tau_1^+$ \\
		\hline \expandableinput tab_sleptons13000.tex 
		\hline \expandableinput tab_sleptons13600.tex 
	\end{tabular*}
\end{center}
\caption{aNNLO+NNLL total cross sections for slepton pair production at the LHC, for $\sqrt{S} = \SI{13.0}{TeV}$ (upper) and \SI{13.6}{TeV} (lower). All other SUSY particles are decoupled.} \label{tab:sleptons}
\end{table*}

\begin{table*}
\begin{center}
	\begin{tabular*}{\textwidth}{@{\extracolsep{\fill}}ccccc@{}} 
		$m_{\tilde \chi}$ [GeV]  &  \multicolumn{4}{c}{aNNLO+NNLL$_{-\text{scale}-\text{PDF}}^{+\text{scale}+\text{PDF}}$ [fb]}  \\
		&  $pp\to\tilde \chi_1^0 \tilde \chi_1^+$ & $pp\to\tilde \chi_1^0 \tilde \chi_1^-$& $pp\to\tilde \chi_1^+ \tilde \chi_1^-$& $pp\to\tilde \chi_1^0 \tilde \chi_2^0$ \\
		\hline \expandableinput tab_hinos13000.tex 
        \hline \expandableinput tab_hinos13600.tex 
	 \hline
	\end{tabular*}
\end{center}
\caption{aNNLO+NNLL total cross sections for the production of a pair of mass-degenerate higgsinos at the LHC, for $\sqrt{S} = \SI{13.0}{TeV}$ (upper) and \SI{13.6}{TeV} (lower). Higgsinos are defined as in eq.~\eqref{eq:hino}, and all other SUSY particles are decoupled.}\label{tab:hinos}
\end{table*}

\renewcommand{\arraystretch}{1.50}
\begin{table*}
\begin{center}
	\begin{tabular*}{\textwidth}{@{\extracolsep{\fill}}ccccccc@{}} 
		$m_{\tilde \chi^0_1}$ [GeV] &$m_{\tilde \chi^0_2}$ [GeV] &$m_{\tilde \chi^\pm_1}$ [GeV]  &  \multicolumn{4}{c}{aNNLO+NNLL$_{-\text{scale}-\text{PDF}}^{+\text{scale}+\text{PDF}}$ [fb]}  \\
		&&&  $pp\to\tilde \chi_2^0 \tilde \chi_1^+$ & $pp\to\tilde \chi_2^0 \tilde \chi_1^-$& $pp\to\tilde \chi_1^+ \tilde \chi_1^-$& $pp\to\tilde \chi_1^0 \tilde \chi_2^0$ \\
	    \hline \expandableinput tab_higgsino_split_13000.tex
	 	\hline \expandableinput tab_higgsino_split_13600.tex
\end{tabular*}
\end{center}
\caption{aNNLO+NNLL total cross sections for the production of a pair of compressed higgsino states at the LHC, for $\sqrt{S} = \SI{13.0}{TeV}$ (upper) and \SI{13.6}{TeV} (lower). Higgsinos are defined as in eq.~\eqref{eq:hino2}, and all other SUSY particles are decoupled.} \label{tab:hinos_split}
\end{table*}

\renewcommand{\arraystretch}{1.40}
\begin{table*}
\begin{center}
	\begin{tabular*}{\textwidth}{@{\extracolsep{\fill}}cccc@{}} 
		$m_{\tilde \chi}$ [GeV] &  \multicolumn{3}{c}{aNNLO+NNLL$_{-\text{scale}-\text{PDF}}^{+\text{scale}+\text{PDF}}$ [fb]}  \\
		&  $pp\to\tilde \chi_2^0 \tilde \chi_1^+$ & $pp\to\tilde \chi_2^0 \tilde \chi_1^-$& $pp\to\tilde \chi_1^+ \tilde \chi_1^-$ \\
		\hline \expandableinput tab_winos13000.tex 
        \hline \expandableinput tab_winos13600.tex 
	 \hline
	\end{tabular*}
\end{center}
\caption{aNNLO+NNLL total cross sections for the production of a pair of mass-degenerate winos at the LHC, for $\sqrt{S} = \SI{13.0}{TeV}$ (upper) and \SI{13.6}{TeV} (lower). Winos are defined as in eq.~\eqref{eq:gino}, and all other SUSY particles are decoupled.} \label{tab:winos}
\end{table*}

\begin{table*}
\begin{center}
	\begin{tabular*}{\textwidth}{@{\extracolsep{\fill}}ccccc@{}} 
		$m_{\squark}$ [GeV] & $m_{\tilde \chi}$ [GeV] & \multicolumn{3}{c}{aNNLO+NNLL$_{-\text{scale}-\text{PDF}}^{+\text{scale}+\text{PDF}}$ [fb]}  \\
		&&  $pp\to\tilde \chi_2^0 \tilde \chi_1^+$ & $pp\to\tilde \chi_2^0 \tilde \chi_1^-$& $pp\to\tilde \chi_1^+ \tilde \chi_1^-$ \\
		\hline \expandableinput tab_winosq13000.tex
	    \hline \expandableinput tab_winosq13600.tex
	\end{tabular*}
\end{center}
\caption{
	aNNLO+NNLL total cross sections for the production of a pair of mass-degenerate winos at the LHC, for $\sqrt{S} = \SI{13.0}{TeV}$ (upper) and \SI{13.6}{TeV} (lower). Winos are defined as in eq.~\eqref{eq:gino}, and the spectrum features mass-degenerate first- and second-generation squarks. All other SUSY particles are decoupled.}\label{tab:wino_sq}
\end{table*}


\bibliographystyle{JHEP} 
\bibliography{References}

\providecommand{\href}[2]{#2}\begingroup\raggedright\begin{thebibliography}{10}

\bibitem{Nilles:1983ge}
H.~P. Nilles, \emph{{Supersymmetry, Supergravity and Particle Physics}},
  \href{http://dx.doi.org/10.1016/0370-1573(84)90008-5}{\emph{Phys. Rept.} {\bf
  110} (1984) 1--162}.

\bibitem{Haber:1984rc}
H.~E. Haber and G.~L. Kane, \emph{{The Search for Supersymmetry: Probing
  Physics Beyond the Standard Model}},
  \href{http://dx.doi.org/10.1016/0370-1573(85)90051-1}{\emph{Phys. Rept.} {\bf
  117} (1985) 75--263}.

\bibitem{Fuks:2013vua}
B.~Fuks, M.~Klasen, D.~R. Lamprea and M.~Rothering, \emph{{Precision
  predictions for electroweak superpartner production at hadron colliders with
  Resummino}},
  \href{http://dx.doi.org/10.1140/epjc/s10052-013-2480-0}{\emph{Eur. Phys. J.
  C} {\bf 73} (2013) 2480}, [\href{http://arxiv.org/abs/1304.0790}{{\tt
  1304.0790}}].

\bibitem{ATLAS:2019lff}
{\scshape ATLAS} collaboration, G.~Aad et~al., \emph{{Search for electroweak
  production of charginos and sleptons decaying into final states with two
  leptons and missing transverse momentum in $\sqrt{s}=13$ TeV $pp$ collisions
  using the ATLAS detector}},
  \href{http://dx.doi.org/10.1140/epjc/s10052-019-7594-6}{\emph{Eur. Phys. J.
  C} {\bf 80} (2020) 123}, [\href{http://arxiv.org/abs/1908.08215}{{\tt
  1908.08215}}].

\bibitem{ATLAS:2021yqv}
{\scshape ATLAS} collaboration, G.~Aad et~al., \emph{{Search for charginos and
  neutralinos in final states with two boosted hadronically decaying bosons and
  missing transverse momentum in $pp$ collisions at $\sqrt {s}$ = 13\,\,TeV
  with the ATLAS detector}},
  \href{http://dx.doi.org/10.1103/PhysRevD.104.112010}{\emph{Phys. Rev. D} {\bf
  104} (2021) 112010}, [\href{http://arxiv.org/abs/2108.07586}{{\tt
  2108.07586}}].

\bibitem{CMS-PAS-SUS-21-008}
{\scshape CMS} collaboration, \emph{{Combined search for electroweak production
  of winos, binos, higgsinos, and sleptons in proton-proton collisions at
  $sqrt{s}=$ 13 TeV}},
  \href{https://cds.cern.ch/record/2853345}{CMS-PAS-SUS-21-008}.

\bibitem{Barger:1983wc}
V.~D. Barger, R.~W. Robinett, W.-Y. Keung and R.~J.~N. Phillips,
  \emph{{Production of Gauge Fermions at Colliders}},
  \href{http://dx.doi.org/10.1016/0370-2693(83)90519-1}{\emph{Phys. Lett. B}
  {\bf 131} (1983) 372}.

\bibitem{Dawson:1983fw}
S.~Dawson, E.~Eichten and C.~Quigg, \emph{{Search for Supersymmetric Particles
  in Hadron - Hadron Collisions}},
  \href{http://dx.doi.org/10.1103/PhysRevD.31.1581}{\emph{Phys. Rev. D} {\bf
  31} (1985) 1581}.

\bibitem{Beenakker:1999xh}
W.~Beenakker, M.~Klasen, M.~Kramer, T.~Plehn, M.~Spira and P.~M. Zerwas,
  \emph{{The Production of charginos / neutralinos and sleptons at hadron
  colliders}},
  \href{http://dx.doi.org/10.1103/PhysRevLett.100.029901}{\emph{Phys. Rev.
  Lett.} {\bf 83} (1999) 3780--3783},
  [\href{http://arxiv.org/abs/hep-ph/9906298}{{\tt hep-ph/9906298}}].

\bibitem{Baglio:2016rjx}
J.~Baglio, B.~J\"ager and M.~Kesenheimer, \emph{{Electroweakino pair production
  at the LHC: NLO SUSY-QCD corrections and parton-shower effects}},
  \href{http://dx.doi.org/10.1007/JHEP07(2016)083}{\emph{JHEP} {\bf 07} (2016)
  083}, [\href{http://arxiv.org/abs/1605.06509}{{\tt 1605.06509}}].

\bibitem{Frixione:2019fxg}
S.~Frixione, B.~Fuks, V.~Hirschi, K.~Mawatari, H.-S. Shao, P.~A. Sunder et~al.,
  \emph{{Automated simulations beyond the Standard Model: supersymmetry}},
  \href{http://dx.doi.org/10.1007/JHEP12(2019)008}{\emph{JHEP} {\bf 12} (2019)
  008}, [\href{http://arxiv.org/abs/1907.04898}{{\tt 1907.04898}}].

\bibitem{Debove:2010kf}
J.~Debove, B.~Fuks and M.~Klasen, \emph{{Threshold resummation for gaugino pair
  production at hadron colliders}},
  \href{http://dx.doi.org/10.1016/j.nuclphysb.2010.08.016}{\emph{Nucl. Phys. B}
  {\bf 842} (2011) 51--85}, [\href{http://arxiv.org/abs/1005.2909}{{\tt
  1005.2909}}].

\bibitem{Fuks:2012qx}
B.~Fuks, M.~Klasen, D.~R. Lamprea and M.~Rothering, \emph{{Gaugino production
  in proton-proton collisions at a center-of-mass energy of 8 TeV}},
  \href{http://dx.doi.org/10.1007/JHEP10(2012)081}{\emph{JHEP} {\bf 10} (2012)
  081}, [\href{http://arxiv.org/abs/1207.2159}{{\tt 1207.2159}}].

\bibitem{Fiaschi:2018hgm}
J.~Fiaschi and M.~Klasen, \emph{{Neutralino-chargino pair production at NLO+NLL
  with resummation-improved parton density functions for LHC Run II}},
  \href{http://dx.doi.org/10.1103/PhysRevD.98.055014}{\emph{Phys. Rev. D} {\bf
  98} (2018) 055014}, [\href{http://arxiv.org/abs/1805.11322}{{\tt
  1805.11322}}].

\bibitem{Fiaschi:2020udf}
J.~Fiaschi and M.~Klasen, \emph{{Higgsino and gaugino pair production at the
  LHC with aNNLO+NNLL precision}},
  \href{http://dx.doi.org/10.1103/PhysRevD.102.095021}{\emph{Phys. Rev. D} {\bf
  102} (2020) 095021}, [\href{http://arxiv.org/abs/2006.02294}{{\tt
  2006.02294}}].

\bibitem{Baer:1993ew}
H.~Baer, C.-h. Chen, F.~Paige and X.~Tata, \emph{{Detecting Sleptons at Hadron
  Colliders and Supercolliders}},
  \href{http://dx.doi.org/10.1103/PhysRevD.49.3283}{\emph{Phys. Rev. D} {\bf
  49} (1994) 3283--3290}, [\href{http://arxiv.org/abs/hep-ph/9311248}{{\tt
  hep-ph/9311248}}].

\bibitem{Jager:2012hd}
B.~Jager, A.~von Manteuffel and S.~Thier, \emph{{Slepton pair production in the
  POWHEG BOX}}, \href{http://dx.doi.org/10.1007/JHEP10(2012)130}{\emph{JHEP}
  {\bf 10} (2012) 130}, [\href{http://arxiv.org/abs/1208.2953}{{\tt
  1208.2953}}].

\bibitem{Bozzi:2007qr}
G.~Bozzi, B.~Fuks and M.~Klasen, \emph{{Threshold Resummation for Slepton-Pair
  Production at Hadron Colliders}},
  \href{http://dx.doi.org/10.1016/j.nuclphysb.2007.03.052}{\emph{Nucl. Phys. B}
  {\bf 777} (2007) 157--181}, [\href{http://arxiv.org/abs/hep-ph/0701202}{{\tt
  hep-ph/0701202}}].

\bibitem{Fuks:2013lya}
B.~Fuks, M.~Klasen, D.~R. Lamprea and M.~Rothering, \emph{{Revisiting slepton
  pair production at the Large Hadron Collider}},
  \href{http://dx.doi.org/10.1007/JHEP01(2014)168}{\emph{JHEP} {\bf 01} (2014)
  168}, [\href{http://arxiv.org/abs/1310.2621}{{\tt 1310.2621}}].

\bibitem{Fiaschi:2018xdm}
J.~Fiaschi and M.~Klasen, \emph{{Slepton pair production at the LHC in NLO+NLL
  with resummation-improved parton densities}},
  \href{http://dx.doi.org/10.1007/JHEP03(2018)094}{\emph{JHEP} {\bf 03} (2018)
  094}, [\href{http://arxiv.org/abs/1801.10357}{{\tt 1801.10357}}].

\bibitem{Fiaschi:2019zgh}
J.~Fiaschi, M.~Klasen and M.~Sunder, \emph{{Slepton pair production with
  aNNLO+NNLL precision}},
  \href{http://dx.doi.org/10.1007/JHEP04(2020)049}{\emph{JHEP} {\bf 04} (2020)
  049}, [\href{http://arxiv.org/abs/1911.02419}{{\tt 1911.02419}}].

\bibitem{Feng:2005gj}
J.~L. Feng, S.~Su and F.~Takayama, \emph{{Lower limit on dark matter production
  at the large hadron collider}},
  \href{http://dx.doi.org/10.1103/PhysRevLett.96.151802}{\emph{Phys. Rev.
  Lett.} {\bf 96} (2006) 151802},
  [\href{http://arxiv.org/abs/hep-ph/0503117}{{\tt hep-ph/0503117}}].

\bibitem{Bai:2010hh}
Y.~Bai, P.~J. Fox and R.~Harnik, \emph{{The Tevatron at the Frontier of Dark
  Matter Direct Detection}},
  \href{http://dx.doi.org/10.1007/JHEP12(2010)048}{\emph{JHEP} {\bf 12} (2010)
  048}, [\href{http://arxiv.org/abs/1005.3797}{{\tt 1005.3797}}].

\bibitem{Berger:1999mc}
E.~L. Berger, M.~Klasen and T.~M.~P. Tait, \emph{{Associated production of
  gauginos and gluinos at hadron colliders in next-to-leading order SUSY QCD}},
  \href{http://dx.doi.org/10.1016/S0370-2693(99)00617-6}{\emph{Phys. Lett. B}
  {\bf 459} (1999) 165--170}, [\href{http://arxiv.org/abs/hep-ph/9902350}{{\tt
  hep-ph/9902350}}].

\bibitem{Berger:2000iu}
E.~L. Berger, M.~Klasen and T.~M.~P. Tait, \emph{{Next-to-leading order SUSY
  QCD predictions for associated production of gauginos and gluinos}},
  \href{http://dx.doi.org/10.1103/PhysRevD.67.099901}{\emph{Phys. Rev. D} {\bf
  62} (2000) 095014}, [\href{http://arxiv.org/abs/hep-ph/0212306}{{\tt
  hep-ph/0212306}}].

\bibitem{Fuks:2016vdc}
B.~Fuks, M.~Klasen and M.~Rothering, \emph{{Soft gluon resummation for
  associated gluino-gaugino production at the LHC}},
  \href{http://dx.doi.org/10.1007/JHEP07(2016)053}{\emph{JHEP} {\bf 07} (2016)
  053}, [\href{http://arxiv.org/abs/1604.01023}{{\tt 1604.01023}}].

\bibitem{Fiaschi:2022odp}
J.~Fiaschi, B.~Fuks, M.~Klasen and A.~Neuwirth, \emph{{Soft gluon resummation
  for associated squark-electroweakino production at the LHC}},
  \href{http://dx.doi.org/10.1007/JHEP06(2022)130}{\emph{JHEP} {\bf 06} (2022)
  130}, [\href{http://arxiv.org/abs/2202.13416}{{\tt 2202.13416}}].

\bibitem{Gavin:2010az}
R.~Gavin, Y.~Li, F.~Petriello and S.~Quackenbush, \emph{{FEWZ 2.0: A code for
  hadronic Z production at next-to-next-to-leading order}},
  \href{http://dx.doi.org/10.1016/j.cpc.2011.06.008}{\emph{Comput. Phys.
  Commun.} {\bf 182} (2011) 2388--2403},
  [\href{http://arxiv.org/abs/1011.3540}{{\tt 1011.3540}}].

\bibitem{Gavin:2012sy}
R.~Gavin, Y.~Li, F.~Petriello and S.~Quackenbush, \emph{{W Physics at the LHC
  with FEWZ 2.1}},
  \href{http://dx.doi.org/10.1016/j.cpc.2012.09.005}{\emph{Comput. Phys.
  Commun.} {\bf 184} (2013) 208--214},
  [\href{http://arxiv.org/abs/1201.5896}{{\tt 1201.5896}}].

\bibitem{Fuks:2007gk}
B.~Fuks, M.~Klasen, F.~Ledroit, Q.~Li and J.~Morel, \emph{{Precision
  predictions for $Z^\prime$ - production at the CERN LHC: QCD matrix elements,
  parton showers, and joint resummation}},
  \href{http://dx.doi.org/10.1016/j.nuclphysb.2008.01.017}{\emph{Nucl. Phys. B}
  {\bf 797} (2008) 322--339}, [\href{http://arxiv.org/abs/0711.0749}{{\tt
  0711.0749}}].

\bibitem{Jezo:2014wra}
T.~Jezo, M.~Klasen, D.~R. Lamprea, F.~Lyonnet and I.~Schienbein, \emph{{NLO+NLL
  limits on $W'$ and $Z'$ gauge boson masses in general extensions of the
  Standard Model}},
  \href{http://dx.doi.org/10.1007/JHEP12(2014)092}{\emph{JHEP} {\bf 12} (2014)
  092}, [\href{http://arxiv.org/abs/1410.4692}{{\tt 1410.4692}}].

\bibitem{Bozzi:2004qq}
G.~Bozzi, B.~Fuks and M.~Klasen, \emph{{Slepton production in polarized hadron
  collisions}},
  \href{http://dx.doi.org/10.1016/j.physletb.2005.01.060}{\emph{Phys. Lett. B}
  {\bf 609} (2005) 339--350}, [\href{http://arxiv.org/abs/hep-ph/0411318}{{\tt
  hep-ph/0411318}}].

\bibitem{Bozzi:2007me}
G.~Bozzi, B.~Fuks, B.~Herrmann and M.~Klasen, \emph{{Squark and gaugino
  hadroproduction and decays in non-minimal flavour violating supersymmetry}},
  \href{http://dx.doi.org/10.1016/j.nuclphysb.2007.05.031}{\emph{Nucl. Phys. B}
  {\bf 787} (2007) 1--54}, [\href{http://arxiv.org/abs/0704.1826}{{\tt
  0704.1826}}].

\bibitem{Debove:2008nr}
J.~Debove, B.~Fuks and M.~Klasen, \emph{{Model-independent analysis of
  gaugino-pair production in polarized and unpolarized hadron collisions}},
  \href{http://dx.doi.org/10.1103/PhysRevD.78.074020}{\emph{Phys. Rev. D} {\bf
  78} (2008) 074020}, [\href{http://arxiv.org/abs/0804.0423}{{\tt 0804.0423}}].

\bibitem{Sterman:1986aj}
G.~F. Sterman, \emph{{Summation of Large Corrections to Short Distance Hadronic
  Cross-Sections}},
  \href{http://dx.doi.org/10.1016/0550-3213(87)90258-6}{\emph{Nucl. Phys. B}
  {\bf 281} (1987) 310--364}.

\bibitem{Catani:1989ne}
S.~Catani and L.~Trentadue, \emph{{Resummation of the QCD Perturbative Series
  for Hard Processes}},
  \href{http://dx.doi.org/10.1016/0550-3213(89)90273-3}{\emph{Nucl. Phys. B}
  {\bf 327} (1989) 323--352}.

\bibitem{Catani:1990rp}
S.~Catani and L.~Trentadue, \emph{{Comment on QCD exponentiation at large x}},
  \href{http://dx.doi.org/10.1016/0550-3213(91)90506-S}{\emph{Nucl. Phys. B}
  {\bf 353} (1991) 183--186}.

\bibitem{Kidonakis:1997gm}
N.~Kidonakis and G.~F. Sterman, \emph{{Resummation for QCD hard scattering}},
  \href{http://dx.doi.org/10.1016/S0550-3213(97)00506-3}{\emph{Nucl. Phys. B}
  {\bf 505} (1997) 321--348}, [\href{http://arxiv.org/abs/hep-ph/9705234}{{\tt
  hep-ph/9705234}}].

\bibitem{Kidonakis:1998bk}
N.~Kidonakis, G.~Oderda and G.~F. Sterman, \emph{{Threshold resummation for
  dijet cross-sections}},
  \href{http://dx.doi.org/10.1016/S0550-3213(98)00243-0}{\emph{Nucl. Phys. B}
  {\bf 525} (1998) 299--332}, [\href{http://arxiv.org/abs/hep-ph/9801268}{{\tt
  hep-ph/9801268}}].

\bibitem{Vogt:2000ci}
A.~Vogt, \emph{{Next-to-next-to-leading logarithmic threshold resummation for
  deep inelastic scattering and the Drell-Yan process}},
  \href{http://dx.doi.org/10.1016/S0370-2693(00)01344-7}{\emph{Phys. Lett. B}
  {\bf 497} (2001) 228--234}, [\href{http://arxiv.org/abs/hep-ph/0010146}{{\tt
  hep-ph/0010146}}].

\bibitem{Kramer:1996iq}
M.~Kramer, E.~Laenen and M.~Spira, \emph{{Soft gluon radiation in Higgs boson
  production at the LHC}},
  \href{http://dx.doi.org/10.1016/S0550-3213(97)00679-2}{\emph{Nucl. Phys. B}
  {\bf 511} (1998) 523--549}, [\href{http://arxiv.org/abs/hep-ph/9611272}{{\tt
  hep-ph/9611272}}].

\bibitem{Catani:2001ic}
S.~Catani, D.~de~Florian and M.~Grazzini, \emph{{Higgs production in hadron
  collisions: Soft and virtual QCD corrections at NNLO}},
  \href{http://dx.doi.org/10.1088/1126-6708/2001/05/025}{\emph{JHEP} {\bf 05}
  (2001) 025}, [\href{http://arxiv.org/abs/hep-ph/0102227}{{\tt
  hep-ph/0102227}}].

\bibitem{Kulesza:2002rh}
A.~Kulesza, G.~F. Sterman and W.~Vogelsang, \emph{{Joint resummation in
  electroweak boson production}},
  \href{http://dx.doi.org/10.1103/PhysRevD.66.014011}{\emph{Phys. Rev. D} {\bf
  66} (2002) 014011}, [\href{http://arxiv.org/abs/hep-ph/0202251}{{\tt
  hep-ph/0202251}}].

\bibitem{Almeida:2009jt}
L.~G. Almeida, G.~F. Sterman and W.~Vogelsang, \emph{{Threshold Resummation for
  Di-hadron Production in Hadronic Collisions}},
  \href{http://dx.doi.org/10.1103/PhysRevD.80.074016}{\emph{Phys. Rev. D} {\bf
  80} (2009) 074016}, [\href{http://arxiv.org/abs/0907.1234}{{\tt 0907.1234}}].

\bibitem{Debove:2009ia}
J.~Debove, B.~Fuks and M.~Klasen, \emph{{Transverse-momentum resummation for
  gaugino-pair production at hadron colliders}},
  \href{http://dx.doi.org/10.1016/j.physletb.2010.04.013}{\emph{Phys. Lett. B}
  {\bf 688} (2010) 208--211}, [\href{http://arxiv.org/abs/0907.1105}{{\tt
  0907.1105}}].

\bibitem{Bozzi:2006fw}
G.~Bozzi, B.~Fuks and M.~Klasen, \emph{{Transverse-momentum resummation for
  slepton-pair production at the CERN LHC}},
  \href{http://dx.doi.org/10.1103/PhysRevD.74.015001}{\emph{Phys. Rev. D} {\bf
  74} (2006) 015001}, [\href{http://arxiv.org/abs/hep-ph/0603074}{{\tt
  hep-ph/0603074}}].

\bibitem{Collins:1981va}
J.~C. Collins and D.~E. Soper, \emph{{Back-To-Back Jets: Fourier Transform from
  B to K-Transverse}},
  \href{http://dx.doi.org/10.1016/0550-3213(82)90453-9}{\emph{Nucl. Phys. B}
  {\bf 197} (1982) 446--476}.

\bibitem{Collins:1981uk}
J.~C. Collins and D.~E. Soper, \emph{{Back-To-Back Jets in QCD}},
  \href{http://dx.doi.org/10.1016/0550-3213(81)90339-4}{\emph{Nucl. Phys. B}
  {\bf 193} (1981) 381}.

\bibitem{Collins:1984kg}
J.~C. Collins, D.~E. Soper and G.~F. Sterman, \emph{{Transverse Momentum
  Distribution in Drell-Yan Pair and W and Z Boson Production}},
  \href{http://dx.doi.org/10.1016/0550-3213(85)90479-1}{\emph{Nucl. Phys. B}
  {\bf 250} (1985) 199--224}.

\bibitem{Debove:2011xj}
J.~Debove, B.~Fuks and M.~Klasen, \emph{{Joint Resummation for Gaugino Pair
  Production at Hadron Colliders}},
  \href{http://dx.doi.org/10.1016/j.nuclphysb.2011.03.015}{\emph{Nucl. Phys. B}
  {\bf 849} (2011) 64--79}, [\href{http://arxiv.org/abs/1102.4422}{{\tt
  1102.4422}}].

\bibitem{Bozzi:2007tea}
G.~Bozzi, B.~Fuks and M.~Klasen, \emph{{Joint resummation for slepton pair
  production at hadron colliders}},
  \href{http://dx.doi.org/10.1016/j.nuclphysb.2007.10.021}{\emph{Nucl. Phys. B}
  {\bf 794} (2008) 46--60}, [\href{http://arxiv.org/abs/0709.3057}{{\tt
  0709.3057}}].

\bibitem{Li:1998is}
H.-n. Li, \emph{{Unification of the k(T) and threshold resummations}},
  \href{http://dx.doi.org/10.1016/S0370-2693(99)00350-0}{\emph{Phys. Lett. B}
  {\bf 454} (1999) 328--334}, [\href{http://arxiv.org/abs/hep-ph/9812363}{{\tt
  hep-ph/9812363}}].

\bibitem{Laenen:2000de}
E.~Laenen, G.~F. Sterman and W.~Vogelsang, \emph{{Higher order QCD corrections
  in prompt photon production}},
  \href{http://dx.doi.org/10.1103/PhysRevLett.84.4296}{\emph{Phys. Rev. Lett.}
  {\bf 84} (2000) 4296--4299}, [\href{http://arxiv.org/abs/hep-ph/0002078}{{\tt
  hep-ph/0002078}}].

\bibitem{Laenen:2000ij}
E.~Laenen, G.~F. Sterman and W.~Vogelsang, \emph{{Recoil and threshold
  corrections in short distance cross-sections}},
  \href{http://dx.doi.org/10.1103/PhysRevD.63.114018}{\emph{Phys. Rev. D} {\bf
  63} (2001) 114018}, [\href{http://arxiv.org/abs/hep-ph/0010080}{{\tt
  hep-ph/0010080}}].

\bibitem{Kinoshita:1962ur}
T.~Kinoshita, \emph{{Mass singularities of Feynman amplitudes}},
  \href{http://dx.doi.org/10.1063/1.1724268}{\emph{J. Math. Phys.} {\bf 3}
  (1962) 650--677}.

\bibitem{Lee:1964is}
T.~D. Lee and M.~Nauenberg, \emph{{Degenerate Systems and Mass Singularities}},
  \href{http://dx.doi.org/10.1103/PhysRev.133.B1549}{\emph{Phys. Rev.} {\bf
  133} (1964) B1549--B1562}.

\bibitem{Collins:1989gx}
J.~C. Collins, D.~E. Soper and G.~F. Sterman, \emph{{Factorization of Hard
  Processes in QCD}},
  \href{http://dx.doi.org/10.1142/9789814503266_0001}{\emph{Adv. Ser. Direct.
  High Energy Phys.} {\bf 5} (1989) 1--91},
  [\href{http://arxiv.org/abs/hep-ph/0409313}{{\tt hep-ph/0409313}}].

\bibitem{Beenakker:2013mva}
W.~Beenakker, T.~Janssen, S.~Lepoeter, M.~Kr\"amer, A.~Kulesza, E.~Laenen
  et~al., \emph{{Towards NNLL resummation: hard matching coefficients for
  squark and gluino hadroproduction}},
  \href{http://dx.doi.org/10.1007/JHEP10(2013)120}{\emph{JHEP} {\bf 10} (2013)
  120}, [\href{http://arxiv.org/abs/1304.6354}{{\tt 1304.6354}}].

\bibitem{Martin:2009iq}
A.~D. Martin, W.~J. Stirling, R.~S. Thorne and G.~Watt, \emph{{Parton
  distributions for the LHC}},
  \href{http://dx.doi.org/10.1140/epjc/s10052-009-1072-5}{\emph{Eur. Phys. J.
  C} {\bf 63} (2009) 189--285}, [\href{http://arxiv.org/abs/0901.0002}{{\tt
  0901.0002}}].

\bibitem{Dulat:2015mca}
S.~Dulat, T.-J. Hou, J.~Gao, M.~Guzzi, J.~Huston, P.~Nadolsky et~al.,
  \emph{{New parton distribution functions from a global analysis of quantum
  chromodynamics}},
  \href{http://dx.doi.org/10.1103/PhysRevD.93.033006}{\emph{Phys. Rev. D} {\bf
  93} (2016) 033006}, [\href{http://arxiv.org/abs/1506.07443}{{\tt
  1506.07443}}].

\bibitem{Hou:2019efy}
T.-J. Hou et~al., \emph{{New CTEQ global analysis of quantum chromodynamics
  with high-precision data from the LHC}},
  \href{http://dx.doi.org/10.1103/PhysRevD.103.014013}{\emph{Phys. Rev. D} {\bf
  103} (2021) 014013}, [\href{http://arxiv.org/abs/1912.10053}{{\tt
  1912.10053}}].

\bibitem{Bailey:2020ooq}
S.~Bailey, T.~Cridge, L.~A. Harland-Lang, A.~D. Martin and R.~S. Thorne,
  \emph{{Parton distributions from LHC, HERA, Tevatron and fixed target data:
  MSHT20 PDFs}},
  \href{http://dx.doi.org/10.1140/epjc/s10052-021-09057-0}{\emph{Eur. Phys. J.
  C} {\bf 81} (2021) 341}, [\href{http://arxiv.org/abs/2012.04684}{{\tt
  2012.04684}}].

\bibitem{NNPDF:2021njg}
{\scshape NNPDF} collaboration, R.~D. Ball et~al., \emph{{The path to proton
  structure at 1\% accuracy}},
  \href{http://dx.doi.org/10.1140/epjc/s10052-022-10328-7}{\emph{Eur. Phys. J.
  C} {\bf 82} (2022) 428}, [\href{http://arxiv.org/abs/2109.02653}{{\tt
  2109.02653}}].

\bibitem{Contopanagos:1993yq}
H.~Contopanagos and G.~F. Sterman, \emph{{Principal value resummation}},
  \href{http://dx.doi.org/10.1016/0550-3213(94)90358-1}{\emph{Nucl. Phys. B}
  {\bf 419} (1994) 77--104}, [\href{http://arxiv.org/abs/hep-ph/9310313}{{\tt
  hep-ph/9310313}}].

\bibitem{Catani:1996yz}
S.~Catani, M.~L. Mangano, P.~Nason and L.~Trentadue, \emph{{The Resummation of
  soft gluons in hadronic collisions}},
  \href{http://dx.doi.org/10.1016/0550-3213(96)00399-9}{\emph{Nucl. Phys. B}
  {\bf 478} (1996) 273--310}, [\href{http://arxiv.org/abs/hep-ph/9604351}{{\tt
  hep-ph/9604351}}].

\bibitem{cmake}
CMake, ``{CMake build tool}.'' \url{http://www.cmake.org/}, 2023.

\bibitem{Gough2009-ji}
B.~Gough, ed., \emph{{GNU} scientific library reference manual}.
\newblock Network Theory, Bristol, England, 3~ed., Jan., 2009.

\bibitem{boost}
Boost, ``{Boost C++ Libraries}.'' \url{http://www.boost.org/}, 2022.

\bibitem{Lepage:1980dq}
G.~P. Lepage, \emph{{VEGAS: An Adaptive Multidimensional Integration Program}},
   CLNS-80/447.

\bibitem{10.2307/43633451}
K.~LEVENBERG, \emph{A method for the solution of certain non-linear problems in
  least squares}, {\emph{Quarterly of Applied Mathematics} {\bf 2} (1944)
  164--168}.

\bibitem{10.2307/2098941}
D.~W. Marquardt, \emph{An algorithm for least-squares estimation of nonlinear
  parameters}, {\emph{Journal of the Society for Industrial and Applied
  Mathematics} {\bf 11} (1963) 431--441}.

\bibitem{slhaea}
SLHAea, ``{SLHAea - containers for SUSY Les Houches Accord input/output}.''
  \url{https://github.com/fthomas/slhaea}, 2022.

\bibitem{Skands:2003cj}
P.~Z. Skands et~al., \emph{{SUSY Les Houches accord: Interfacing SUSY spectrum
  calculators, decay packages, and event generators}},
  \href{http://dx.doi.org/10.1088/1126-6708/2004/07/036}{\emph{JHEP} {\bf 07}
  (2004) 036}, [\href{http://arxiv.org/abs/hep-ph/0311123}{{\tt
  hep-ph/0311123}}].

\bibitem{Allanach:2008qq}
B.~C. Allanach et~al., \emph{{SUSY Les Houches Accord 2}},
  \href{http://dx.doi.org/10.1016/j.cpc.2008.08.004}{\emph{Comput. Phys.
  Commun.} {\bf 180} (2009) 8--25}, [\href{http://arxiv.org/abs/0801.0045}{{\tt
  0801.0045}}].

\bibitem{lhapdf}
A.~Buckley, J.~Ferrando, S.~Lloyd, K.~Nordstr\"om, B.~Page, M.~R\"ufenacht
  et~al., \emph{{LHAPDF6: parton density access in the LHC precision era}},
  \href{http://dx.doi.org/10.1140/epjc/s10052-015-3318-8}{\emph{Eur. Phys. J.
  C} {\bf 75} (2015) 132}, [\href{http://arxiv.org/abs/1412.7420}{{\tt
  1412.7420}}].

\bibitem{Hahn:1998yk}
T.~Hahn and M.~Perez-Victoria, \emph{{Automatized one loop calculations in
  four-dimensions and D-dimensions}},
  \href{http://dx.doi.org/10.1016/S0010-4655(98)00173-8}{\emph{Comput. Phys.
  Commun.} {\bf 118} (1999) 153--165},
  [\href{http://arxiv.org/abs/hep-ph/9807565}{{\tt hep-ph/9807565}}].

\bibitem{ParticleDataGroup:2022pth}
{\scshape Particle Data Group} collaboration, R.~L. Workman et~al.,
  \emph{{Review of Particle Physics}},
  \href{http://dx.doi.org/10.1093/ptep/ptac097}{\emph{PTEP} {\bf 2022} (2022)
  083C01}.

\bibitem{Butterworth:2015oua}
J.~Butterworth et~al., \emph{{PDF4LHC recommendations for LHC Run II}},
  \href{http://dx.doi.org/10.1088/0954-3899/43/2/023001}{\emph{J. Phys. G} {\bf
  43} (2016) 023001}, [\href{http://arxiv.org/abs/1510.03865}{{\tt
  1510.03865}}].

\bibitem{Beenakker:1996ed}
W.~Beenakker, R.~Hopker and M.~Spira, \emph{{PROSPINO: A Program for the
  production of supersymmetric particles in next-to-leading order QCD}},
  \href{http://arxiv.org/abs/hep-ph/9611232}{{\tt hep-ph/9611232}}.

\bibitem{PDF4LHCWorkingGroup:2022cjn}
{\scshape PDF4LHC Working Group} collaboration, R.~D. Ball et~al., \emph{{The
  PDF4LHC21 combination of global PDF fits for the LHC Run III}},
  \href{http://dx.doi.org/10.1088/1361-6471/ac7216}{\emph{J. Phys. G} {\bf 49}
  (2022) 080501}, [\href{http://arxiv.org/abs/2203.05506}{{\tt 2203.05506}}].

\bibitem{Fuks:2017rio}
B.~Fuks, M.~Klasen, S.~Schmiemann and M.~Sunder, \emph{{Realistic simplified
  gaugino-higgsino models in the MSSM}},
  \href{http://dx.doi.org/10.1140/epjc/s10052-018-5695-2}{\emph{Eur. Phys. J.
  C} {\bf 78} (2018) 209}, [\href{http://arxiv.org/abs/1710.09941}{{\tt
  1710.09941}}].

\bibitem{CMS:2018kag}
{\scshape CMS} collaboration, A.~M. Sirunyan et~al., \emph{{Search for new
  physics in events with two soft oppositely charged leptons and missing
  transverse momentum in proton-proton collisions at $\sqrt{s}=$ 13 TeV}},
  \href{http://dx.doi.org/10.1016/j.physletb.2018.05.062}{\emph{Phys. Lett. B}
  {\bf 782} (2018) 440--467}, [\href{http://arxiv.org/abs/1801.01846}{{\tt
  1801.01846}}].

\bibitem{ATLAS:2019lng}
{\scshape ATLAS} collaboration, G.~Aad et~al., \emph{{Searches for electroweak
  production of supersymmetric particles with compressed mass spectra in
  $\sqrt{s}=$ 13 TeV $pp$ collisions with the ATLAS detector}},
  \href{http://dx.doi.org/10.1103/PhysRevD.101.052005}{\emph{Phys. Rev. D} {\bf
  101} (2020) 052005}, [\href{http://arxiv.org/abs/1911.12606}{{\tt
  1911.12606}}].

\bibitem{ATLAS:2021moa}
{\scshape ATLAS} collaboration, G.~Aad et~al., \emph{{Search for
  chargino\textendash{}neutralino pair production in final states with three
  leptons and missing transverse momentum in $\sqrt{s} = 13$~TeV pp collisions
  with the ATLAS detector}},
  \href{http://dx.doi.org/10.1140/epjc/s10052-021-09749-7}{\emph{Eur. Phys. J.
  C} {\bf 81} (2021) 1118}, [\href{http://arxiv.org/abs/2106.01676}{{\tt
  2106.01676}}].

\bibitem{NNPDF:2017mvq}
{\scshape NNPDF} collaboration, R.~D. Ball et~al., \emph{{Parton distributions
  from high-precision collider data}},
  \href{http://dx.doi.org/10.1140/epjc/s10052-017-5199-5}{\emph{Eur. Phys. J.
  C} {\bf 77} (2017) 663}, [\href{http://arxiv.org/abs/1706.00428}{{\tt
  1706.00428}}].

\bibitem{Ball:2022qtp}
R.~D. Ball, A.~Candido, S.~Forte, F.~Hekhorn, E.~R. Nocera, J.~Rojo et~al.,
  \emph{{Parton distributions and new physics searches: the
  Drell\textendash{}Yan forward\textendash{}backward asymmetry as a case
  study}}, \href{http://dx.doi.org/10.1140/epjc/s10052-022-11133-y}{\emph{Eur.
  Phys. J. C} {\bf 82} (2022) 1160},
  [\href{http://arxiv.org/abs/2209.08115}{{\tt 2209.08115}}].

\bibitem{Fiaschi:2022wgl}
J.~Fiaschi, F.~Giuli, F.~Hautmann, S.~Moch and S.~Moretti, \emph{{Z'-boson
  dilepton searches and the high-x quark density}},
  \href{http://dx.doi.org/10.1016/j.physletb.2023.137915}{\emph{Phys. Lett. B}
  {\bf 841} (2023) 137915}, [\href{http://arxiv.org/abs/2211.06188}{{\tt
  2211.06188}}].

\bibitem{Alekhin:2017kpj}
S.~Alekhin, J.~Bl\"umlein, S.~Moch and R.~Placakyte, \emph{{Parton distribution
  functions, $\alpha_s$, and heavy-quark masses for LHC Run II}},
  \href{http://dx.doi.org/10.1103/PhysRevD.96.014011}{\emph{Phys. Rev. D} {\bf
  96} (2017) 014011}, [\href{http://arxiv.org/abs/1701.05838}{{\tt
  1701.05838}}].

\bibitem{Tarasov:1980au}
O.~V. Tarasov, A.~A. Vladimirov and A.~Y. Zharkov, \emph{{The Gell-Mann-Low
  Function of QCD in the Three Loop Approximation}},
  \href{http://dx.doi.org/10.1016/0370-2693(80)90358-5}{\emph{Phys. Lett. B}
  {\bf 93} (1980) 429--432}.

\bibitem{Larin:1993tp}
S.~A. Larin and J.~A.~M. Vermaseren, \emph{{The Three loop QCD Beta function
  and anomalous dimensions}},
  \href{http://dx.doi.org/10.1016/0370-2693(93)91441-O}{\emph{Phys. Lett. B}
  {\bf 303} (1993) 334--336}, [\href{http://arxiv.org/abs/hep-ph/9302208}{{\tt
  hep-ph/9302208}}].

\bibitem{Catani:1996vz}
S.~Catani and M.~H. Seymour, \emph{{A General algorithm for calculating jet
  cross-sections in NLO QCD}},
  \href{http://dx.doi.org/10.1016/S0550-3213(96)00589-5}{\emph{Nucl. Phys. B}
  {\bf 485} (1997) 291--419}, [\href{http://arxiv.org/abs/hep-ph/9605323}{{\tt
  hep-ph/9605323}}].

\bibitem{Catani:2002hc}
S.~Catani, S.~Dittmaier, M.~H. Seymour and Z.~Trocsanyi, \emph{{The Dipole
  formalism for next-to-leading order QCD calculations with massive partons}},
  \href{http://dx.doi.org/10.1016/S0550-3213(02)00098-6}{\emph{Nucl. Phys. B}
  {\bf 627} (2002) 189--265}, [\href{http://arxiv.org/abs/hep-ph/0201036}{{\tt
  hep-ph/0201036}}].

\bibitem{Catani:2000ef}
S.~Catani, S.~Dittmaier and Z.~Trocsanyi, \emph{{One loop singular behavior of
  QCD and SUSY QCD amplitudes with massive partons}},
  \href{http://dx.doi.org/10.1016/S0370-2693(01)00065-X}{\emph{Phys. Lett. B}
  {\bf 500} (2001) 149--160}, [\href{http://arxiv.org/abs/hep-ph/0011222}{{\tt
  hep-ph/0011222}}].

\bibitem{Gavin:2013kga}
R.~Gavin, C.~Hangst, M.~Kr\"amer, M.~M\"uhlleitner, M.~Pellen, E.~Popenda
  et~al., \emph{{Matching Squark Pair Production at NLO with Parton Showers}},
  \href{http://dx.doi.org/10.1007/JHEP10(2013)187}{\emph{JHEP} {\bf 10} (2013)
  187}, [\href{http://arxiv.org/abs/1305.4061}{{\tt 1305.4061}}].

\end{thebibliography}\endgroup

\end{document}